\documentclass[a4paper,11pt]{article}
\pdfoutput=1 

\usepackage{jheppub} 
\usepackage{amsmath,amssymb}
\usepackage{graphicx}
\usepackage[percent]{overpic}
\usepackage[dvipsnames]{xcolor}
\usepackage{bm}
\usepackage{dashrule}

\def\be{\begin{equation}}
\def\ee{\end{equation}}
\def\bea{\begin{eqnarray}}

\def\bal{\begin{align}}
\def\eal{\end{align}}

\def\Tr{{\rm{Tr}}}
\def\MCG{{\rm{MCG}}}

\newcommand\CN{\mathcal{N}}
\newcommand\CG{\mathcal{G}}
\newcommand\CM{\mathcal{M}}
\newcommand\CB{\mathcal{B}}
\newcommand\CK{\mathcal{K}}

\newcommand\bCG{\overline{\mathcal{G}}}
\newcommand\bc{\mathbf{c}}
\newcommand\fOmega{\overline{\underline{\Omega}}}

\renewcommand{\(}{\left(}
\renewcommand{\)}{\right)}

\newcommand{\CS}{\mathcal{S}}

\newcommand{\CW}{\mathcal{W}}

\newcommand{\IC}{\mathbb{C}}
\newcommand{\IZ}{\mathbb{Z}}

\newcommand{\IR}{\rm IR}
\newcommand{\UV}{\rm UV}

\title{S duality and Framed BPS States via BPS Graphs}

\author[a]{Dongmin Gang,}
\author[b]{Pietro Longhi,}
\author[c]{and Masahito Yamazaki}

\affiliation[a]{Center for Theoretical Physics, Seoul National University, Seoul 08826, Korea}
\affiliation[b]{Department of Physics and Astronomy, Uppsala University, Box 516, SE-75120 Uppsala, Sweden}
\affiliation[c]{Kavli Institute for the Physics and Mathematics of the Universe (WPI), University of Tokyo, Chiba 277-8583, Japan}

\emailAdd{{arima275@snu.ac.kr}}
\emailAdd{pietro.longhi@physics.uu.se}
\emailAdd{masahito.yamazaki@ipmu.jp}

\preprint{UUITP-42/17, IPMU-17-0152}

\abstract{
We study a realization of S dualities of four-dimensional $\mathcal{N}=2$ class $\mathcal{S}$ theories based on BPS graphs.
S duality transformations of the UV curve are explicitly expressed as a sequence 
of topological transitions of the graph, and translated into cluster transformations of the algebra 
associated to the dual BPS quiver. 
Our construction applies to generic class $\mathcal{S}$ theories, including those with non-maximal flavor symmetry, generalizing 
previous results based on higher triangulations.
We study the the action of S duality on UV line operators, and show that it matches precisely 
with the mapping class group, by a careful analysis of framed wall-crossing.
We comment on the implications of our results for the computation of three-manifold 
invariants via cluster partition functions.
}

\begin{document} 

\maketitle
\flushbottom

\section{Introduction and discussion}

Twisted compactifications of the six-dimensional $(2,0)$ theory provide a valuable laboratory to explore  various nonperturbative aspects of quantum field theories.
The archetypal example is perhaps the identification of generalized S-duality for four-dimensional $\CN=2$ dualities
with the mapping class groups of Riemann surfaces \cite{Gaiotto:2009we}, whose implications have reverberated into the studies of partition functions, BPS line operators, and three-dimensional compactifications, to name a few \cite{Alday:2009aq, Drukker:2009tz, Terashima:2011qi, Dimofte:2011ju}.

In this paper we study S-duality from a low-energy perspective, by going to the Coulomb branch of a four-dimensional class $\CS$ theory and asking how dualities act on the BPS spectrum.
One advantage of the IR setting is that it gives us a good control on certain observables, allowing us to perform explicit computations.
On the other hand, by going the Coulomb branch we apparently lose connection with the UV description of the theory, including the geometric interpretation of S-duality as a mapping class group transformation of the UV curve $C$.
The resolution of this issue comes from studying protected quantities, such as BPS states and their generalizations in presence of line and surface defects, which retain some information about the UV physics.
For example, it is known that framed BPS states encode enough information to characterize the algebra of UV line operators \cite{Gaiotto:2010be, Cordova:2013bza}.
There is in fact a precise map between low-energy line operators and their UV counterparts, developed on spectral networks in \cite{Gaiotto:2012rg}, where it is called ``nonabelianization map''.
This result provides the conceptual foundation for our approach, explaining how operations on certain low-energy observables can encode UV dualities.

\begin{figure}[htbp]
\begin{center}
\includegraphics[width=0.75\textwidth]{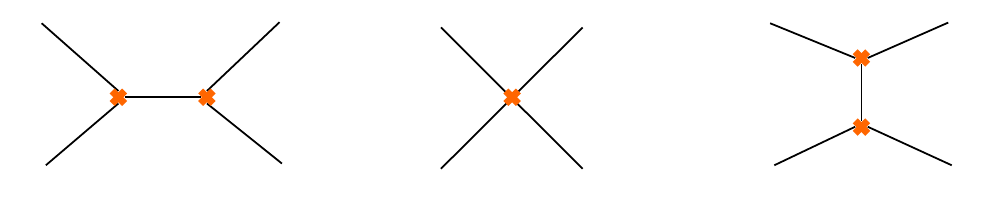}\\
\hdashrule[0.5ex]{0.75\textwidth}{0.4pt}{1pt}
\includegraphics[width=0.75\textwidth]{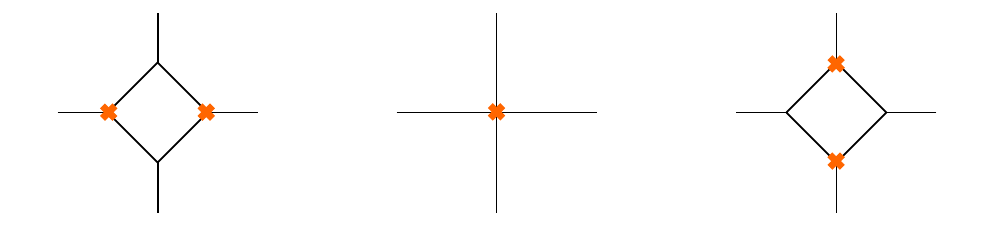}
\caption{The \emph{flip} move is shown on top, the \emph{cootie} move at the bottom. Vertices of BPS graphs come in two types: \emph{branch points} are marked by a yellow cross, \emph{joints} are unmarked. 
}
\label{fig:flip-cootie}
\end{center}
\end{figure}

We derive a low-energy characterization of the mapping class group $\MCG(C)$ of $C$ based on \emph{BPS graphs}, which are graphs embedded in $C$. 
BPS graphs arise from a degenerate limit of spectral networks at points in the Coulomb branch where the phases of central charges are maximally aligned, and they encode both the BPS quiver and all the BPS spectra of a theory \cite{Gabella:2017hpz, Longhi:2016wtv}.
There is a whole equivalence class of BPS graphs associated to a given theory, generated by two basic moves shown in Figure \ref{fig:flip-cootie}. 
The topology of a BPS graph $\CG$ is characterized by the type of each vertex (see Figure \ref{fig:flip-cootie}), by the adjacency matrix of its edges, and by a cyclic ordering of edges at each vertex. 
We identify a mapping class group transformation of $C$ with a sequence of elementary moves $\kappa$ which takes $\CG$ to a new graph $\CG'$ with the same topology.
In general $\CG$ and $\CG'$ need not wrap $C$ in the same way, instead they will wrap the UV curve in ways related by an element  $g_\kappa\in\MCG(C)$.
There is a natural map $\kappa\to g_\kappa$, that arises from using the BPS graph to characterize the action of $\kappa$ on $H_1(C,\IZ)$.

Our construction of $\MCG(C)$ moreover has a natural interpretation in the context of cluster algebras, because any BPS graph $\CG$ is dual to a BPS quiver $Q$  \cite{Gabella:2017hpz}---any sequence $\kappa$ translates naturally into a sequence of ``quiver mutations'' of the quiver $Q$,
and into associated sequential changes of cluster variables.
The existence of a relation between mapping class groups and cluster algebras has been known for some time \cite{FockGoncharovHigher,FST1}, however its explicit characterization was limited to Riemann surfaces decorated by ``full'' punctures, \emph{i.e.} punctures encoding maximal flavor symmetry.
Our construction via BPS graphs agrees with these previous results, but further extends to Riemann surfaces with more general types of punctures. 
The only fundamental requirement for our construction is the existence of a BPS graph for the theory.

The sequence of moves $\kappa$ that generates a mapping class group transformation can be interpreted as a path in the moduli space of BPS graphs, along which some edges shrink to a point and subsequently grow again.
However, finding such a region in the physical moduli space of a theory, which may include both Coulomb and UV parameters, can be challenging.
This is both one of the main shortcomings and advantages of BPS graphs. On the one hand, it is difficult to find the region of moduli space where a BPS graph arises from spectral networks.
On the other hand, the BPS graph of a theory is typically so much simpler than the generic network.
Based on the experience gained from some examples, we can then make educated guesses for the 
BPS graphs of more complicated ones.
In this paper we follow this approach, and construct the BPS graphs of $A_{N-1}$ theories of class $\CS$ defined by a torus with a ``simple'' puncture using an ansatz inspired by methods of \cite{Gabella:2017hpz}.
We thus obtain candidates for the BPS quivers of the so-called $SU(N)$ $\CN=2^*$ theories, and check that the map from $\MCG(C)$ to the cluster algebra is a homomorphism.

A key property of our map from $\MCG(C)$ to cluster algebras is its that it reproduces the action of S duality on UV line operators \cite{Kapustin:2005py, Drukker:2009tz, Alday:2009fs, Drukker:2009id}. 
The duality relating BPS graphs to quivers leads to an identification between VEVs of IR line operators and cluster coordinates.\footnote{More precisely, there is a relation between the two, but they are not quite identical. For example, they have different transformation properties as we explain in the main body of the paper.}
Through this relation $\kappa$ translates into a transformation of the IR line operators of the theory, which can be further mapped into an action on the UV line operators, thanks to a relation between the two sets of observables characterized by {framed BPS states} \cite{Gaiotto:2010be}.
Since a honest BPS graph originates from a spectral network $\bCG$, we can always use the latter to compute the spectrum of {framed BPS states}, via the nonabelianization map \cite{Gaiotto:2012rg}.
In this paper we focus on UV line operators $L_\wp$ labeled by closed paths $\wp\subset C$, whose VEVs are identified with traces of holonomies of a flat connection on $C$.
A detailed analysis shows that acting with $\kappa$ on the underlying spectral network $\bCG$ induces a sequence of framed wall-crossing phenomena. 
These jumps of framed BPS states translate into a transformation $L_\wp\to L_{g_\kappa(\wp)}$, that  maps a UV line operator wrapping $\wp$ into a new one wrapping $g_\kappa(\wp)$.
We give a general derivation of this property, which provides a strong check that our construction of S duality via cluster algebras acts in the expected way on UV line operators. 
We also provide explicit checks of this property for two simple cases: the $A_1$ theories defined by a once-punctured torus and the four-punctured sphere, as a byproduct we illustrate for the first time computations of VEVs of UV line operators (and framed BPS states) using BPS graphs.\footnote{More precisely, the computation always relies on the data of the underlying spectral network, and we recover previous results obtained by different, but related, techniques. The point is that it is often easier to work with the BPS graph than producing the generic spectral network, this can make computations more accessible depending on the context.}

The main novelty of the relation between S duality and cluster algebras uncovered in this paper is the fact that it extends to Riemann surfaces with generic types of punctures. This generalizes previous relations based on (higher) ideal triangulations \cite{FockGoncharovHigher}, and relies only on the existence of the BPS graph.

One reason why such a generalization is interesting comes from applications to the study of three-manifold invariants computed by cluster partition functions \cite{Terashima:2013fg, Gang:2015wya}.
A Riemann surface $C$ together with an element $\kappa$ of its mapping class group define a three-manifold known as a mapping torus $M = C\times_\kappa S^1$. Under suitable conditions  on $\kappa$, $M$ is a link complement in $S^3$.
The 3d--3d correspondence \cite{Terashima:2011qi, Dimofte:2011ju, Dimofte:2011py, Lee:2013ida,Yagi:2013fda, Cordova:2013cea} associates a 3d $\CN=2$ theory $T[M]$ to a three-manifold, and in the case when $M$ is a mapping torus, there is a natural quiver $Q$ associated with the theory $T[M]$.
$Q$ coincides in fact with the BPS quiver of the 4d $\CN=2$ class $\CS$ theory defined by the Riemann surface $C$, which is dual to the BPS graph. 
The cluster partition function is a versatile computational tool for studying $SL(N)$ Chern-Simons partition functions on $M$, whose definition relies precisely on the cluster algebra representation of $\MCG(C)$.

The results of this paper provide the necessary ingredients to compute cluster partition functions for mapping tori fibered by Riemann surfaces with non-maximal punctures.
From the viewpoint of Chern-Simons theory, the types of punctures on $C$ enter the definition of the path integral, as they specify the conjugacy class for the holonomy around a cycle along the link (\emph{e.g.} the longitudinal cycle), therefore they characterize the types of topological invariants encoded by the partition function. 
In particular, for the Chern-Simons path integral on a knot complement with non-generic holonomy along the knot, it is important to sum over saddle points that include several conjugacy classes of holonomies with fixed eigenvalues \cite{Romo:2017ruv}.
Very little is known about these invariants in the case of non-maximal punctures, therefore it will be very interesting to construct them using the cluster algebra realization of mapping class groups developed in this paper; this is the subject of our upcoming work \cite{future}. 
For the case of a torus with a simple puncture $[N-1,1]$ and an element $\varphi$ of the mapping class group $SL(2, \mathbb{Z})$ (see Section \ref{sec:N-torus}),
the resulting cluster partition function can also be compared with the partition function of  $\textrm{Tr}(T[SU(N), \varphi])$---this is a theory obtained by gauging the diagonal $SU(N)$ subgroup of the duality domain wall non-Abelian gauge theory $T[SU(N), \varphi]$,
studied in \cite{Terashima:2011qi}. Schematically, we have
\begin{align}
Z^{\textrm{cluster partition function}}_{\textrm{Tr}(\varphi)}=Z^{\textrm{3d $\mathcal{N}=2$ theory}}_{\textrm{Tr}(T[SU(N), \varphi])}
\end{align}
and this will provide one of the most stringent checks of the BPS graphs and their S duality action proposed in this paper \cite{future}.

\bigskip

This paper is organized as follows. Section \ref{sec:mcg} contains the characterization of S-duality groups (\emph{i.e.} mapping class groups) based on BPS graphs, and the map to cluster algebras.
In Section \ref{sec:line-operators} we analyze the framed wall-crossing that is induced by the action of the mapping class group, and show that it reproduces the expected action of S duality on UV line operators. Section \ref{sec:evidence} contains examples of our construction together with various checks.

\section{Mapping class group from BPS graphs}
\label{sec:mcg}

The aim of this section is to explain how a representation of the mapping class group of a Riemann surface can be derived using BPS graphs.

\subsection{BPS graphs}
\label{subsec:bps-graphs}

A BPS graph $\CG$ is a graph embedded in the UV curve $C$ of a class $\CS$ theory, and arises as a  maximally degenerate spectral network \cite{Gabella:2017hpz}
(see also the related works \cite{Hollands:2013qza, Longhi:2016wtv, Gabella:2017hxq}).
The shape of the spectral network reflects the geometry of the Seiberg-Witten curve $\Sigma$, which is presented as a $N$-sheeted ramified covering of $C$, and depends on a choice of Coulomb vacuum and UV moduli.
$\CG$ appears at a special locus within the moduli space, a.k.a. the \emph{Roman locus}, where central charges of BPS particles all have the same phase $\vartheta_c$ (anti-particles have phase $\vartheta_c+\pi$).
An important feature of BPS graphs is that they are quite simple, compared to the generic spectral network of a theory.
In fact, it is sometimes possible to deduce or guess the BPS graph of a theory without plotting the actual spectral network.\footnote{This option is important because it can be in practice challenging to find the Roman locus within the moduli space.}
In this paper we will mostly take this route, \emph{i.e.} we will adopt an ansatz for the BPS graph of a theory, and will assume that it arises from a honest spectral network at some point of the moduli space. 
The validity of this assumption is crucial for some of our considerations, and our ansatze for BPS graphs will be supported by several types of checks.

\begin{figure}[htbp]
\begin{center}
\includegraphics[width=0.4\textwidth]{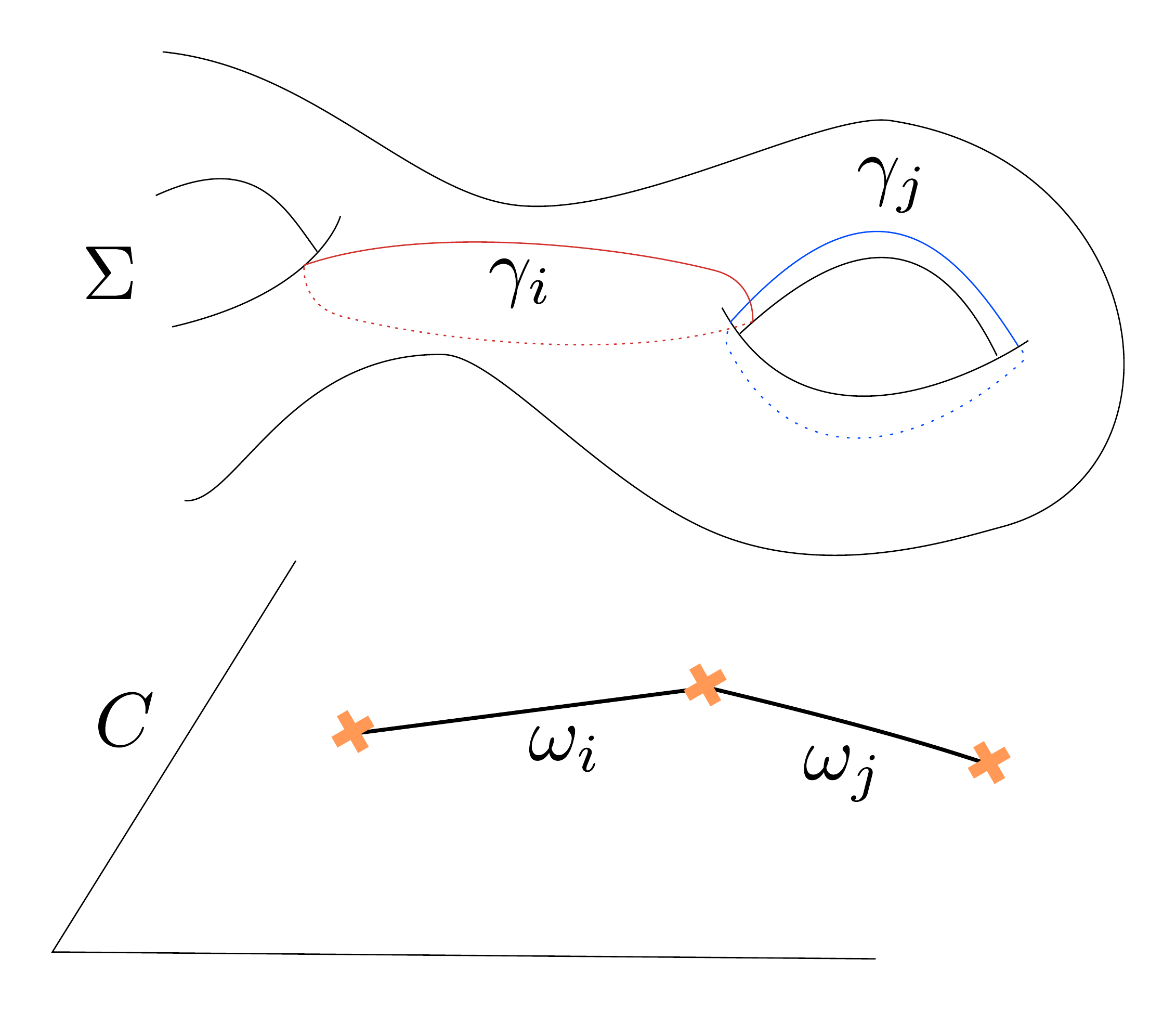}
\caption{Two elementary webs $\omega_i, \omega_j$ made of a single edge each, and their lifts to closed cycles on $\Sigma$ ($\gamma_i$ and $\gamma_j$).
}
\label{fig:h-map}
\end{center}
\end{figure}

For our purposes, a BPS graph consists of edges $e_i \in E(\CG)$ attached together at vertices $v \in V(\CG)$, which come in two types: \emph{branch points} or \emph{joints}. 
The topological data defining $\CG$ includes an embedding in $C$ up to homotopy, the adjacency matrix of its edges, and a cyclic ordering $\sigma_v$ of edges at each vertex $v$.
$\CG$ is naturally divided into smaller connected sets of edges, called \emph{elementary webs}, 
defined as the connected components of $\CG$ after cutting the graph at the branch points (indicated by crosses in figures).\footnote{This definition of elementary webs is not entirely accurate, but will be appropriate for the BPS graphs studied in this paper. For more details see \cite{Gabella:2017hpz}.}
An elementary web $\omega$ may consist of a single edge, or of several edges connected together at joints.
A BPS graph comes equipped with a map from the set of elementary webs to homology cycles on~$\Sigma$
\be\label{eq:h-map}
	h \, : \, \omega \mapsto \gamma  \in H_1(\Sigma,\IZ)\,.
\ee
This map is inherited from the spectral network: $\gamma$ is the class of a cycle arising as a lift of $\omega$ from $C$ to $\Sigma$, see Figure \ref{fig:h-map}.
The cyclic ordering $\sigma_v$ of the edges at each vertex $v$ encodes the intersection pairing $\langle \gamma,\gamma'\rangle$ of cycles associated to webs $\omega,\omega'$ that meet at $v$.

\begin{figure}[htbp]
\begin{center}
\includegraphics[width=0.6\textwidth]{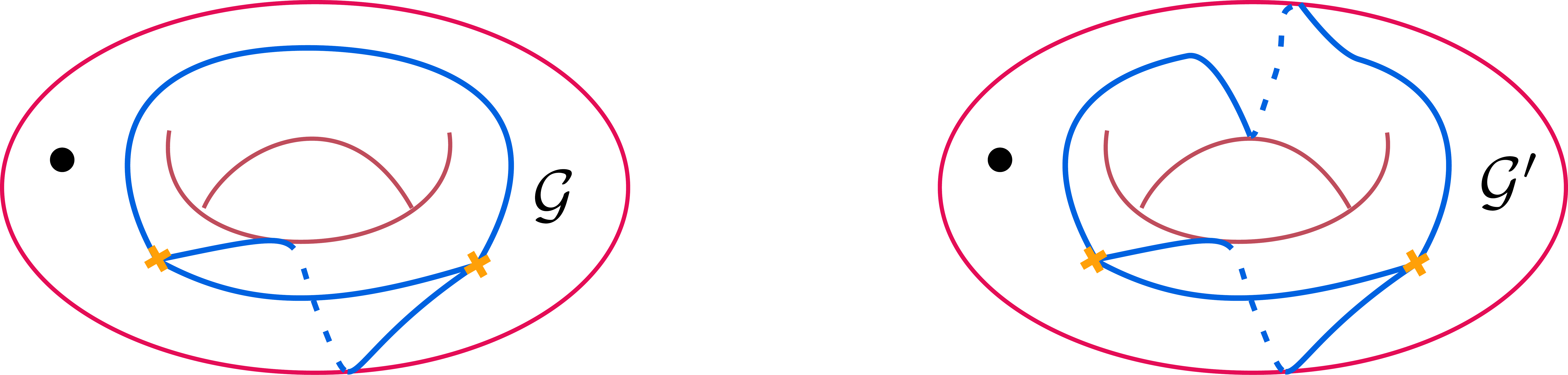}
\caption{Two BPS graphs on the punctured torus, wrapping it in two different ways related by a mapping class group transformation (a Dehn twist).}
\label{fig:mcg-graph}
\end{center}
\end{figure}

Two BPS graphs with the same topological data are regarded as the same graph $\CG$.
On the other hand, if the embeddings of $\CG,\CG'$ in $C$ are not homotopy equivalent, we say that $\CG\simeq\CG'$ are {equivalent} \emph{as abstract graphs} if there is a 1-1 map $f$ which takes $V(\CG) \to V(\CG')$ and $E(\CG) \to E(\CG')$ and respects $\sigma_v$ at each vertex $v$
\be
	f(\sigma_v) = \sigma_{f(v)} \,.
\ee
This equivalence relation implies that $\CG,\CG'$ have the same adjacency matrix and same cyclic ordering of edges at each vertex, regardless of how $\CG$ and $\CG'$ are placed on $C$. 
An example of equivalent graphs with different embeddings is shown in Figure \ref{fig:mcg-graph}.

\subsection{Mapping class group}
\label{subsec:mcg}

BPS graphs come in families generated by two topological transitions: the \emph{flip} and the \emph{cootie} moves shown in Figure \ref{fig:flip-cootie}.
We will be interested in pairs of equivalent graphs $(\CG,\CG')$, both embedded in the same Riemann surface $C$, that are related by a sequence of flips and cootie moves.
Let $\kappa_s$ be a sequence of such moves, which takes $\CG$ to an {equivalent} graph $\CG'$, up to a relabeling of the edges $\kappa_r$
\be
	\CG' \equiv \kappa_r \circ\kappa_s \(\CG\) \quad \simeq \quad \CG\,.
\ee
We define $\kappa_r$ as the relabeling $e_i\to e'_{i'}$ such that $f(e_i) = e'_i$, thereby fixing $f$ in terms of $\kappa_r\circ\kappa_s$ for the rest of the paper.
$\CG'$ may wrap $C$ in a different way from $\CG$, and through the map $f$ relating them we can define a mapping class group transformation for $C$, under certain conditions which we now specify.

Choose a basis for $H_1(C,\IZ)$, together with a representative of each basis element made of an oriented closed chain of edges $(e_{i_1},\dots,e_{i_k})$ of  $\CG$. 
The map $f$ takes this to a new chain $(e'_{i_1},\dots,e'_{i_k})\subset E(\CG')$ which is also closed, since $\CG\simeq\CG'$. The new chain defines a new element of $H_1(C,\IZ)$.
In order to define a honest mapping class group transformation, we require that any two chains of edges in $E(\CG)$ in the same homology class
\be
	\left[(e_{i_1},\dots,e_{i_k})\right]  = \left[ (e_{j_1},\dots,e_{j_\ell})\right]  
\ee
must be mapped to two chains in $E(\CG')$ which are also in the same homology class.
\be
	\left[(e'_{i_1},\dots,e'_{i_k})\right]  = \left[ (e'_{j_1},\dots,e'_{j_\ell})\right]  \, .
\ee
If this condition is satisfied, $f$ acts as an endomorphism of $H_1(C,\IZ)$, and can be identified with a mapping class group transformation.
We will henceforth restrict our attention to sequences $\kappa_s$ composed with relabelings $\kappa_r$ which lead to equivalent graphs $\CG,\CG'$ related by a map $f$ satisfying this requirement.
Since we fixed $f$ in terms of $\kappa_r\circ\kappa_s$ we will leave $f$ implicit in the following, and simply refer to the operation $\kappa = \kappa_r\circ\kappa_s$ as a mapping class group transformation.

\begin{figure}[htbp]
\begin{center}
\includegraphics[width=0.99\textwidth]{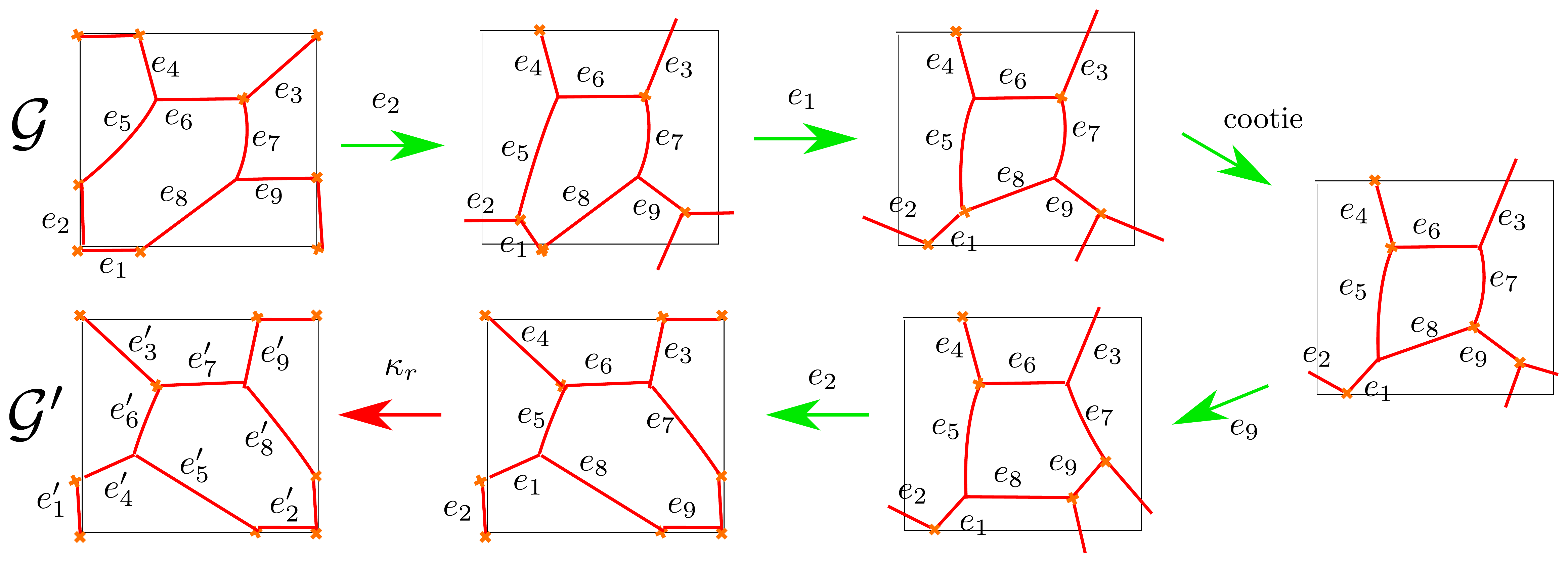}
\caption{The BPS graph $\CG$ is placed on a torus with one puncture, whose fundamental domain is depicted as a square, with the puncture placed at the corners. 
$\CG$ undergoes a sequence of moves $\kappa_s$ (green arrows) followed by the relabeling $\kappa_r$.}
\label{fig:full-mcg-example}
\end{center}
\end{figure}

This construction can be made quite explicit: given $\kappa$ satisfying the consistency conditions, it can be identified with a specific element $g_\kappa$ of the mapping class group.
For example let us consider the sequence of moves on the BPS graph shown in Figure \ref{fig:full-mcg-example}.
$\kappa_s$ consists of the following moves: flip  $e_2$, flip $e_1$, cootie on $(e_5, e_6, e_7, e_8)$, flip $e_9$, flip $e_2$.
The graph obtained after applying $\kappa_s$ has a $\IZ_3$ symmetry, so there are three inequivalent relabelings $\kappa_r,\kappa_r',\kappa_r''$ that can be used to produce a new BPS graph $\CG'$ equivalent to the original one. 
We choose $\kappa_r$ as 
\be\label{eq:21-Smove-relabel}
	\kappa_r : \left\{\begin{array}{l}
		e_2\to e_1\to e_4 \to e_3\to e_9 \to e_2 \\
		e_5\to e_6 \to e_7 \to e_8 \to e_5
		\end{array}\right. \,,
\ee
meaning that $\kappa_r(e_2) = e_1'$, et cetera.
Next we choose generators for $H_1(C,\IZ)$ as the homology classes of the following sequences of edges of $\CG$
\be\label{eq:21AB}
	A: [(e_1, e_8, e_9, e_2)] \,,\qquad B: [(e_2, e_5, e_4, e_1)]\,.
\ee
The orientation of a cycle is understood as \emph{left to right}, when reading each sequence. 
As the moves $\kappa_s$ are applied to $\CG$, we keep track of these edges, and finally apply the relabeling $\kappa_r$. This leads to a new pair of cycles, defined by the new sequences of edges identified by the equivalence $\CG\simeq \CG'$
\be
	A': [(e'_1, e'_8, e'_9, e'_2)]\,,\qquad B': [(e'_2, e'_5, e'_4, e'_1)]\,.
\ee
As homology classes, they are related to the original ones by
\be\label{eq:21-S-hom-action}
	A' = B \,,\qquad B' = -A\,,
\ee
therefore we identify $\kappa$ with the following generator of $\MCG(C)\simeq SL(2,\IZ)$
\be
	g_\kappa = S^{-1} =\left(\begin{array}{cc}  0 & -1 \\ 1 & 0  \end{array}\right)\,.
\ee
More generally, a transformation which takes $A,B$ to
$A' = d A + c B$, $B'=b A + a B$ corresponds to the following element of $SL(2,\IZ)$
\be
	g_\kappa = \left(\begin{array}{cc}  a & b \\ c & d  \end{array}\right)\,.
\ee
This representation of $g_\kappa$ obviously depends on the choice of homology basis on $C$, and different choices should be related by conjugation. 
Therefore upon fixing a basis, any two sequences $\kappa,\kappa'$ must satisfy
\be
	g_{\kappa'}g_\kappa = g_{\kappa'\circ\kappa}\,.
\ee

\subsection{Map to cluster algebra}
\label{subsec:cluster-algebra}

\begin{figure}[htbp]
\begin{center}
\includegraphics[width=0.4\textwidth]{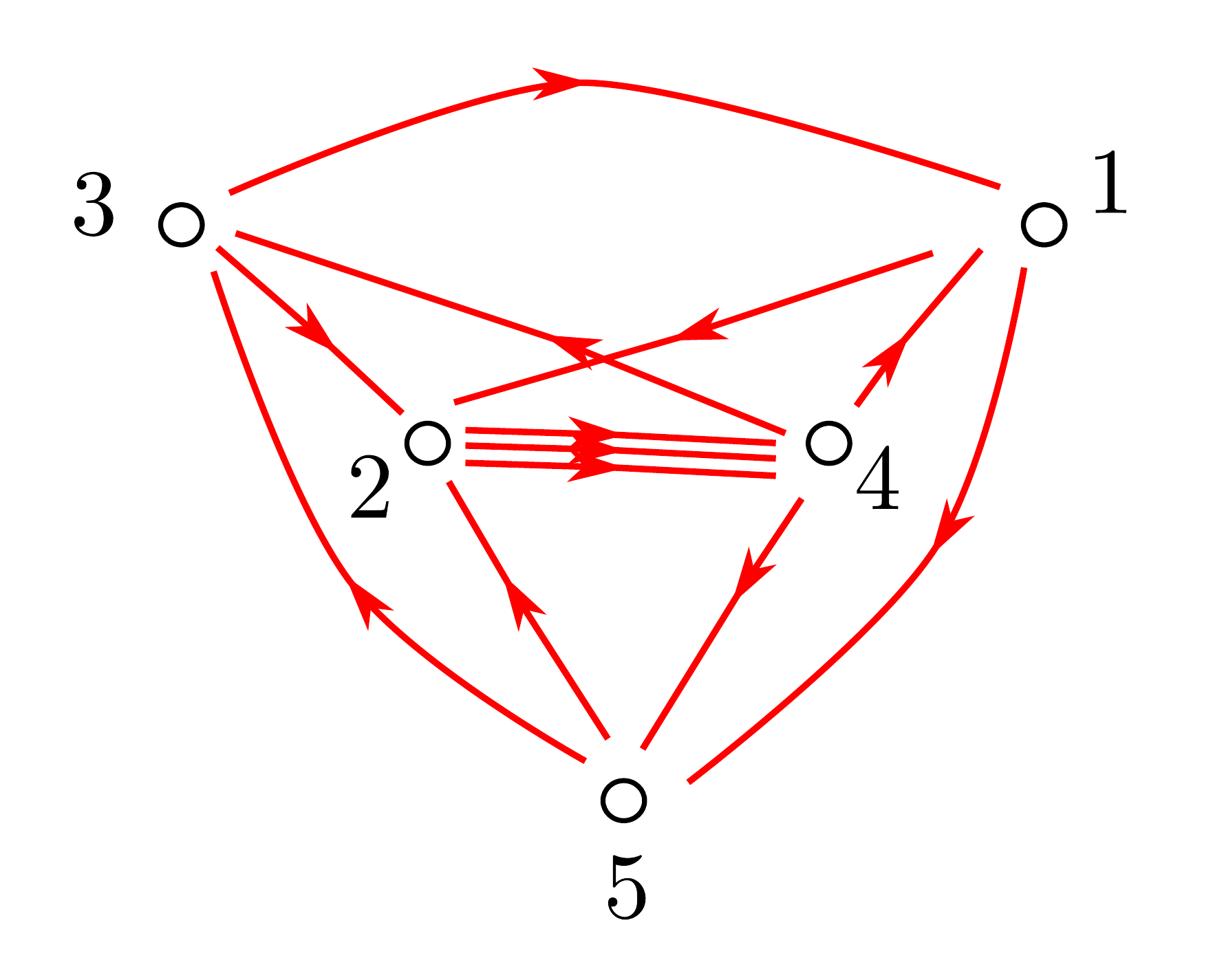}
\caption{BPS quiver dual to the graphs $\CG$ and $\CG'$ of Figure \ref{fig:full-mcg-example}.}
\label{fig:21quiver}
\end{center}
\end{figure}

A BPS graph $\CG$ is dual to a BPS quiver $Q$: an oriented graph consisting of nodes $Q_0$ connected by arrows $Q_1$ \cite{Alim:2011ae,Alim:2011kw}.
Nodes are in 1-1 correspondence with elementary webs of $\CG$, we associate to the $i$-th node the generator $\gamma_i=h(\omega_i)$ of the charge lattice identified by (\ref{eq:h-map}).
There are $b_{ij}= -\langle\gamma_i,\gamma_j\rangle$ arrows oriented from node $i$ to node $j$, when counted with signs.\footnote{
Negative values of $b_{ij}$ mean that the arrows go from $j$ to $i$. 
This definition assumes that quivers do not contain two-cycles or loops, and is appropriate for the purposes of this paper. A more general dictionary between quivers and BPS graphs can be found in \cite{Gabella:2017hpz}.
}
Figure \ref{fig:21quiver} shows the BPS quiver dual to the initial BPS graph of Figure \ref{fig:full-mcg-example}, node label $i$ stands for $\gamma_i$. 
Since the pairing $\langle\gamma_i,\gamma_j\rangle $ is determined by the adjacency matrix of $\CG$ and by the cyclic orderings $\{\sigma_v\}_{v\in V(\CG)}$,  
two equivalent graphs $\CG\simeq\CG'$ are dual to the same BPS quiver.

\begin{figure}[htbp]
\begin{center}
\includegraphics[width=0.6\textwidth]{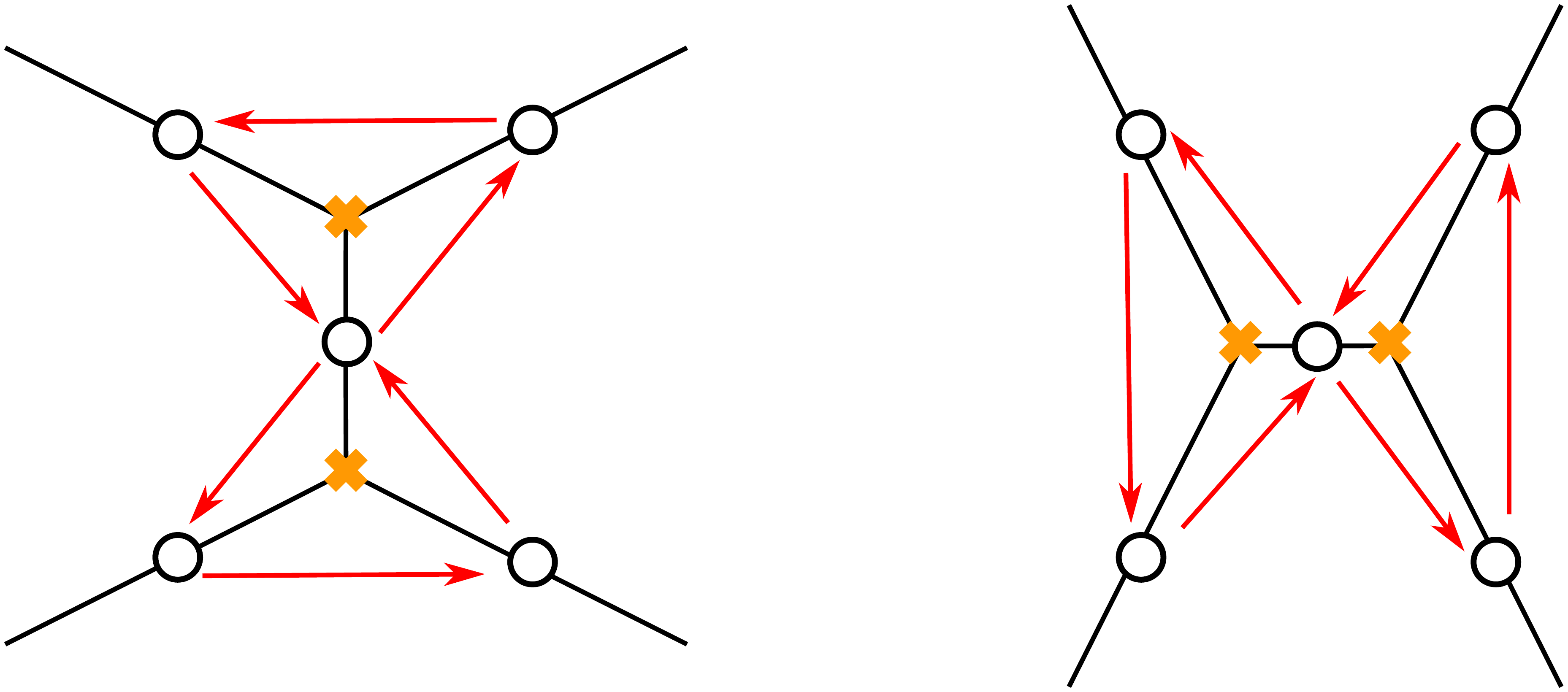}
\caption{The flip of an edge of the graph and the mutation it induces of the dual BPS quiver.}
\label{fig:quiver-dual}
\end{center}
\end{figure}

A quiver further encodes the information of an associated cluster algebra \cite{FominZelevinsky1,FominZelevinsky4}. 
Let $y_i$ be a set of variables associated to each node of $Q$. The pair $(Q, \{y_i\})$ defines a \emph{seed} for the cluster algebra.
The algebra is generated by an elementary operation on the seed, known as a mutation.
A mutation $\mu_k$ on node $k$ produces a new quiver with the same set of nodes but with new arrows 
\be\label{eq:bij-jump}
b'_{ij} = \left\{  
\begin{array}{ll} 
- b_{ij} & \text{if $i=k$ or $j=k$\,,} \\
b_{ij} + \frac{1}{2} \(|b_{ik}| b_{kj} + b_{ik}  |b_{kj}| \) \qquad & \text{otherwise}\,.
\end{array}
\right.
\ee
The mutation also acts on the cluster variables\footnote{More precisely cluster $y$-variables \cite{FominZelevinsky4}. In this paper a cluster variable always means a cluster $y$-variable.} by
\be\label{eq:yi-mutation}
	y_i' =  \, y_i y_k^{[-b_{ik}]_+} \(  1+ y_k^{-1} \)^{-b_{ik}} \,,
\ee
where $[a]_+$ is zero if $a$ is negative, and equal to $a$ otherwise.
The flip move on a BPS graph corresponds to a mutation on the dual quiver, performed on the node corresponding to the shrinking edge, see Figure \ref{fig:quiver-dual}.
The cootie move leaves the quiver invariant  instead.

Going back to the mapping class group, let us consider the action of a sequence ${\kappa:\CG\mapsto\CG'}$ on the quiver.
The sequence $\kappa_s$ translates into a sequence of mutations, while the relabeling $\kappa_r$ maps to a simultaneous reshuffling of the quiver nodes and cluster variables.
Overall, the composite operation $\kappa=\kappa_r\circ\kappa_s$ must take $Q$ back to itself since equivalent BPS graphs have identical quivers.
Nevertheless, the resulting transformation on the cluster variables needs not be trivial, and we take it as the definition of the action of $\kappa$ on the cluster algebra.
In this sense, through BPS graphs we have given a representation of the mapping class group of $C$ in the cluster algebra.

The fact that mapping class groups of surfaces admit a cluster algebra representation is not new, in fact such maps have been constructed by several other authors, see for example \cite{FockGoncharovHigher,FST1,Nagao:2011aa,Terashima:2011xe,KitayamaTerashima,2016arXiv160705228G}.
However, most of the previous constructions are limited to cluster varieties associated to Riemann surfaces decorated by \emph{full} punctures, \emph{i.e.} corresponding to maximal flavor symmetry (exceptions include \cite{Xie:2012dw,Gang:2015wya}). 
We both re-derive these previous constructions via BPS graphs, and extend them to cases including partially higgsed punctures, for which BPS graphs can be defined. 
In Section \ref{sec:evidence} we provide several examples of this construction of mapping class groups from cluster algebras, in higher rank theories of class $\CS$ defined by Riemann surfaces decorated with simple punctures, \emph{i.e.} with minimal flavor symmetry.

\section{S-duality on framed BPS states}
\label{sec:line-operators}

This section is devoted to studying the action of the generalized S-duality group on UV line operators. We will show that our construction of the mapping class group of $C$ based on transitions of BPS graphs is consistent with the action of dualities on a certain class of UV observables.

A crucial assumption in our derivation will be that the BPS graph always arises from a degenerate limit of spectral networks. 
This analysis leads to two consequences. If a BPS graph is known to arise from a degenerate spectral network, it is guaranteed that the representation of $\MCG(C)$ from Section \ref{sec:mcg} acts correctly on a certain class of UV line operators.
On the other hand, if $\CG$ is a conjectural BPS graph, and the corresponding mapping class group action behaves appropriately (in a sense that will be defined in this section), this provides a strong consistency check that $\CG$ actually arises as a degenerate limit of spectral networks, and is therefore the physical BPS graph of the theory.

\subsection{Connecting IR and UV line operators}
\label{subsec:non-ab}

The Hitchin system defined by the class $\CS$ data $(A_{N-1}, C, D)$ is a hyper-K\"ahler manifold, that in complex structure $J_\zeta$ (with $\zeta\in\IC^*$) can be viewed as a moduli space of flat $GL(N)$ connections over $C$.
There is a standard set of coordinates on this moduli space, namely traces of holonomies along cycles in  $H_1(C,\IZ)$. 
The traces of these holonomies bear the interpretation of expectation values of a certain class of BPS line operators in the gauge theory \cite{Gaiotto:2010be, Drukker:2009id, Alday:2009fs, Drukker:2009tz, Coman:2015lna}. 
Let $\wp$ be a closed path on $C$, we will denote the corresponding UV line operator by $L^{(\UV)}_\wp$.

On the Coulomb branch $\CB$, the gauge symmetry is broken to an Abelian torus $U(1)^r$, and the set of line operators of the IR theory is therefore quite different from that of the UV theory.
Line operators in the IR are classified by electromagnetic charges $\gamma$ valued in $H_1(\Sigma,\IZ)$, where $\Sigma$ is the spectral curve of the Hitchin system in a fixed vacuum $u\in \CB$. We will denote the corresponding line operator by $L^{(\IR)}_\gamma$. Just like for UV line operators, the expectation value of $L_\gamma^{(\IR)}$ can be interpreted as the holonomy of a $GL(1)$ connection over $\Sigma$.

The dictionaries between UV/IR line operators and holonomies can be used to establish a relation between the two sets of observables, based on a relation between the two moduli spaces of flat connections. 
This goes by the name of \emph{nonabelianization map}, and can be characterized via spectral networks \cite{Gaiotto:2012rg}.
A spectral network $\CW$ on $C$ defines a map 
\be
	\Psi_\CW \ : \ \CM_{\rm flat}(GL(1), \Sigma) \ \longrightarrow \ \CM_{\rm flat}(GL(N), C)\,,
\ee
that associates to any smooth closed path $\wp$ on $C$ a formal parallel transport $F(\wp;\CW)$ for the flat $GL(N)$ connection on $C$.
The trace of holonomy can be expanded as follows
\be\label{eq:formal-F-canonical}
	\Tr F(\wp;\CW) = \sum_\gamma \fOmega(\CW,\gamma,\wp) X_\gamma \,,
\ee
where $X_\gamma$ are formal variables representing $GL(1)$ holonomies along cycles $\gamma\in H_1(\Sigma,\IZ)$.
The coefficients $\fOmega(\CW,\gamma,\wp)$ depend on $\wp$ only through its homotopy class, 
this highly nontrivial property justifies the interpretation of $F(\wp,\CW)$ as the parallel transport of a flat $GL(N)$ connection.
Physically $\fOmega(\CW,\gamma,\wp)$ is an index which counts \emph{framed BPS states}, semiclassically these can be viewed as supersymmetric boundstates of BPS particles and the line operator \cite{Gaiotto:2010be,  Moore:2014jfa, Moore:2014gua, Moore:2015szp, Moore:2015qyu, Brennan:2016znk}, whereas 
mathematically they encode the relation between the two sets of holonomies. 
By the dictionary relating holonomies to VEVs of line operators
\be
	 \langle L^{(\UV)}\rangle \sim \Tr F(\wp, \CW)\,, 
	 \qquad 
	 \langle L^{(\IR)}\rangle \sim X_\gamma \,,
\ee
the expansion in framed BPS states (\ref{eq:formal-F-canonical}) therefore encodes the relation between the VEVs. 

This relation between UV and IR line operators will play a key role towards our goal of studying the action of $\MCG(C)$ on the former.
Recall that a BPS graph $\CG$ provides a basis for the IR charge lattice through the map (\ref{eq:h-map}), at least locally in some patch of the Coulomb branch.
On the other hand, $\CG$ should first and foremost arise as a degenerate spectral network, and therefore can be used to compute framed BPS states.
In the rest of this section we will explain how to formulate the nonabelianization map (\ref{eq:formal-F-canonical}) entirely in terms of the data associated to a BPS graph.
In particular, we propose a relation between the formal variables $X_\gamma$ and the cluster coordinates $y_i$ \cite{Gaiotto:2012rg, Gabella:2017hpz}, and use it to define the action of $\MCG(C)$ on UV line operators via nonabelianization.

\subsection{Resolved BPS graphs}
\label{subsec:resolved-BPS}

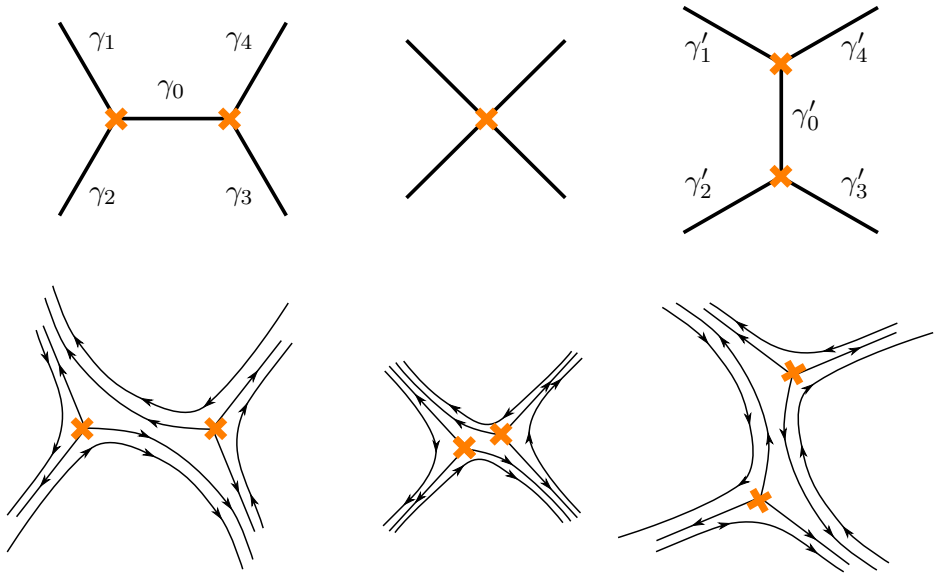
\begin{figure}[htbp]
\begin{center}
        \begin{overpic}[width=0.85\textwidth]{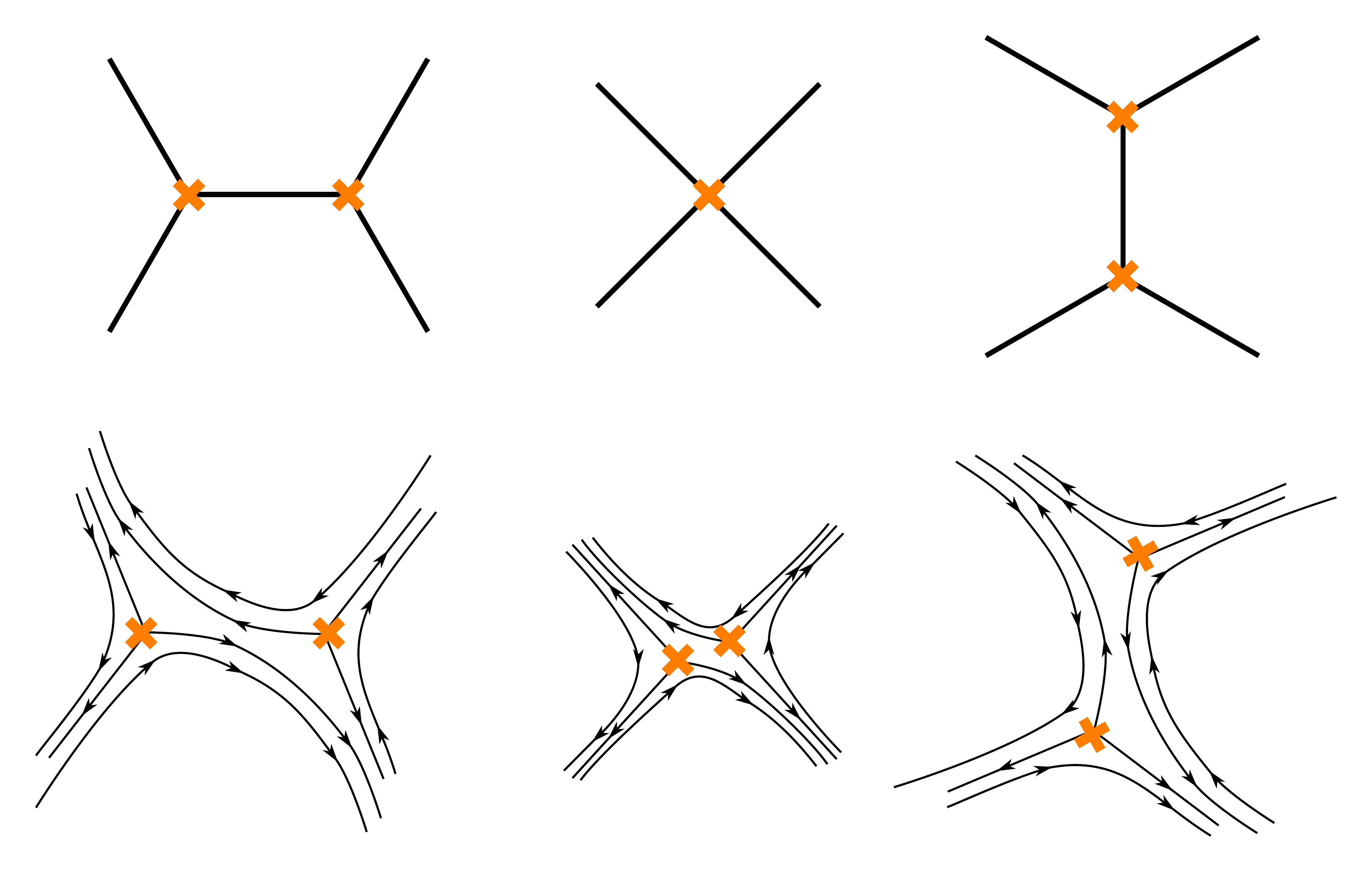}
 	\put (11, 57) {$\gamma_1$} \put (25, 57) {$\gamma_4$}
	\put (18, 52) {$\gamma_0$} 
	\put (11, 41) {$\gamma_2$} \put (25, 41) {$\gamma_3$}
	 \put (72, 56) {$\gamma_1'$} \put (88, 56) {$\gamma_4'$}
	\put (83, 49) {$\gamma_0'$} 
	\put (72, 42) {$\gamma_2'$} \put (88, 42) {$\gamma_3'$}
\end{overpic} 
\caption{The flip move on a BPS graph $\CG$, and the corresponding deformation of the spectral network corresponding to the American resolution of $\CG$. The resulting spectral network \emph{does not} coincide with the American resolution of the BPS graph after the flip.}
\label{fig:flip-resolved}
\end{center}
\end{figure}

Let  $\CG$ be a BPS graph on  $C$ arising from a maximally degenerate spectral network $\CW(u,\vartheta_c)$ at a point $u$ on the Roman locus, and critical phase $\vartheta_c$.\footnote{
Note that the spectral network $\CW_c$, which also looks like a graph on $C$, may contain more edges than $\CG$. See for example \cite[Fig. 19]{Gabella:2017hpz}. It is important to retain all the edges of the network.
}
There are two canonical resolutions of $\bCG$, corresponding to positive or negative perturbations of the phase $\vartheta_c$ involved in the definition of a network spectral network. These two options are known as the \emph{American} and the \emph{British} resolution \cite{Gaiotto:2012rg}.\footnote{Apologies to the rest of the world.}
Note that resolving the spectral network is a necessary condition for the nonabelianization map (\ref{eq:formal-F-canonical}) to be well-defined.
We choose to resolve by going to a phase $\vartheta_c-\epsilon$, known as the \emph{American resolution} of $\CW_c$, and denote the resolved network by $\bCG$ to stress its relation to the BPS graph $\CG$.
In practice, resolving $\CW_c \to \bCG$ amounts to replacing the unoriented edges with families of oriented edges running ``on the right'', as shown in the bottom-left frame of Figure \ref{fig:flip-resolved}.
The oriented edges of a spectral network can be sourced either at branch points or at joints, and their precise shape is determined by the geometry of the spectral curve $\Sigma$.
Since $\bCG$ is the American resolution of a BPS graph, for small $\epsilon$ the oriented edges will typically run very close to the original shape of $\CG$, although they may eventually veer off and be captured by punctures, after a very long time.

Next let us consider two equivalent BPS graphs $\CG,\, \CG'$ related by a sequence $\kappa = \kappa_r\circ \kappa_s$.
First of all, it is important to realize that $\CG,\CG'$ are generally related to two different spectral curves $\Sigma,\Sigma'$. 
In other words, if we assume that $\CG,\CG'$ arise from actual spectral networks on $C$, they would occur in different regions of the moduli space of the theory.\footnote{This would mean different loci on the Coulomb branch, but possibly also different Coulomb branches, related by deformations of UV moduli like masses and couplings.} 
Therefore homology classes encoded by the respective elementary webs through the map (\ref{eq:h-map}) belong to distinct homology lattices $H_1(\Sigma,\IZ)$ and  $H_1(\Sigma',\IZ)$.
In order to compare charges $\gamma$ between $\CG$ and $\CG'$ we must specify a parallel transport for the homology lattice, through the sequence of transitions $\kappa$ that takes $\CG$ to $\CG'$.
Happily, there is a canonical way to do this suggested by our choice of a resolution, which allows to smoothly deform homology classes throughout $\kappa$.
In physical terms, working with {resolved} spectral networks $\bCG$ instead of the actual BPS graph, we always avoid singularities of the moduli space where branch points collide.
At these singularities a cycle of $\Sigma$ shrinks, and the Picard-Lefshetz monodromy of the charge lattice would introduce an ambiguity.
Instead of colliding branch points, we deform the network in a way that they ``scatter off'' each other, as shown in Figure \ref{fig:flip-resolved}.
In keeping with the choice of American resolution, we will adopt the convention that after the scattering each branch point veers off to its right.
Overall the sequence of moves $\kappa_s$, together with this convention on the motion of branch points through flips, unambiguously fixes a parallel transport for $H_1(\Sigma,\IZ)$. 
We stress that using this parallel transport is crucial for making sense of any relation between homology cycles $\gamma_i$ and $ \gamma_i'$ associated respectively to elementary webs of $\CG$ and  $\CG'$.
This identification of charges by parallel transport in the moduli space will henceforth be understood for the rest of the discussion.

Having settled the question of comparing homology lattices of different spectral curves, there is one additional subtlety to confront, in order to relate the charges $\gamma_i$ to $\gamma_i'$ associated to elementary webs of $\CG$ and $\CG'$.
This is the fact that the map (\ref{eq:h-map}) which associates homology classes to edges of $\CG$ actually \emph{jumps} in correspondence of a flip.
This jump is clearly necessary, because a flip on $\CG$ acts by a mutation on the dual quiver $Q$, and its arrows correspond to the intersection pairing $b_{ij} = \langle\gamma_j,\gamma_i\rangle$ of the charges assigned by $h$ to elementary webs.
Referring to Figure \ref{fig:flip-resolved} the jump of $h$ reads%
\footnote{
This is one place where the choice of American resolution is relevant. In British resolution we should employ a different transformation.
} 
\be\label{eq:h-jump}
	\gamma_0' = -\gamma_0 \;, \quad 
	\gamma_1'=\gamma_1+\langle \gamma_1,\gamma_0\rangle \gamma_0  \;,\quad 
	\gamma_2'=\gamma_2 \;,\quad 
	\gamma_3'=\gamma_3+\langle \gamma_3,\gamma_0\rangle \gamma_0 \;,\quad 
	\gamma_4'=\gamma_4\;,
\ee
in agreement with (\ref{eq:bij-jump}).\footnote{The transformation \eqref{eq:h-jump} coincides with the transformation properties of (the logarithm of) the so-called tropical $y$-variables in the cluster algebras.}

A mutation of the quiver induces a cluster transformation  (\ref{eq:yi-mutation}) on the cluster variables $y_i$.
Therefore cluster variables $y_i, y_i'$ of quivers $Q,Q'$ dual to $\CG,\CG'$ must be related by a sequence of cluster mutations, possibly composed with a permutation, corresponding respectively to the flips in $\kappa_s$ and to relabelings in $\kappa_r$. 
By a slight abuse of notation we will denote this relation by 
\be\label{eq:coord-change}
	y_i' = \kappa(y_i)\,.
\ee

\begin{figure}[htbp]
\begin{center}
\includegraphics[width=0.7\textwidth]{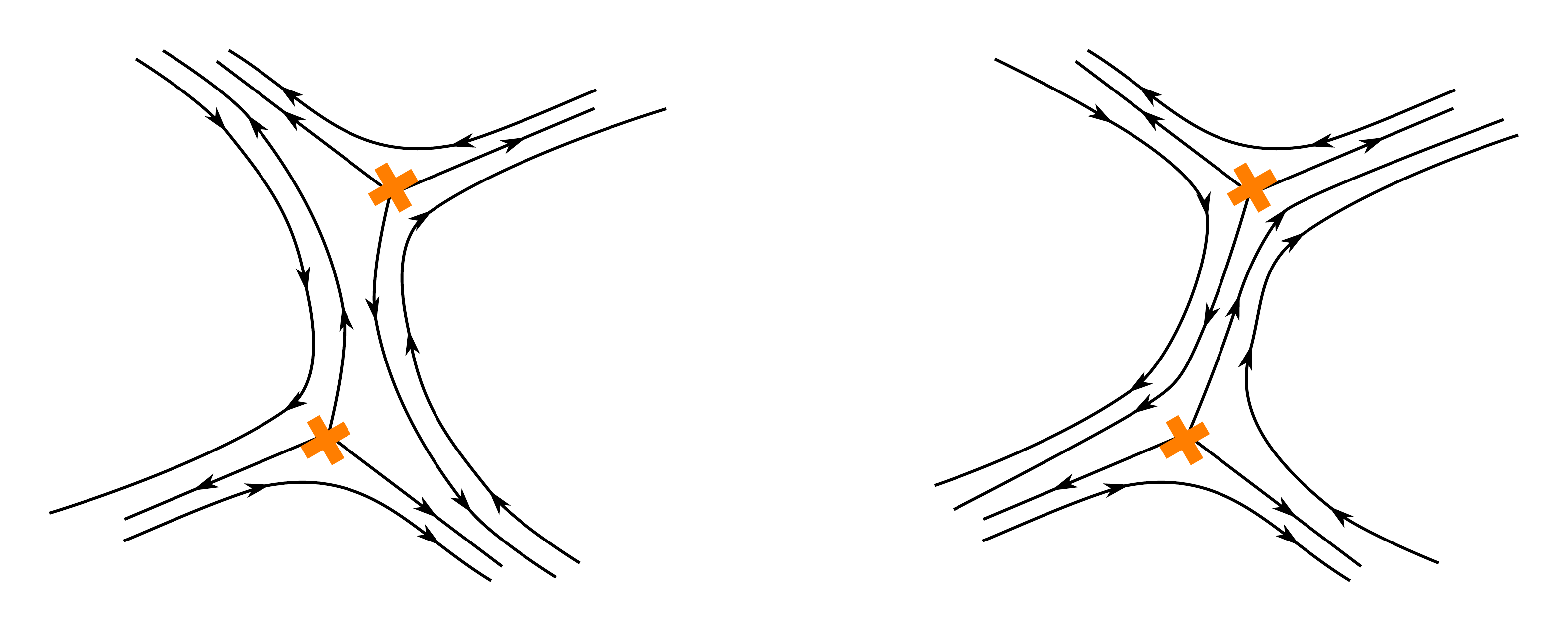}
\caption{An inverse $\CK$-wall transition of a spectral network.}
\label{fig:K-wall}
\end{center}
\end{figure}

The jump of $h$ may appear in contradiction with the continuity of the transport of the charge lattice described previously, instead there is a rather subtle interplay between the two. 
The parallel transport of the homology lattice is defined by choosing a resolution $\bCG$ of the {initial} BPS graph $\CG$.
However a flip transition on $\bCG$ will \emph{not} produce a network that is a resolution of the new BPS graph, as made evident in Figure \ref{fig:flip-resolved}. 
In this example the new spectral network in the bottom-right frame differs from the American resolution of the BPS graph in the top-right frame, since the two vertical oriented edges in the middle run ``on the left''.
In fact, there is a precise relation between this spectral network and the resolution of the new BPS graph, which is known as a \emph{$\CK$-wall transformation} and shown in Figure \ref{fig:K-wall}.
This is a jumps of the spectral network which involves a change of its topology, and induces a transformation of both the $\fOmega$ and the $X_\gamma$ appearing in (\ref{eq:formal-F-canonical}) \cite{Gaiotto:2012rg}.

In conclusion, we \emph{define} the parallel transport of the homology lattice by moving branch points through a flip transition, as shown in Figure \ref{fig:flip-resolved}. 
Note that the deformation of the spectral curve, as reflected by the motion of branch points over $C$, does not imply any choice about what we do with the spectral network.
On the other hand this deformation produces non-canonical spectral networks, in the sense that the resulting network doesn't correspond to the resolution of a BPS graph, but is related to one precisely by $\CK$-wall transitions.
We therefore perform a $\CK$-wall transformation on the network $\bCG$, to compensate for this mismatch.
At the same time, we also introduce a jump (\ref{eq:h-jump}) for the map $h$ which associates homology cycles $\gamma_i$ to elementary webs of $\CG$. 
This change of basis for the charge lattice will be denoted by
\be\label{eq:charge-rotation}
	\gamma_i' = \kappa(\gamma_i)
\ee
by a small abuse of notation. It is understood that $\kappa=\kappa_r\circ\kappa_s$ also includes the effect of the relabeling $\kappa_r$, which acts by a permutation on the basis.

\subsection{Nonabelianization for BPS graphs}
\label{subsec:non-ab-BPS}
Let us recall some key properties of the computation of framed BPS indices $\fOmega(\bCG,\gamma,\wp)$ using spectral networks. 
The $\fOmega$ are entirely determined by how the path $\wp$ intersects the spectral network $\bCG$, and by the topology of the network.
The contribution of each intersection is determined by a combinatorial problem formulated in terms of the overall topology of $\bCG$.
When the network arises as the resolution of a BPS graph, the combinatorial data, a.k.a. 2d-4d soliton data, can be computed directly in terms of topological data of $\bCG$ which includes a cyclic ordering of edges at each vertex \cite{Longhi:2016wtv}.
A bit more precisely, the construction of $F(\wp,\bCG)$ depends on which oriented edges are crossed by $\wp$, but the final result is actually invariant under homotopy of the latter, including deformations across branch points and joints of $\bCG$.

Through the map (\ref{eq:h-map}) the elementary webs provide a basis  $\gamma_i\in H_1(\Sigma,\IZ)$ for the IR charge lattice. 
Any charge $\gamma$ therefore admits a unique decomposition $\gamma = \sum_i c_i \gamma_i$ and can be represented by a lattice vector $\bc=(c_1,\dots, c_d)$,\footnote{This is known as the $c$-vector in cluster algebras \cite{FominZelevinsky4}.} where $d$ is the number of elementary webs of $\CG$ and coincides with the rank of the charge lattice.
We can therefore reformulate the nonabelianization map entirely in terms of data of $\bCG$ as 
follows
\be\label{eq:tr-hol-BPS}
\begin{split}
	\Tr F(\wp;\bCG) & =  \sum_{\bc} \fOmega(\bCG, \bc,\wp) X_\bc 
\end{split}
\ee
without any explicit reference to $H_1(\Sigma,\IZ)$.
It is always possible to bring this expression back to the form (\ref{eq:formal-F-canonical}), through the map (\ref{eq:h-map}).
This expression of holonomies is especially convenient for studying the action of $\MCG(C)$ derived in Section \ref{sec:mcg}, we turn to this next.

\subsection{Mapping class group action on holonomies}
\label{subsec:MCG-hol}

Consider two equivalent BPS graphs $\CG$ and $\CG'$ related by a sequence $\kappa$, 
we would like to study how the respective holonomies (\ref{eq:tr-hol-BPS}) are related.
Recall that to compute these holonomies one should work with non-degenerate spectral networks, and we will choose  $\bCG, \bCG'$ as defined previously.\footnote{It is important to distinguish from the situation of Figure \ref{fig:flip-resolved}. Here we always take the \emph{actual} American resolution both for $\CG$ and for $\CG'$. As explained previously, we perform $\CK$-wall transitions on the network at each flip, in order to preserve this property.}
The computation of $\fOmega$ depends on two pieces of information: how $\wp$ intersects  $\bCG$, and certain combinatorial data computed from $\bCG$.
The latter is known as \emph{soliton data}, and is entirely determined by the topology of the spectral network.
By definition $\CG$ and $\CG'$ have the same topology, just different embeddings in $C$, and the same will be assumed of $\bCG$ and $\bCG'$. 
This implies that the soliton data of $\bCG$ and $\bCG'$ must be essentially identical, the only difference between the two will be in the map $h$, which plays the role of translating  combinatorial data on a BPS graph into homology classes on $\Sigma$.

A bit more precisely, the soliton data of $\bCG$ consists of {relative} homology classes, counting paths on $\Sigma$ which run ``above'' edges of $\bCG$ according to the projection $\pi:\Sigma\to C$ \cite{Gaiotto:2012rg}.
This data is determined by a set of equations, which can be formulated entirely in terms of the adjacency matrix of $\bCG$ and of the cyclic ordering of edges at its vertices \cite{Longhi:2016wtv}. 
Therefore the equations that determine soliton data are formally identical for $\bCG$ and for $\bCG'$, and there is a 1-1 correspondence between them.
However the spectral networks $\bCG,\bCG'$ encode different framed BPS states. 
This is because the respective elementary webs of $\CG$ and $\CG'$ define different bases of the IR charge lattice, related by jumps of the map (\ref{eq:h-map}).

The fact that soliton data sets are formally identical implies that the integers
\be\label{eq:framed-identity}
	\fOmega(\bCG,\bc,\wp) = \fOmega(\bCG',\bc',\wp')
\ee
coincide if 
\be
	\CG' = \kappa(\CG),\quad \bc'=\bc,\quad \wp' \simeq  g_\kappa(\wp)\,,
\ee
as a direct consequence of the construction of (\ref{eq:tr-hol-BPS}).
The first requirement states that $\CG\simeq \CG'$ must be equivalent as abstract graphs, and are moreover related by a sequence of moves $\kappa = \kappa_r\circ\kappa_s$. 
The second relation is simply an identity of vectors in $\IZ^n$.
The third relation denotes equivalence as homology classes $[\wp]$ and $[\wp']$, related by the mapping class group transformation $g_\kappa$ associated to $\kappa$.

Combining the identity (\ref{eq:framed-identity}) with the change of basis for the charge lattice  (\ref{eq:charge-rotation}) we arrive the following formula, which relates the holonomy computed along $\wp$ using $\bCG'$ to the one along $\wp$ computed using $\CG$
\be\label{eq:crazy}
\begin{split}
	\Tr F(\wp'; \bCG') & = \sum_{\bc'} \fOmega(\bCG',\bc',\wp') X_{c'_i\gamma_i'} \\
	& = \sum_{\bc} \fOmega(\bCG,\bc,\wp) X_{c_i \gamma_i'} \\
	& = \sum_{\bc} \fOmega(\bCG,\bc,\wp) X_{c_i \kappa(\gamma_i)}\,.
\end{split}
\ee
We can also ask how does the holonomy around a fixed path $\wp'$ change, as the network undergoes a sequence of flips and cooties $\kappa_s$. 
Recall that we smoothly deform the spectral network from $\bCG$ to $\bCG'$, except for a $\CK$-wall jump at each flip, therefore 
\be
	\Tr F(\wp', \bCG') = \CK_n^{-1}\cdots  \CK_1^{-1}\(\Tr F(\wp', \bCG)\)\,,
\ee
provided that the homology cycles $\gamma$ appearing in the expansion of either side are identified according to the parallel transport described in Subsection \ref{subsec:resolved-BPS}. 
Here $ \CK_n^{-1}\cdots  \CK_1^{-1}$ denotes a sequence of (inverse) $\CK$-wall jumps of the network, corresponding precisely to the flips in $\kappa$, ordered from right to left.
A $\CK$ wall corresponding to a mutation on node $k$ acts on the $X_\gamma$ as follows
\be
	\CK^{-1}(X_\gamma) =   X_\gamma (1+ X_{-\gamma_k})^{\langle \gamma_k,\gamma\rangle}\,.
\ee
Note the appearance of $-\gamma_k$ as opposed to $\gamma_k$, this is because the $\CK$-wall jump involves the edge of the new BPS graph \emph{after} the flip, see Figures \ref{fig:flip-resolved} and \ref{fig:K-wall} and equation (\ref{eq:h-jump}).
Combining this identity with (\ref{eq:crazy}) gives
\be\label{eq:partial-1}
	\Tr F(\wp', \bCG) = \sum_{\bc} \fOmega(\bCG,\gamma,\wp) \,  \CK_1\cdots  \CK_n \(X_{\kappa(\gamma)}\) \,.
\ee

We would like to express the r.h.s. in terms of the holonomy $\Tr F(\wp,\bCG)$. In order to achieve that, we first need a technical result, whose proof can be found in Appendix \ref{app:contravariance}.
Introduce a map $\rho$ defined as follows
\be\label{eq:rho-map}
	\rho_\CG :  X_{\gamma_i}\to y_i\,,
\ee
this definition is formulated in terms of the basis elements of the initial BPS graph $\CG$ and its cluster variables $y_i$. 
$\rho$ extends to other homology classes by multiplication $X_{\gamma+\gamma'}=X_{\gamma}X_{\gamma'}$.
We claim that 
\be\label{eq:contravariance}
	\CK_{1}\dots\CK_{n}(X_{\kappa(\gamma_i)})  = \rho^{-1} \circ \kappa \circ \rho (X_{\gamma_i}) \,,
\ee
where $\CK_{1}\dots\CK_{n}$ is a sequence of $\CK$-wall transformations, applied in the \emph{opposite} order compared to how they occur on the BPS graph ($\CK_1$ happens first, but its transformation is applied last). 

Using this result in (\ref{eq:partial-1}) gives
\be
\begin{split}
	\Tr F(\wp', \bCG) 
	& = \sum_{\bc} \fOmega(\bCG,\gamma,\wp) \, \rho^{-1} \circ \kappa \circ \rho \(X_{\gamma}\) \,.
\end{split}
\ee
Finally, we go back to cluster variables by applying $\rho$ on each side, and come to the main result of this section
\be\label{eq:mcg-action-hol}
\begin{split}
	\rho\(\Tr F(\wp', \bCG)\) 
	& = \kappa\circ \rho\( \Tr F(\wp, \bCG)  \) \,.
\end{split}
\ee
This identity holds for a generic sequence of moves $\kappa=\kappa_r\circ\kappa_s$ made of flips, cooties and relabelings, provided that it corresponds to an element of $\MCG(C)$, subject to the consistency conditions described in Section \ref{sec:mcg}.

Equation (\ref{eq:mcg-action-hol}) shows that the action by $\kappa$ on cluster variables $y_i$ corresponds to turning the holonomy along $\wp$ into the holonomy along a new path $\wp'$. 
Moreover $\wp'$ and  $\wp$ are related precisely by the mapping class group transformation $g_\kappa$ acting on $H_1(C,\IZ)$.
This relation provides strong evidence that our construction of the mapping class group via the cluster algebra and BPS graphs acts as expected on the moduli space of flat $GL(K)$ connections on $C$.
On the other hand, if the mapping class group action generated by a BPS graph is shown to satisfy this relation, this provides evidence that $\CG$ really arises from a degenerate spectral network.

\section{Examples}
\label{sec:evidence}

In this section we work out the construction of the mapping class group in several examples, and perform several checks, including the group relations of generators of $\MCG(C)$ and the action on UV line operators.

\subsection{\texorpdfstring{$SU(2)$ $\CN=2^*$ theory}{SU(2) N=2* theory}}
\label{subsec:A1-torus}

The $SU(2)$ gauge theory with a massive adjoint hypermultiplet is realized  as a class $\CS$ theory by taking the $A_1$ Hitchin system with $C$ a torus with a regular puncture.
The mapping class group action for this example does not require the machinery of BPS graphs, and has been previously studied  using other techniques \cite{Terashima:2011qi,Dimofte:2011jd,Terashima:2011xe,Nagao:2011aa,Gang:2013sqa}. Nevertheless we include it both for completeness and for pedagogical purposes.

\begin{figure}[htbp]
\begin{center}
\includegraphics[width=0.75\textwidth]{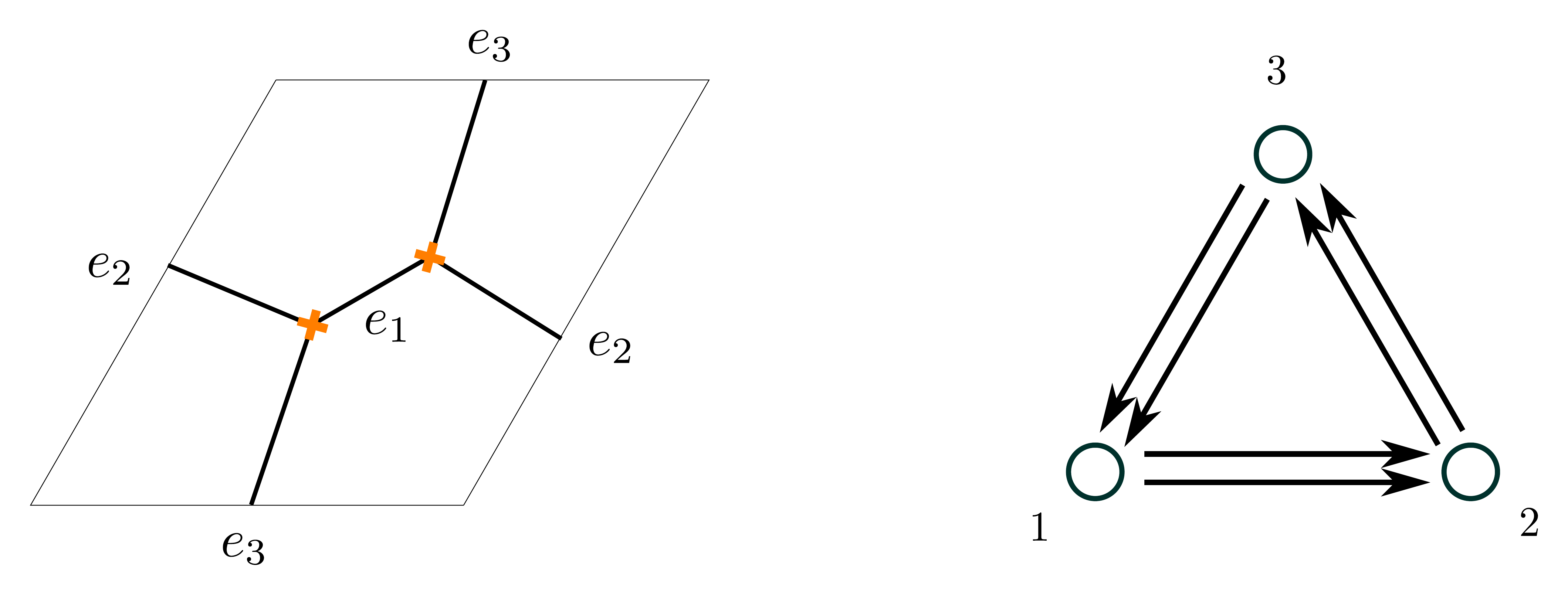}
\caption{The BPS graph of $SU(2)$ $\CN=2^*$ theory, and the dual quiver.}
\label{fig:N2star}
\end{center}
\end{figure}

The spectral curve has genus two and has two punctures, however the physical charge lattice is just three dimensional, as a result of a quotient \cite{Gaiotto:2009hg}.
The BPS graph of the theory is shown in Figure \ref{fig:N2star}, it is made of three edges $e_i$. 
Each edge corresponds to an elementary web, the intersection pairing of the respective homology cycles $\gamma_i = h(e_i)$ is
\be
	\langle \gamma_i ,\gamma_{i+1}\rangle = -2 \,.
\ee
Accordingly the quiver has $b_{i,i+1}=2$ arrows connecting node $i$ to $i+1$.
Note that $\gamma_1+\gamma_2+\gamma_3$ is a flavor charge, it has vanishing intersection with any other cycle.
We choose to represent the generators of the homology lattice of $C$ as
\be
	A = [(e_1,e_2)] \,,\qquad B = [(e_1,e_3)]\,,
\ee
note that the orientation of each is ambiguous, since the cyclic ordering is preserved by reversing the ordering. This is a peculiarity of the $SU(2)$ theory, and will bring specific consequences.
We choose $A$ oriented to the right, and $B$ oriented upwards.

\begin{figure}[htbp]
\begin{center}
\includegraphics[width=0.95\textwidth]{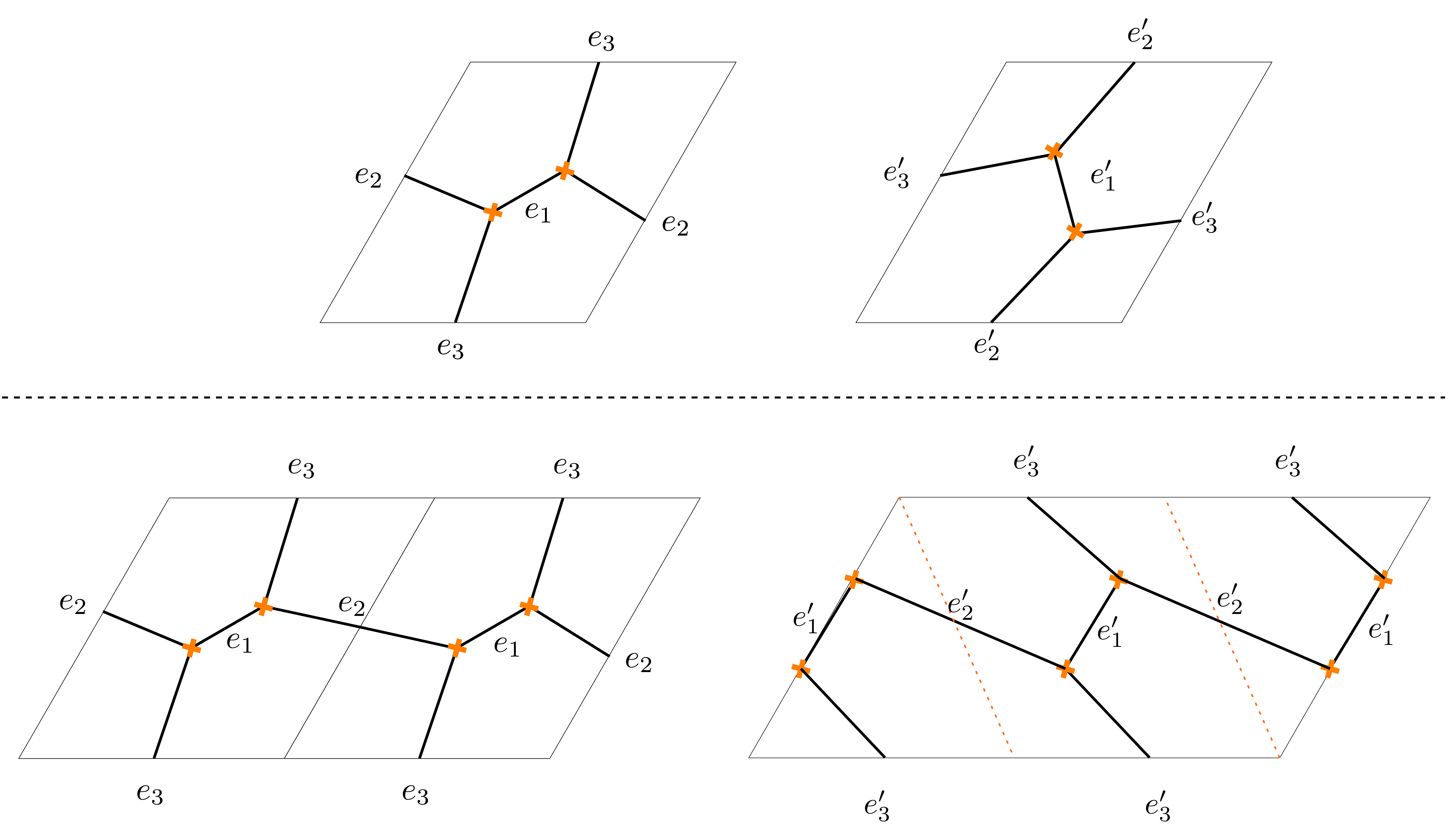}
\caption{Top: the $S^{-1}$ move. Bottom: the $T^{-1}$ move.}
\label{fig:N2star-moves}
\end{center}
\end{figure}

The $S^{-1}$ transformation is obtained by performing a single flip $\kappa_s$ on edge $e_1$ followed by a relabeling $\kappa_r : e_2\leftrightarrow e_3$, see Figure \ref{fig:N2star-moves}.
As a check, note that
\be
	A' = [(e_1', e_2')] = B\,, \qquad B' = [(e_1', e_3')] = -A
\ee
which indeed corresponds to the action of $S^{-1}$.\footnote{
We chose $A'$ oriented upwards, and $B'$ oriented left, preserving the relative orientation of $A,B$.}
Let us therefore denote by $\kappa_{S^{-1}}=\kappa_r\circ\kappa_s $.
On the cluster variables, this transformation acts by the mutation $\mu_1$ on node $1$,  followed by a  permutation of nodes $2\leftrightarrow 3$
\be\label{eq:N2star-S-generator}
	\kappa_{S^{-1}}(y_1, y_2, y_3) = \( y_1^{-1} ,  y_3 (1+y_1^{-1})^{-2} , y_2 y_1^2 (1+y_1^{-1})^2 \) \,.
\ee

Another generator of $\MCG(C)$ can be chosen to be the flip on edge $e_2$ followed by the permutation of $e_1\leftrightarrow e_2$, see  Figure \ref{fig:N2star-moves}.
The homology basis on  $C$ gets mapped to
\be
	A ' = [(e_1', e_2')] = A\,, \qquad B' = [(e_1', e_3')] = B-A\,,
\ee
therefore we recognize this as the action of
\be
	T^{-1} = \(\begin{array}{cc} 1 & -1 \\ 0 & 1\end{array}\)\,.
\ee
On the cluster variables this acts by composition of a mutation $\mu_2$ on node $2$, followed by a permutation of nodes $1\leftrightarrow 2$
\be
	\kappa_{T^{-1}}(y_1, y_2, y_3) =\( y_2^{-1} , y_1  (1+y_2^{-1})^{-2} ,y_3  (1+y_2)^2 \) \,.
\ee
As a check that the identifications are consistent among themselves, we can verify that the identity $\(S^{-1} T^{-1}\)^3 = 1$
\be
\begin{split}
	(y_1, y_2, y_3) 
	&\stackrel{\kappa_{T^{-1}}}{\longrightarrow} 
	\left(
	{y_2^{-1}}, {y_1}\left(1 + y_2^{-1}\right)^{-2}, y_3 \left(1+y_2\right)^2 \right)\\
	&\stackrel{\kappa_{S^{-1}}}{\longrightarrow} 
	(y_2, y_3, y_1) \\
	&\stackrel{\kappa_{T^{-1}}}{\longrightarrow} 
	\left({y_3^{-1}},{y_2}\left(1 + y_3^{-1}\right)^{-2}, y_1 \left(1+y_3\right){}^2\right)\\
	&\stackrel{\kappa_{S^{-1}}}{\longrightarrow} 
	(y_3, y_1, y_2) \\
	&\stackrel{\kappa_{T^{-1}}}{\longrightarrow} 
	\left({y_1^{-1}},{y_3}\left(1 + y_1^{-1}\right)^{-2},  y_2 \left(1+y_1\right)^2\right)\\
	&\stackrel{\kappa_{S^{-1}}}{\longrightarrow} 
	(y_1, y_2, y_3) \,.
\end{split}
\ee
It is also straightforward to check that $(S^{-1})^2=1$, this suggests that we found a representation of the orientation-preserving mapping class group $PSL(2,\IZ)$ (as opposed to its double cover $SL(2, \IZ)$, which allows for orientation change). This appears to be a special feature of $N=2$, and is due to the fact that the orientation of $A$ and $B$ cycles is ambiguous: each is composed of just two edges of $\CG$, therefore their cyclic ordering is invariant under orientation reversal. This ambiguity will be absent for higher $N$.

Notice that the graph has an obvious $\IZ_3$ symmetry, and this raises the possibility of choosing $\kappa_r$ differently. For instance, considering the permutation $\kappa_r : e_1\leftrightarrow e_3$ after the flip on edge $e_2$ would have given $A'=B-A, B'=2 A + B$. This corresponds to the transformation 
\be
	\left(\begin{array}{cc}  1 & -2 \\ 1 & -1  \end{array}\right) = S^{-1} T^{-1} S^{-1} T^{-2} S^{-2}\,.
\ee
Indeed, we checked that performing this transformation corresponds precisely to acting with the sequence $\kappa_{S^{-1}}\kappa_{T^{-1}}\kappa_{S^{-1}} \kappa_{T^{-1}} \kappa_{T^{-1}} \kappa_{S^{-1}}\kappa_{S^{-1}}$ on the cluster variables. This match provides a strong consistency check on the identification between mapping class group and cluster algebra.
A nice property of this construction is how manifest the relation between mapping class group action and cluster algebra is. 
On the one hand the action on the BPS graph resembles directly that of $\MCG(C)$, on the other we read off directly on the dual quiver what is the corresponding cluster transformation.

\subsubsection{Line operators}
We will now illustrate the action of $\MCG(C)$ on the UV line operators of the theory, to show how the general statements of Section \ref{sec:line-operators} are realized in this theory. 
As a byproduct, we also work out the first computation of framed BPS states using BPS graphs, which involves some subtleties due to the occurrence of fractional charges.

\begin{figure}[htbp]
\begin{center}
\includegraphics[width=0.45\textwidth]{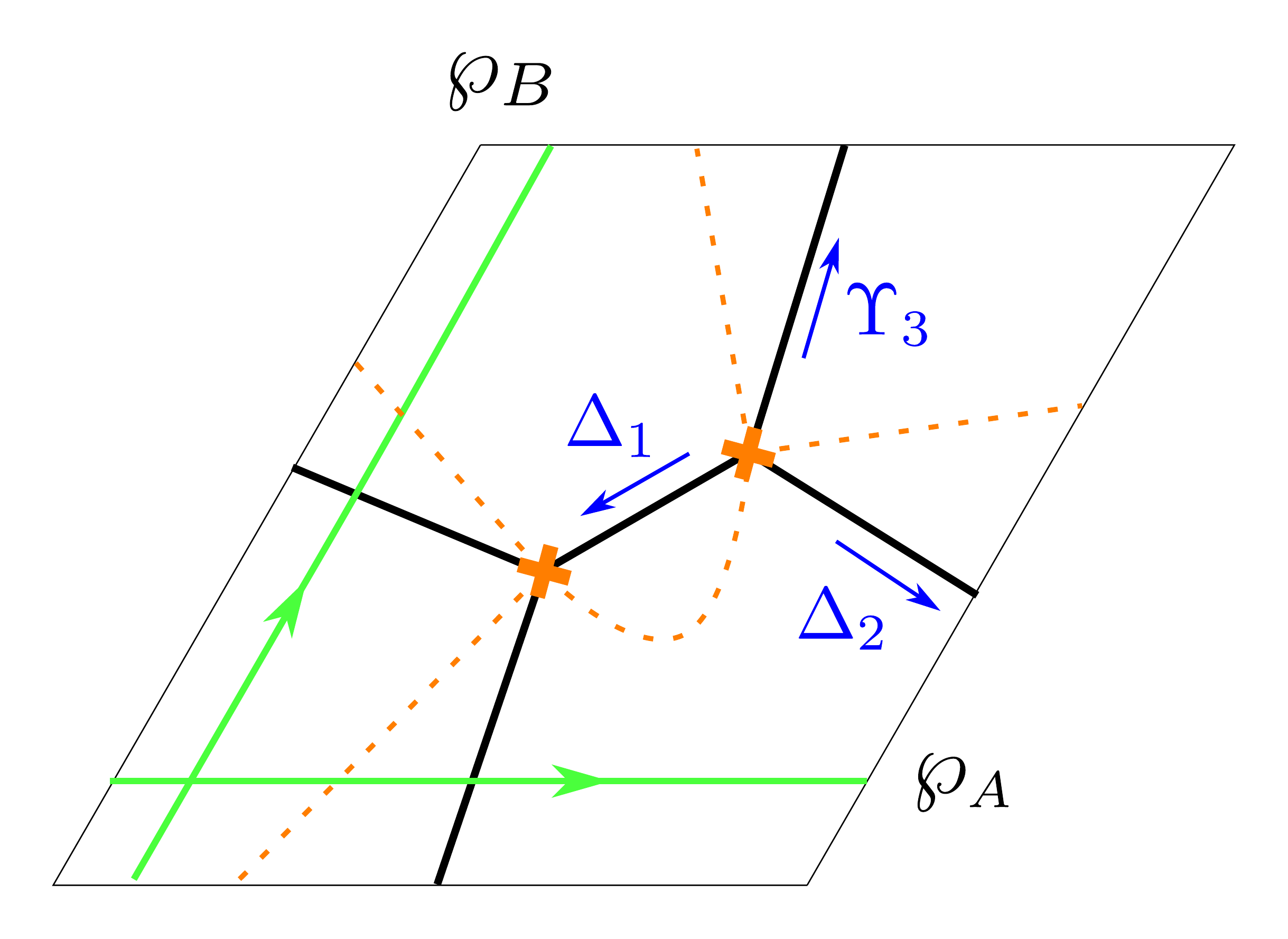}
\caption{A choice of trivialization for the spectral curve, branch cuts are dashed lines. The cycles $\wp_A,\wp_B$ are indicated in green. Blue arrows indicate the direction of the flow of $21$-walls of the spectral network arising as the American resolution of the BPS graph.}
\label{fig:N2star-line}
\end{center}
\end{figure}

Let us consider line operators labeled by the paths $\wp_A$ and $\wp_B$ in Figure \ref{fig:N2star-line}.
First of all we need to compute the holonomies along each path, which in turn requires us to compute the soliton data on the BPS graph. 
Recall that computing soliton data (and BPS states) requires a non-degenerate spectral network, we choose the American resolution of $\CG$, denoted by $\bCG$.
The soliton data for the resolved network has been computed in \cite{Longhi:2016wtv} from which we borrow the results, see the reference for details on the computation.\footnote{Note that in the reference, the British resolution was adopted. This explains the slightly different expressions.}
Let $\Upsilon_i/\Delta_i$ be the generating functions of up/down-going solitons on edge $e_i$.
With the choice of branch cuts shown in Figure \ref{fig:N2star-line}, the generating functions $\Delta_1, \Delta_2, \Upsilon_3$ count solitons of type $21$ while their counterparts obtained from switching $\Delta\leftrightarrow\Upsilon$ count solitons of type $12$.
These generating functions encode all the soliton data of the network, they are determined by the following equations
\be
	\begin{array}{ccc}
	{\rm NE} & \qquad & {\rm SW
	} \\
	\hline
	\Delta_1 = X_{a_1} + \Delta_3 & &\Upsilon_1 = X_{b_1} + \Upsilon_3 \\
	\Delta_2 = X_{a_2} + \Upsilon_1 & & \Upsilon_2 = X_{b_2} + \Delta_1 \\
	\Upsilon_3 = X_{a_3} + \Upsilon_2 & & \Delta_3 = X_{b_3} + \Delta_2 
	\end{array}
\ee
where $a_i, b_i$ are the shortest soliton paths sourced at the NE/SW branch point, supported on edge $e_i$. The equations are solved by 
\be
\begin{split}
	\Delta_1 & = X_{a_1}\frac{1+X_{\gamma_3}+X_{\gamma_2+\gamma_3}}{1-X_{\gamma_1+\gamma_2+\gamma_3}}\,, \\
	\Upsilon_1 & = X_{b_1}\frac{1+X_{\gamma_3}+X_{\gamma_2+\gamma_3}}{1-X_{\gamma_1+\gamma_2+\gamma_3}} \,,\\
\end{split}
\ee
and other similar expressions obtained by cyclic permutations of indices.

\begin{figure}[htbp]
\begin{center}
\includegraphics[width=0.65\textwidth]{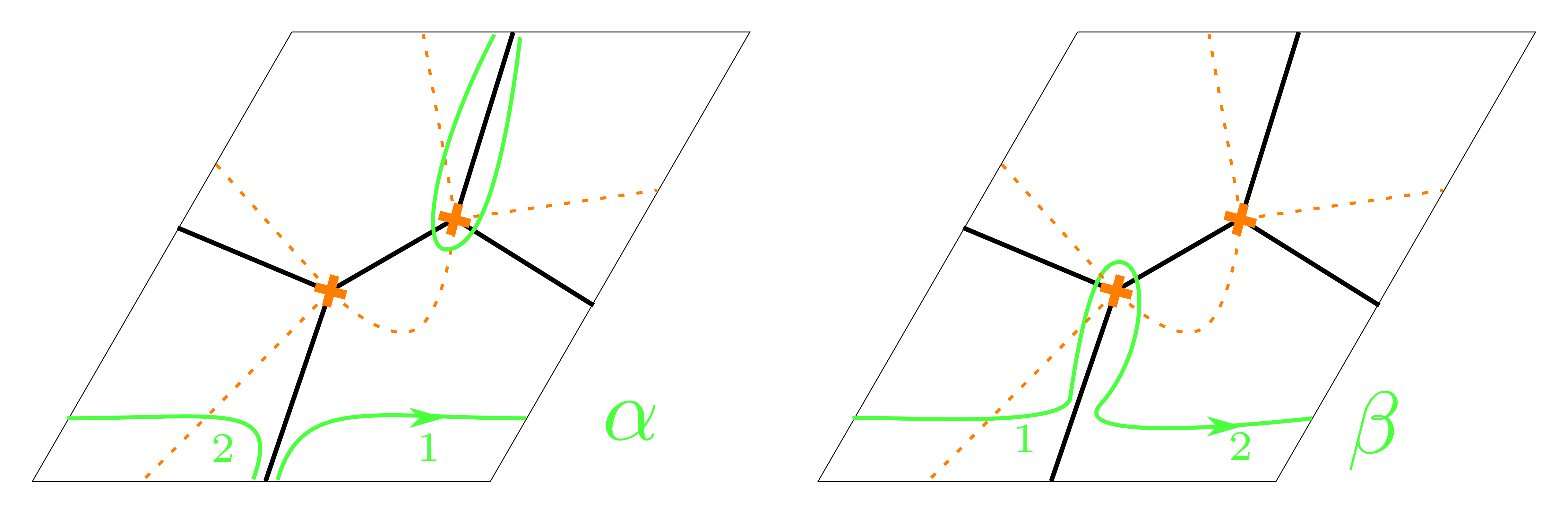}
\caption{The paths $\alpha$ and $\beta$ on the spectral curve $\Sigma$. Labels $1,2$ indicate on which sheet of $\Sigma$ a segment of a path lies. The sheet switches at every crossing with a branch cut (dotted limes).}
\label{fig:N2star-line-alpha-beta}
\end{center}
\end{figure}

The formal parallel transport along the A-cycle can be computed as follows. For convenience, let us split the path into $\wp_A', \wp_A''$, with $\wp'_A$ running from the basepoint (the green dot) to the intersection with edge $e_3$, and $\wp_A''$ its complement. 
The lift of $\wp_A'$ to $\Sigma$ consists of two pieces $A'_1 + A'_2$, running respectively on sheets 1 and 2.  $\wp_A''$ runs through a branch cut, so its lift includes two pieces $A''_{12} + A''_{21}$, which run from sheet 1 to 2 and vice versa. 
The parallel transport is then computed using the standard detour rules of spectral networks \cite{Gaiotto:2012rg}
\be
\begin{split}
	\Tr F(\wp_A,\bCG)  & =\Tr   %
	\(\begin{array}{cc}	 X_{A_1'}	&	\\  	&	X_{A_2'}           \end{array}\) %
	\(\begin{array}{cc}	1 	&	\Delta_3	\\ 	&	1	\end{array}\) %
	\(\begin{array}{cc}	1 	&		\\ \Upsilon_3	&	1	\end{array}\) %
	\(\begin{array}{cc}	 	&	X_{A_{12}''}	\\  	X_{A_{21}''}	&	\end{array}\) \\%
& =  X_{A'_1}  \Delta_3  X_{ A''_{21}} +X_{A'_2}  \Upsilon_3  X_{A''_{12}}  \,.
\end{split}
\ee
To unpack this expression, consider the paths on $\Sigma$ built from the concatenations $\beta = A'_1 + b_3 + A'_{21}$  and $\alpha = A'_2 + a_3 +  A'_{12}$, see Figure \ref{fig:N2star-line-alpha-beta}. 
Both are closed paths, so their homology classes can be expressed in terms of the basis $\gamma_i$.
By drawing representatives for these paths in our choice of trivialization, it is not hard to derive the following relations\footnote{
The first relation follows immediately from the fact that $A'_1+A'_2+A''_{12} + A''_{21}$ is invariant under the exchange of the two sheets, and therefore is projected out of the physical charge lattice \cite{Gaiotto:2009hg} (related to this, note that the period of the Seiberg-Witten 1-form would vanish along this cycle).
To derive the second one, one draws representatives for $\alpha,-\beta$ and deforms each component to join each other. 
In order to do this, one of the paths must pass through the puncture, picking up a cycle running  clockwise around its lift to sheet $1$. This is equal to $\gamma_1+\gamma_2+\gamma_3$, and the remaining paths are homologically trivial.
}
\be
	\alpha + \beta = a_3 + b_3 =  \gamma_3\,, \qquad 
	\alpha - \beta = \gamma_1+\gamma_2+\gamma_3 \,.
\ee

Taken together, these imply that $\alpha$ and $\beta$ are actually fractional charges
\be
	\alpha = \frac{1}{2}(\gamma_1+\gamma_2) +\gamma_3 \,,\qquad \beta = -\frac{1}{2}(\gamma_1+\gamma_2)\,.
\ee
There is also a crucial sign coming from spectral networks rules, which must be taken into account \cite[eq. (4.1)]{Gaiotto:2012rg}. This comes from the extra unit of winding of the tangent vector of the representative of $\alpha$. Overall we get
\be
\begin{split}
	 \Tr F(\wp_A,\bCG) 
	& = (X_{\beta} - X_{\alpha})  \, \frac{1+X_{\gamma_2}+X_{\gamma_1+\gamma_2}}{1-X_{\gamma_1+\gamma_2+\gamma_3}}   \\
	& = X_{\beta} (1-X_{\alpha-\beta})\frac{1+X_{\gamma_2}+X_{\gamma_1+\gamma_2}}{1-X_{\gamma_1+\gamma_2+\gamma_3}}   \\
	& = X_{-\frac{1}{2}(\gamma_1+\gamma_2)} + X_{\frac{1}{2}(\gamma_2-\gamma_1)} + X_{\frac{1}{2}(\gamma_1+\gamma_2)}\,.
\end{split}
\ee
This result agrees with previous computations such as \cite[eq. (10.46)]{Gaiotto:2010be}.

A similar computation can be done for the B-cycle line operator, and we find
\be
\begin{split}
	\Tr F(\wp_B,\bCG)  & = %
	\Tr %
	\(\begin{array}{cc}	 X_{B'_1}	&	\\  	&	X_{B'_2}           \end{array}\) %
	\(\begin{array}{cc}	1 	&		\\ 	\Delta_2	&	1	\end{array}\) %
	\(\begin{array}{cc}	1 	&	\Upsilon_2	\\ 	&	1	\end{array}\) %
	\(\begin{array}{cc}	 	&	X_{B''_{12}}	\\  	X_{B''_{21}}	&	\end{array}\) \\%
	& = X_{-\frac{1}{2}(\gamma_1+\gamma_3)} + X_{\frac{1}{2}(\gamma_1-\gamma_3)} + X_{\frac{1}{2}(\gamma_1+\gamma_3)} \,.
\end{split}
\ee
We can now test the action of the $S^{-1}$ generator derived in (\ref{eq:N2star-S-generator}) according to our formula (\ref{eq:mcg-action-hol}).
Translating holonomies into cluster variables using the map $\rho$ in (\ref{eq:rho-map}), we find
\be
\begin{split}
	\kappa_{S^{-1}} \circ \rho \(\Tr F(\wp_A ,\bCG) \)& =  %
	\kappa_{S^{-1}} \( 	\frac{1}{\sqrt{y_1 y_2}} + \sqrt{\frac{y_2 }{y_1}}+ \sqrt{{y_1}{y_2}}	\) \\
	&= \frac{1+y_1}{\sqrt{y_1 y_3}} + \frac{\sqrt{y_1^3 y_3}}{1+ y_1} +\frac{\sqrt{y_1 y_3}}{1+y_1}  \\
	&= \frac{1}{\sqrt{y_1 y_3}} + \sqrt{\frac{y_1}{y_3}} + \sqrt{y_1 y_3}\\
	& = \rho \(\Tr F(\wp_B ,\bCG) \)\,
\end{split}
\ee
as expected, since $S^{-1}$ takes the $A$ cycle into the $B$ cycle.

\subsection{\texorpdfstring{$SU(3)$ $\CN=2^*$ theory}{SU(3) N=2* theory}}
\label{subsec:A3-torus}

\begin{figure}[htbp]
\begin{center}
\includegraphics[width=\textwidth]{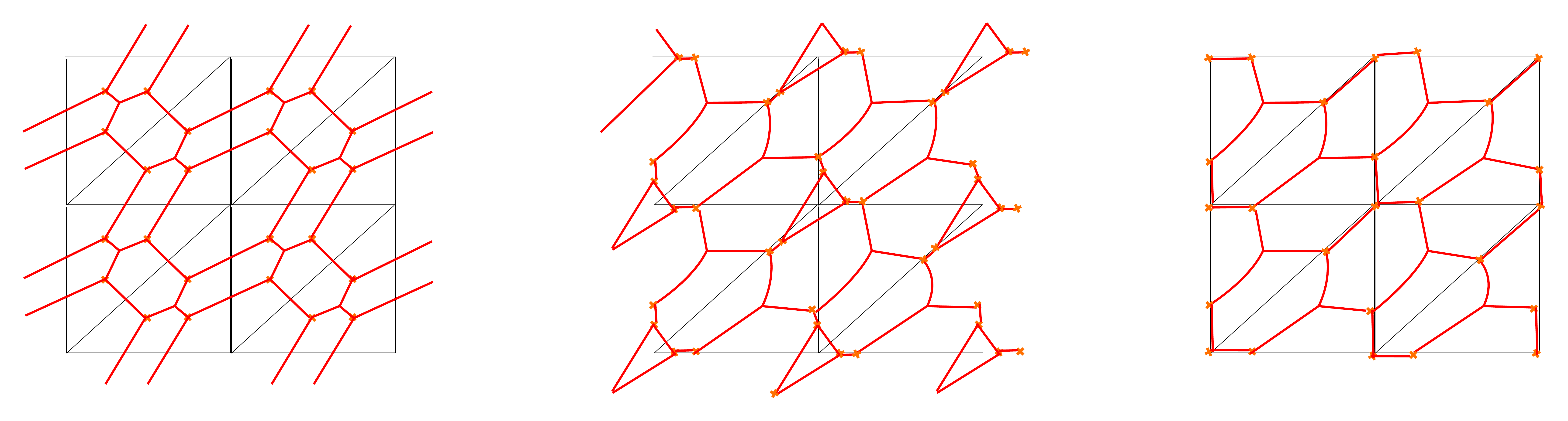}
\caption{Left: the BPS graph for the torus with a full puncture. Center: an equivalent BPS graph obtained by performing flips. Right: the BPS graph for the torus with a $[2,1]$ puncture, after reduction.}
\label{fig:reduction}
\end{center}
\end{figure}

We now move on to higher rank, and consider the class $\CS$ theory corresponding to an $A_2$ Hitchin system on a punctured torus with a minimal (a.k.a. simple) puncture. 
In order to obtain the BPS graph we adopt a procedure described in \cite{Gabella:2017hpz}. 
Let $\CG_0$ be BPS graph of the theory with a \emph{full} puncture, this is conjectured to be dual to a $3$-triangulation of the torus, see Figure \ref{fig:reduction}.
In order to obtain the BPS graph for the simple puncture, we consider a deformation of the mass moduli so as to partially close the puncture. 
This operation is expected to change the shape of the spectral network underlying the BPS graph in a precise way, which we can mimic using the standard flip and cootie moves on $\CG_0$.
The deformation of $\CG_0$ corresponds to the reduction of the puncture shown in Figure \ref{fig:reduction}, and the resulting graph coincides with the one we encountered in Figure \ref{fig:full-mcg-example}.
Note that $\CG$ has a $\IZ_3$ symmetry generated by shifting labels $\gamma_1\to\gamma_5\to \gamma_3$ and by simultaneously rotating the webs of $\gamma_2$ and $\gamma_4$ counter-clockwise by $2\pi/3$, so as to preserve the branch point at which $\gamma_1$ meets $\gamma_2$ and $\gamma_4$, and so on.
The elementary webs of $\CG$ and their respective lifts to homology classes on $\Sigma$ are
\be\label{eq:el-webs-21}
	\gamma_1 =h(e_2)\,,\qquad  \gamma_2=h( e_4,e_5,e_6)\,, \qquad \gamma_3 =h(e_1)\,,  \qquad \gamma_4  = h( e_7, e_8, e_9) \,,\qquad \gamma_5 = h(e_3)\,.
\ee
The dual quiver is shown in Figure \ref{fig:21quiver}, note that it inherits the $\IZ_3$ symmetry of $\CG$.\footnote{This quiver is related by a sequence of mutations to the one proposed in \cite{Gang:2015wya}.}

In order to study the mapping class group
let us fix $A,B$ cycles as in (\ref{eq:21AB}). 
The sequence of moves that generates the $S^{-1}$ transformation is shown in Figure \ref{fig:full-mcg-example}, and reads
\be
	\kappa_s \ : f_2 \circ f_9 \circ  \text{cootie}(e_5,e_6,e_7,e_8)\circ  f_1\circ f_2 \,,
\ee
where $f_i$ denotes a flip of edge $e_i$.
The graph obtained after applying $\kappa_s$ has a $\IZ_3$ symmetry, so there are three inequivalent relabelings $\kappa_r,\kappa_r',\kappa_r''$ which turn it back into the original graph. 
We define $\kappa_r$ as in (\ref{eq:21-Smove-relabel}), providing precisely the $S^{-1}$ transformation of the homology basis in (\ref{eq:21-S-hom-action}).

On the cluster variables $\kappa_s$ translates into a
sequence of mutations, which is followed by a permutation of the nodes corresponding to $\kappa_r$:
\be\label{eq:21-quiver-mcg-action}
\begin{split}
	\kappa_s : & \quad \mu_1 \circ \mu_4 \circ \mu_3 \circ \mu_1 \,,\\
	\kappa_r : & \quad 1\to 3\to 2 \to 5\to4\to 1  \,.
\end{split}
\ee
The action on the cluster variables is rather involved 
\be\label{eq:21-S-cluster-action}
\begin{split}
	y_1' = &
	{y_3 \left(  1+ y_1 \left(1+ y_3\right)\right)} \\
	& \(	
		1 
		+ y_1 2 \left(1+ y_3\right) 
		+ y_1^2 \left(1+ y_3\right) \left(1+ y_3 \left(1+y_4\right)\right) 
	\)^{-1}\,,\\
	y_2' =& 
	{y_1^3 y_3^2 y_4^2}
	\left(
		1 +  y_1 \left(2+y_3+2\right) + y_1^2 \left(1+y_3 \left(1+y_4\right)\right) 
	\right)^{-1} \\
	&
	\left(
		1 
		+ 2 y_1 \left(y_3+1\right) 
		+ y_1^2 \left(1+y_3\right) \left(1+ y_3 \left(1+ y_4\right)\right) 
	\right)^{-1}	\,,\\
	y_3' = & 
	{\left(
		1 + y_1 \left(1+y_3\right)
	\right) 
	\left(
		1 + y_1\left(2 + y_3\right)   +  y_1^2 \left(1+y_3 \left(1+y_4\right)\right) 
	\right)}
	{y_1^{-2} y_3^{-2} y_4^{-1}}\,,\\
	y_4' = & 
	y_5 \left(
		1 
		+ y_1 \left(4 + 3 y_3\right) 
		+y_1^2 \left(2+y_3\right) \left(3 + y_3 \left(3 + y_4\right)\right) 
	\right.\\
	&  
		\left.+ y_1^3 \left(4 + 5 y_3 + y_3^2\right) \left(1+ y_3 \left(1+y_4\right)\right) 
		+ y_1^4 \left(1+y_3\right) \left(1 +y_3 \left(1+ y_4\right)\right)^2 
	\right) \,,\\
	&	\({\left(1+ y_1 \left(1+ y_3\right)\right)^3}\)^{-1},\\
	y_5' = & 
		y_2 \left(1+ y_1 \left(1+ y_3\right)\right) 
		\left( 
			1 
			+2 y_1 \left(1+y_3\right) 
			+ y_1^2 \left(1+y_3\right) \left(1+ y_3 \left(1+y_4\right)\right) 
		\right)\\
	& \left(
		1
		+ y_1 \left(2+ y_3\right) 
		+ y_1^2 \left(1+ y_3 \left(1+ y_4\right)\right) 
	\right)^{-1} \,.
\end{split}
\ee

\begin{figure}[htbp]
\begin{center}
\includegraphics[width=\textwidth]{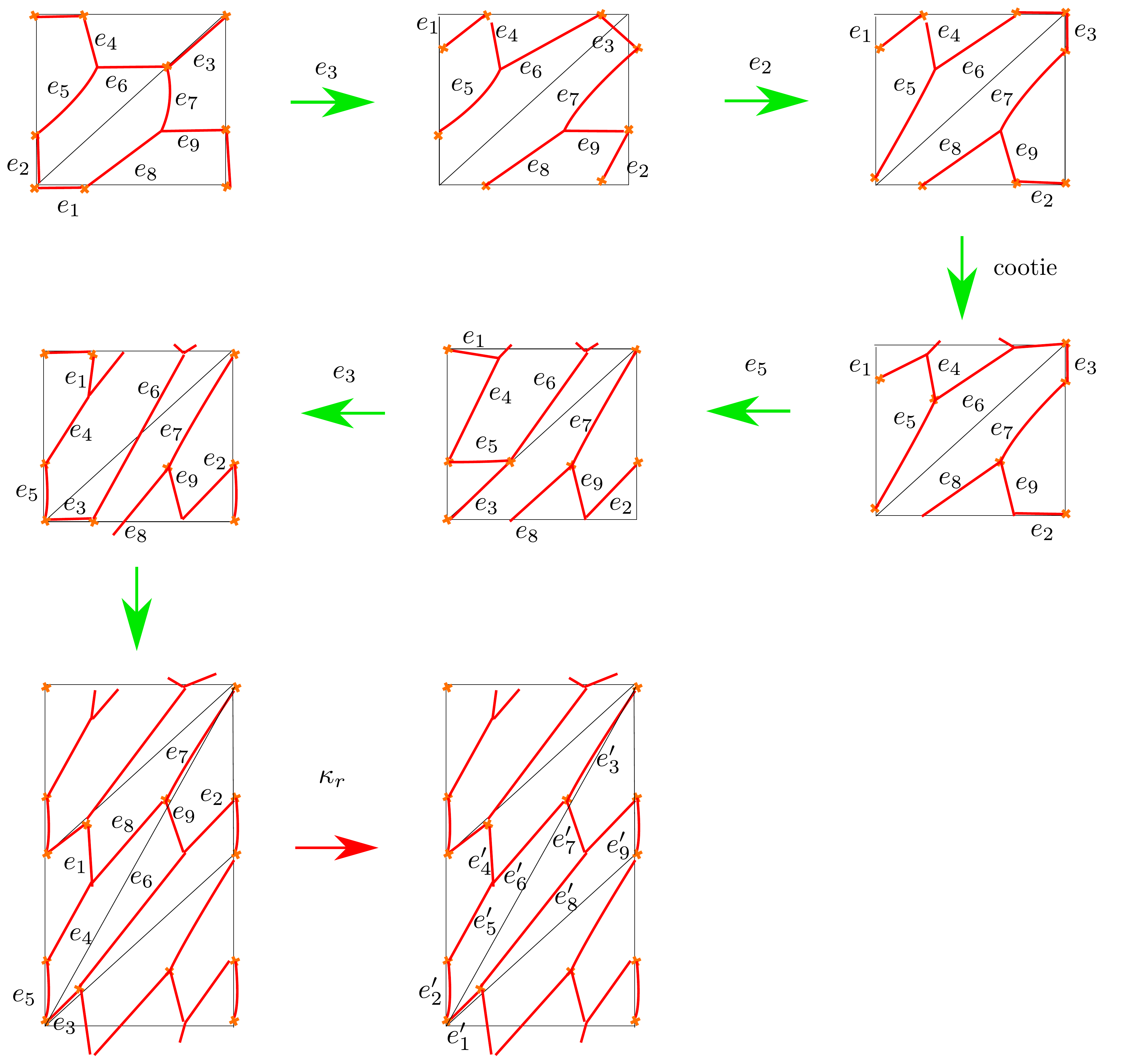}
\caption{The $L$-move for the $[2,1]$-punctured torus}
\label{fig:21-L-move}
\end{center}
\end{figure}

We also found a sequence of moves that corresponds to the $L$ generator of the $\MCG(C)$. 
In this case the sequence of flips and cooties $\kappa_s$ is shown in Figure \ref{fig:21-L-move}. 
We choose a relabeling of edges $\kappa_r$ which is 
\be
	\kappa_r : \quad e_1\to e_4 \to e_5 \to e_2 \to e_9 \to e_7 \to e_3\to e_1 \,.
\ee
The homology basis changes as follows
\be
	A' = A+B \,,\qquad B' = B\,,
\ee
implying that the complex structure of the torus transforms precisely by  the transformation
\be
	L=\(\begin{array}{cc}1 & 0 \\ 1 & 1\end{array}\)\,. 
\ee
The action on cluster variables is $\kappa_r\circ\kappa_s$ with 
\be
\begin{split}
	\kappa_s &: \quad   \mu_5 \circ \mu_2 \circ \mu_1 \circ \mu_5 \,,\\
	\kappa_r & : \quad 1\to 4 \to 5 \to 3 \to 2 \to 1 \,,
\end{split}
\ee
acting as follows
\be\label{eq:21-torus-L}
\begin{split}
	y_1' =  & 
	y_1
	\left(
		1 + y_5\left(1+ y_1\right) 
	\right)^{-1} 
	\left(
		1+ y_2 \left(1+y_1\right)  \left(1+ y_5 \left(1+ y_1\right) \right)
	\right)^{-1},\\
	y_2' =  & 
	y_1 y_2^2 y_3 \left(1 + y_5\left(1+y_1\right) \right)^3 \\
	&
	 \left(
	 	1 + y_2 \left(1+ y_1\right)  \left(1+ y_5 \left(1+ y_1\right) \right)
	\right)^{-1}
	\left(
		1 + y_5 +  y_2 \left(1+ y_5 \left(1+ y_1\right) \right)^2
	\right)^{-1},\\
	y_3' =  & 
	\left(
		1 
		+y_5
		+ y_2 \left(1 + y_5 \left(1+ y_1\right) \right)^2
	\right)
	\left(
		y_1 y_2 \left(1 +  y_5\left(1+y_1\right)\right)
	\right)^{-1},\\
	y_4' =  & 
	y_1^{-1} y_5^{-1} 
	\Big(
		1 
		+y_5
		+y_2 \left(
				2 \left(1+y_5\right)^2
				+ y_1 \left(1 +5 y_5 + 4 y_5^2\right) 
				+ y_5 y_1^2 \left(1+2 y_5\right) 
			\right)
		\\
		&\qquad \quad
	 	+y_2^2 \left(1+y_1\right) \left(1+ y_5\left(1+y_1\right) \right)^3
	\Big),\\
	y_5' =  & 
		 y_1 y_4 y_5^2 \left(1 + y_2 \left(1+ y_1\right)  \left(1 + y_5 \left(1+ y_1\right) \right)\right) \\
	&\left(
		 1+ y_5\left(1 +y_1\right) 
	\right)^{-1}
	\left(
		1 + y_5 + y_2 \left(1 + y_5 \left(1+y_1\right) \right)^2
	\right)^{-1}\,.
\end{split}
\ee
We checked that the $SL(2,\IZ)$ identities
\be
	(L S^{-1})^{3} = 1 \,, \qquad (S^{-1})^4 = 1\,,
\ee
are indeed satisfied by (\ref{eq:21-S-cluster-action}) and (\ref{eq:21-torus-L}).
Unlike for the case of $N=2$, now there is no ambiguity in the orientation of the $A,B$ cycles, simply because they are composed of more than two edges each. In fact, we find that $(S^{-1})^2\neq 1$ in this case, and we have the representation of the mapping class group
$SL(2, \mathbb{Z})$.

\subsection{\texorpdfstring{$SU(N)$ $\CN=2^*$ theory}{SU(N) N=2* theory}}\label{sec:N-torus}

\begin{figure}[htbp]
\begin{center}
\includegraphics[width=.9\textwidth]{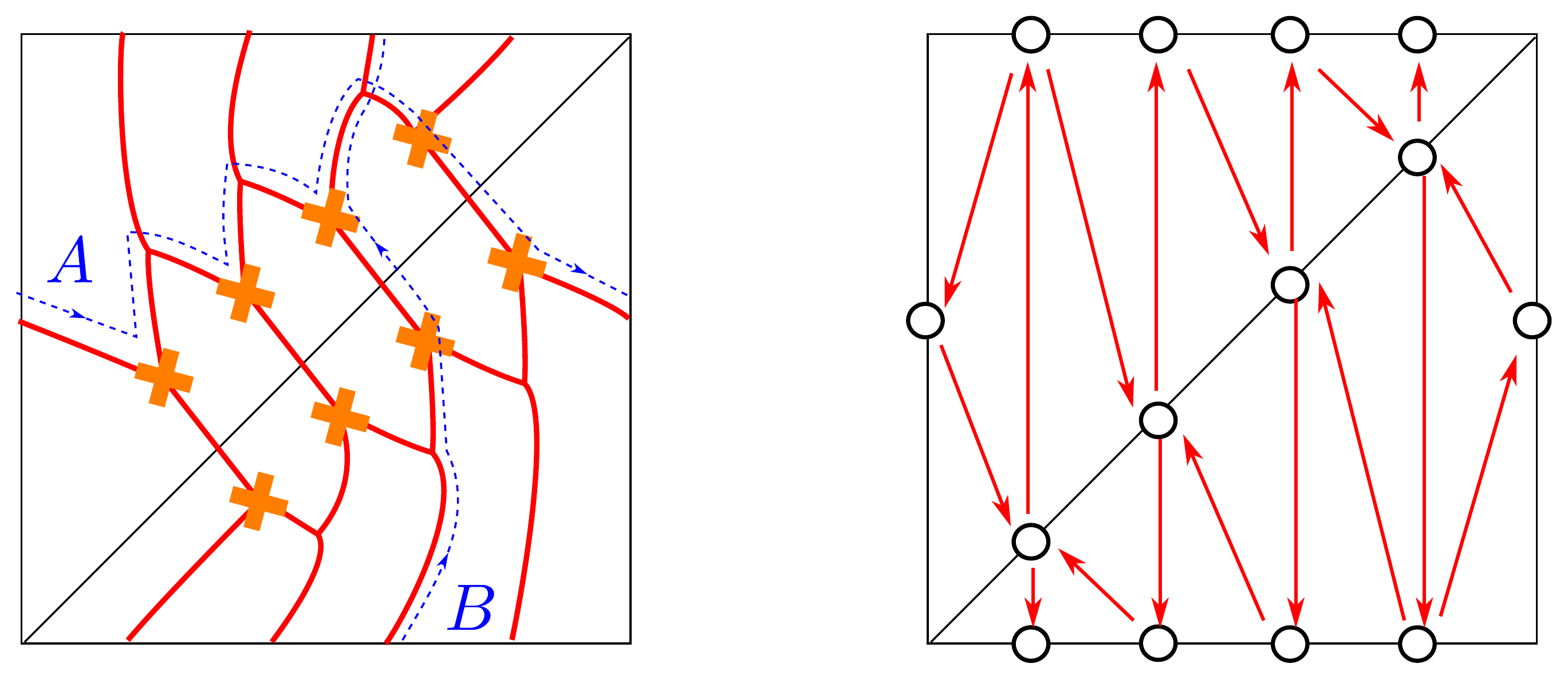}
\caption{BPS graph for the torus with a $[N-1,1]$ puncture, and the dual quiver. There are $N-1$ nodes on the horizontal and diagonal edges.}
\label{fig:selfglued-quiver}
\end{center}
\end{figure}

We wish to generalize the analysis of the previous two subsections to punctured tori with a simple puncture, for Lie algebra $A_{N-1}$.
Finding the BPS graph of these models in full generality is rather challenging, even resorting to the puncture-reduction techniques illustrated in the previous subsection.\footnote{We were able to obtain the BPS graph for the torus with a $[3,1]$ puncture by reducing the BPS graph for the full puncture. We also checked that it is related to the guess of Figure \ref{fig:selfglued-quiver} by a sequence of flips and cooties.}

On the other hand we were able to find a natural a guess for the BPS graphs of these theories, for general $N$
(this is supported by a heuristic argument presented in Appendix \ref{app.degeneration}, which also can deal with
more general punctures of type $[k,1, \dots, 1]$). 
The graph and the dual quiver are shown in Figure \ref{fig:selfglued-quiver}.
The complex dimension of the Coulomb branch is $N-1$ while the flavor symmetry is $U(1)$, therefore the charge lattice has rank $2N-1$, and this coincides with the elementary webs of $\CG$.
The quiver agrees in fact with previous proposals \cite{Xie:2012dw, Gang:2015wya} and, in the case of $N=3$ it is mutation-equivalent to the one in Figure \ref{fig:21quiver}.  
We tested our guess by using the BPS graphs to derive generators of the mapping class group, and checking that they produce consistent representations of $\MCG(C)$ on the cluster algebra for $N=3,4,5$.

The $A$ and $B$ cycles are defined as oriented sequences of edges,  depicted by dashed lines in Figure \ref{fig:selfglued-quiver}.
We were able to find a sequence of moves $\kappa_s$ that corresponds to the element $L$ of the mapping class group. The dual sequence of mutations is 
\be
 	\kappa_s : \quad \prod_{i\in I_H} \mu_i\,,
\ee
where $I_H$ is set of quiver nodes on the horizontal edge in Figure \ref{fig:selfglued-quiver}, and the ordering is irrelevant since these nodes are not connected by arrows. 
Together with these mutations, in order to obtain the $L$ transformation, one must apply the following relabeling of quiver nodes
\be
	\kappa_r : \quad  \left\{%
	\gamma^{(H)}_i \leftrightarrow \gamma^{(D)}_i %
	\right\}_{i=1,\dots,N-1} \,,
\ee
where $\gamma^{(D/H)}$ denote the nodes on the diagonal/horizontal edge respectively. These nodes get exchanged with each other pairwise.
The action of $L$ on the cluster variables is 
\be
	\kappa_L = \kappa_s\circ \kappa_r\,.
\ee

\begin{figure}[htbp]
\begin{center}
\includegraphics[width=.3\textwidth]{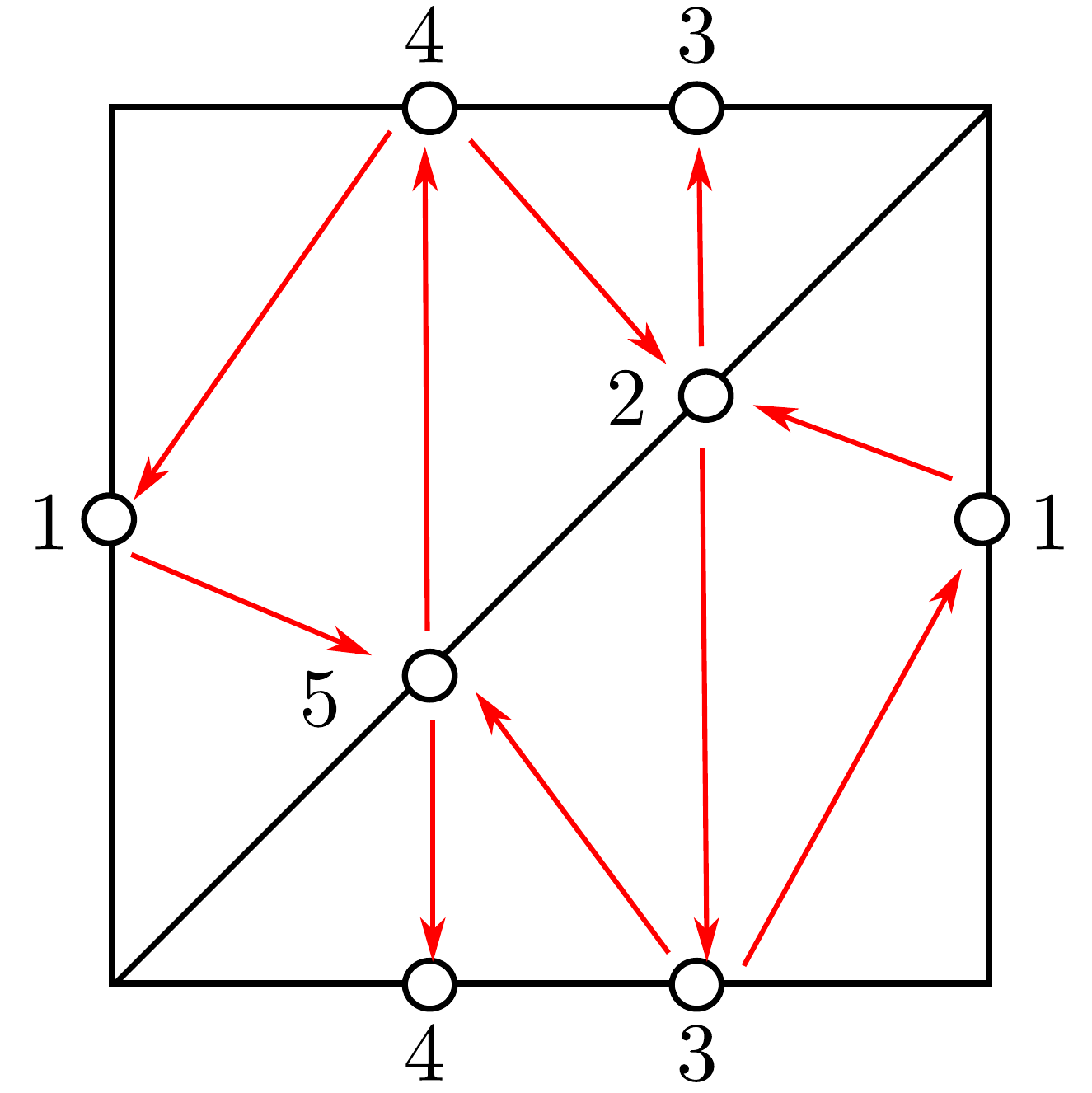}
\includegraphics[width=.3\textwidth]{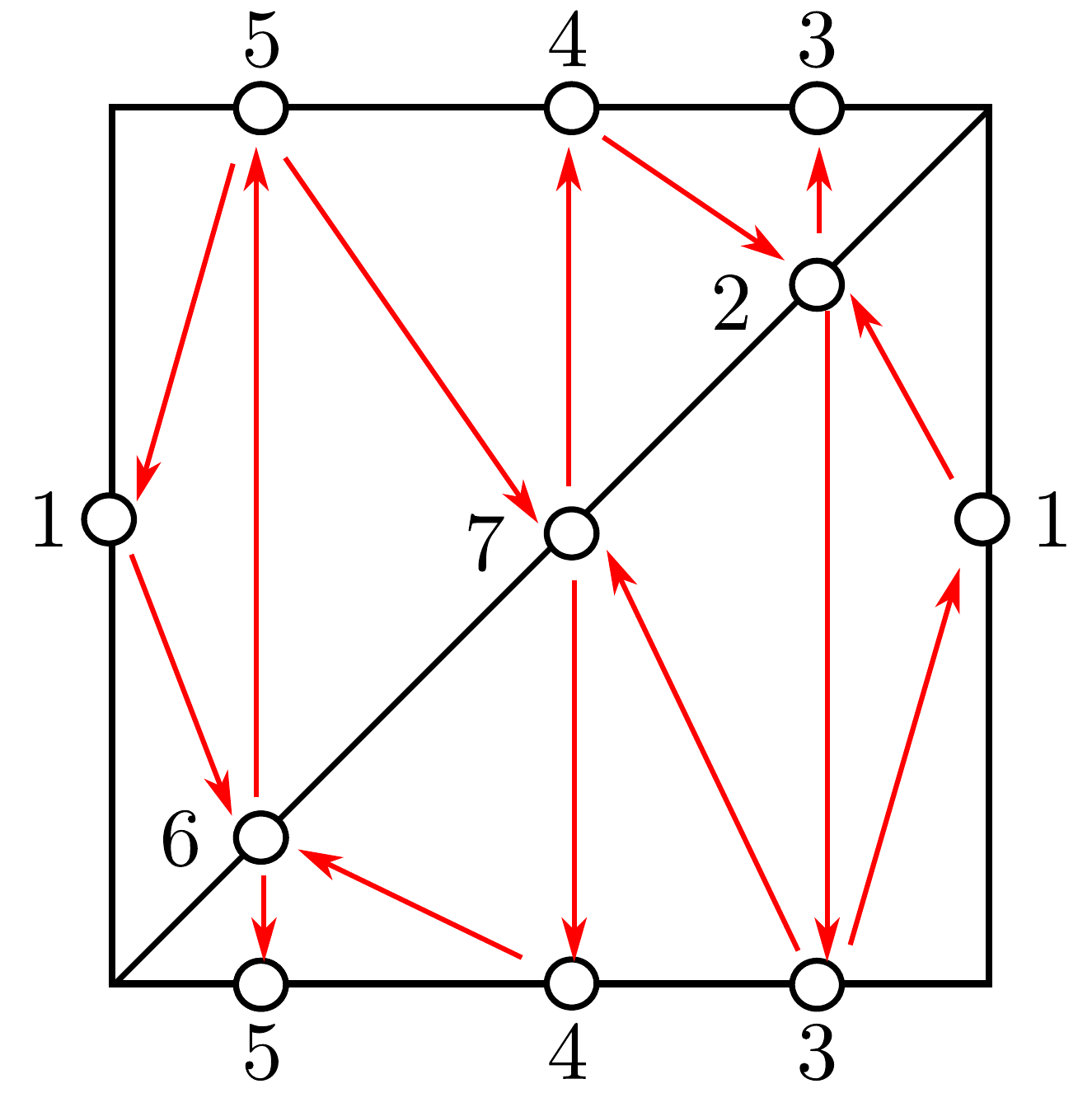}
\includegraphics[width=.3\textwidth]{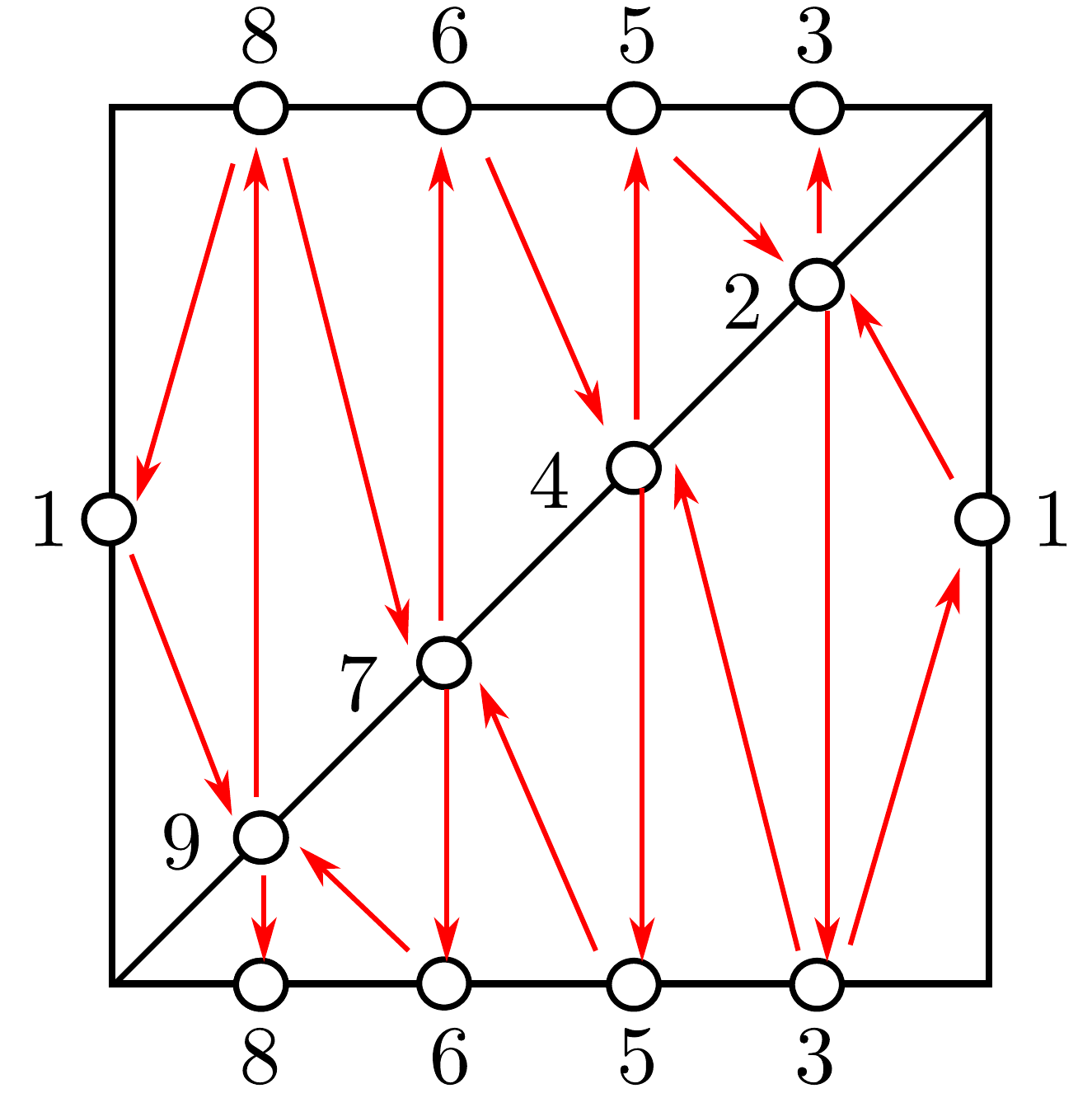}
\end{center}
\caption{BPS quivers for the torus with a minimal puncture, for $N=3,4,5$.}
\label{fig:general-quiver}
\end{figure}

Finding other generators of the mapping class group is feasible, but much more challenging. We were not able to find a general formula valid for all $N$, but for the cases $N=3,4,5$ we found several possible choices for a second generator of $\MCG(C)$. 
They are reported in the tables below, where labels refer to Figure \ref{fig:general-quiver}.
Of course, there may be several mutations corresponding to a single a MCG transformation, and they should all be equivalent (for example, via wall-crossing identities).
In writing the data we adopt the convention that parentheses $(\mu_a\circ\mu_b)$ stand for either $\mu_a\circ\mu_b$ or $\mu_b\circ\mu_a$,
when the ordering does not matter.

\begin{center}
\begin{tabular}{|c|c|l|}
	\hline
	\multicolumn{3}{|c|}{$\bm{N=3}$} \\
	\hline
	$S$ & %
	$\left(\begin{array}{cc} 0 & 1 \\ -1 & 0 \end{array}\right)$ & %
	\begin{tabular}{l}
		$\kappa_s: \, \mu_1$ \\
		$\kappa_r: \,  3\to5\to4\to2\to3$
	\end{tabular}\\
	\hline
	$L$ & %
	$\left(\begin{array}{cc} 1 & 0 \\ 1 & 1 \end{array}\right)$ & %
	\begin{tabular}{l}
		$\kappa_s: \, (\mu_3 \circ \mu_4)$ \\
		$\kappa_r: \, 2\leftrightarrow 3, \ 4\leftrightarrow 5 $
	\end{tabular}\\
	\hline
	$LS$ & %
	$\left(\begin{array}{cc} 0 & 1 \\ -1 & 1 \end{array}\right)$ & %
	\begin{tabular}{l}
		$\kappa_s: \, (\mu_2 \circ \mu_5 )\circ \mu_1$ \\
		$\kappa_r: \, 2\leftrightarrow 5 $
	\end{tabular}\\
	\hline
	$LS^{-1} L^{-1}$ & %
	$\left(\begin{array}{cc} 1 & -1 \\ 2 & -1 \end{array}\right)$ & %
	\begin{tabular}{l}
		$\kappa_s: \, (\mu_3 \circ \mu_4)\circ \mu_1 \circ (\mu_2 \circ \mu_5)$ \\
		$\kappa_r: \, 2\to 3\to 5\to 4\to 2 $
	\end{tabular}\\
	\hline
	$S^{-1} L^{-1}$ & %
	$\left(\begin{array}{cc} 1 & -1 \\ 1 & 0 \end{array}\right)$ & %
	\begin{tabular}{l}
		$\kappa_s: \,  \mu_1 \circ (\mu_2 \circ \mu_5)$ \\
		$\kappa_r: \, 3\leftrightarrow 4$
	\end{tabular}\\
	\hline
	$L^{-1} S^{-1}$ & %
	$\left(\begin{array}{cc} 0 & -1 \\ 1 & 1 \end{array}\right)$ & %
	\begin{tabular}{l}
		$\kappa_s: \,  (\mu_3 \circ \mu_4)\circ \mu_1$ \\
		$\kappa_r: \, 2\leftrightarrow 5$
	\end{tabular}\\
	\hline
	$ SL$ & %
	$\left(\begin{array}{cc} 1 & 1 \\ -1 & 0 \end{array}\right)$ & %
	\begin{tabular}{l}
		$\kappa_s: \, \mu_1\circ (\mu_3 \circ \mu_4) $ \\
		$\kappa_r: \, 2\leftrightarrow 5$
	\end{tabular}\\
	\hline
\end{tabular}

\medskip

\begin{tabular}{|c|c|l|}
	\hline
	\multicolumn{3}{|c|}{$\bm{N=4}$} \\
	\hline
	$S$ & %
	$\left(\begin{array}{cc} 0 & 1 \\ -1 & 0 \end{array}\right)$ & %
	\begin{tabular}{l}
		$\kappa_s: \,    \mu_1 \circ \mu_4 \circ \mu_2\circ \mu_6\circ \mu_4\circ \mu_5\circ \mu_3\circ \mu_1 $ \\
		$\kappa_r: \,   1\leftrightarrow 4, \ 2\to 5\to 6\to 3\to 2  $
	\end{tabular}\\
	\hline
	$L$ & %
	$\left(\begin{array}{cc} 1 & 0 \\ 1 & 1 \end{array}\right)$ & %
	\begin{tabular}{l}
		$\kappa_s: \,    (\mu_3 \circ\mu_5\circ  \mu_4) $ \\
		$\kappa_r: \,    2\leftrightarrow 3,\  4\leftrightarrow 7,\   5\leftrightarrow 6    $
	\end{tabular}\\
	\hline
	$LS$ & %
	$\left(\begin{array}{cc} 0 & 1 \\ -1 & 1 \end{array}\right)$ & %
	\begin{tabular}{l}
		$\kappa_s: \,    \mu_4\circ (\mu_2 \circ \mu_6 )\circ (\mu_3\circ\mu_5)\circ \mu_1 $ \\
		$\kappa_r: \,   1\to 7\to 4\to 1   $
	\end{tabular}\\
	\hline
	$LS^{-1} L^{-1}$ & %
	$\left(\begin{array}{cc} 1 & -1 \\ 2 & -1 \end{array}\right)$ & %
	\begin{tabular}{l}
		$\kappa_s: \,   \mu_1\circ \mu_7\circ (\mu_3 \circ \mu_5)\circ (\mu_2 \circ \mu_6)\circ \mu_7\circ \mu_1   $ \\
		$\kappa_r: \,  1 \leftrightarrow 7, \ 2\to3\to6\to5\to2     $
	\end{tabular}\\
	\hline
	$S^{-1} L^{-1}$ & %
	$\left(\begin{array}{cc} 1 & -1 \\ 1 & 0 \end{array}\right)$ & %
	\begin{tabular}{l}
		$\kappa_s: \,    \mu_7\circ  (\mu_3 \circ \mu_5)\circ (\mu_2 \circ \mu_6)\circ \mu_1 $ \\
		$\kappa_r: \,    1\to 4\to 7\to 1  $
	\end{tabular}\\
	\hline
\end{tabular}

\medskip

\begin{tabular}{|c|c|l|}
	\hline
	\multicolumn{3}{|c|}{$\bm{N=5}$} \\
	\hline
	$S$ & %
	$\left(\begin{array}{cc} 0 & 1 \\ -1 & 0 \end{array}\right)$ & %
	\begin{tabular}{l}
		$\kappa_s: \,$\begin{tabular}{l}
	$\mu_4 \circ \mu_1 \circ \mu_2 \circ \mu_9 \circ \mu_3 \circ \mu_2 \circ \mu_6 \circ \mu_5 \circ \mu_4 \circ \mu_1 \circ \mu_9 \circ \mu_2 \circ \mu_7$ \\
	$ \circ \mu_4 \circ \mu_8 \circ \mu_1 \circ \mu_9 \circ \mu_2 \circ \mu_7 \circ \mu_8 \circ \mu_3 \circ \mu_1 \circ \mu_8 \circ \mu_6 \circ \mu_5 \circ \mu_1$ 
	\end{tabular}\\ 
		$\kappa_r: \,   1\to 6\to 2\to 3\to 8\to 7\to 4\to 5\to 9\to 1   $
	\end{tabular}\\
	\hline
	$L$ & %
	$\left(\begin{array}{cc} 1 & 0 \\ 1 & 1 \end{array}\right)$ & %
	\begin{tabular}{l}
		$\kappa_s: \,   (\mu_3 \circ\mu_5\circ  \mu_6\circ \mu_8)   $ \\
		$\kappa_r: \,     2\leftrightarrow 3, \ 4\leftrightarrow 5, \ 6\leftrightarrow 7, \ 8\leftrightarrow 9$
	\end{tabular}\\
	\hline
	$L^2S^{-1} L^{-1}$ & %
	$\left(\begin{array}{cc} 1 & -1 \\ 3 & -2 \end{array}\right)$ & %
	\begin{tabular}{l}
		$\kappa_s: \, $\begin{tabular}{l}
		$\mu_{1}\circ  \mu_{2}\circ  \mu_{9}\circ  \mu_{1}\circ  \mu_{7}\circ  \mu_{4}\circ  \mu_{6}\circ  \mu_{5}$\\
		$\circ  \mu_{2}\circ  \mu_{9}\circ  \mu_{7}\circ  \mu_{4}\circ  \mu_{8}\circ  \mu_{3}\circ  \mu_{2}\circ  \mu_{9}\circ  \mu_{1}$ 
	\end{tabular}\\
		$\kappa_r: \,     2\leftrightarrow 9, \ 3\to 6\to 4\to 3, \ 5\to 7\to 8\to 5 $
	\end{tabular}\\
	\hline
\end{tabular}
\end{center}

We checked that the cluster transformations generated by these sequences, composed with appropriate permutations of nodes, correctly reproduce the $SL(2,\IZ)$ algebra.

As explained in Section \ref{sec:line-operators},
if our guess for the BPS graph actually arises from a spectral network $\bCG$,
the latter could be used to compute framed BPS spectra, and  our construction of $\MCG(C)$ would act on  the moduli spaces of flat $GL(N)$ connections over $C$.
One way to test this would be simply to draw the ``completion'' of $\CG$ into a spectral network, in the spirit of \emph{general spectral networks} \cite[Section 9]{Gaiotto:2012rg}.
Another possibility would be to test the generators of $\MCG(C)$ on explicit expressions for the VEVs of UV line operators.

\subsection{\texorpdfstring{$SU(2)$ $N_f=4$ theory}{SU(2) Nf=4 theory}}\label{sec:su2nf4}

\begin{figure}[htbp]
\begin{center}
\includegraphics[width=.9\textwidth]{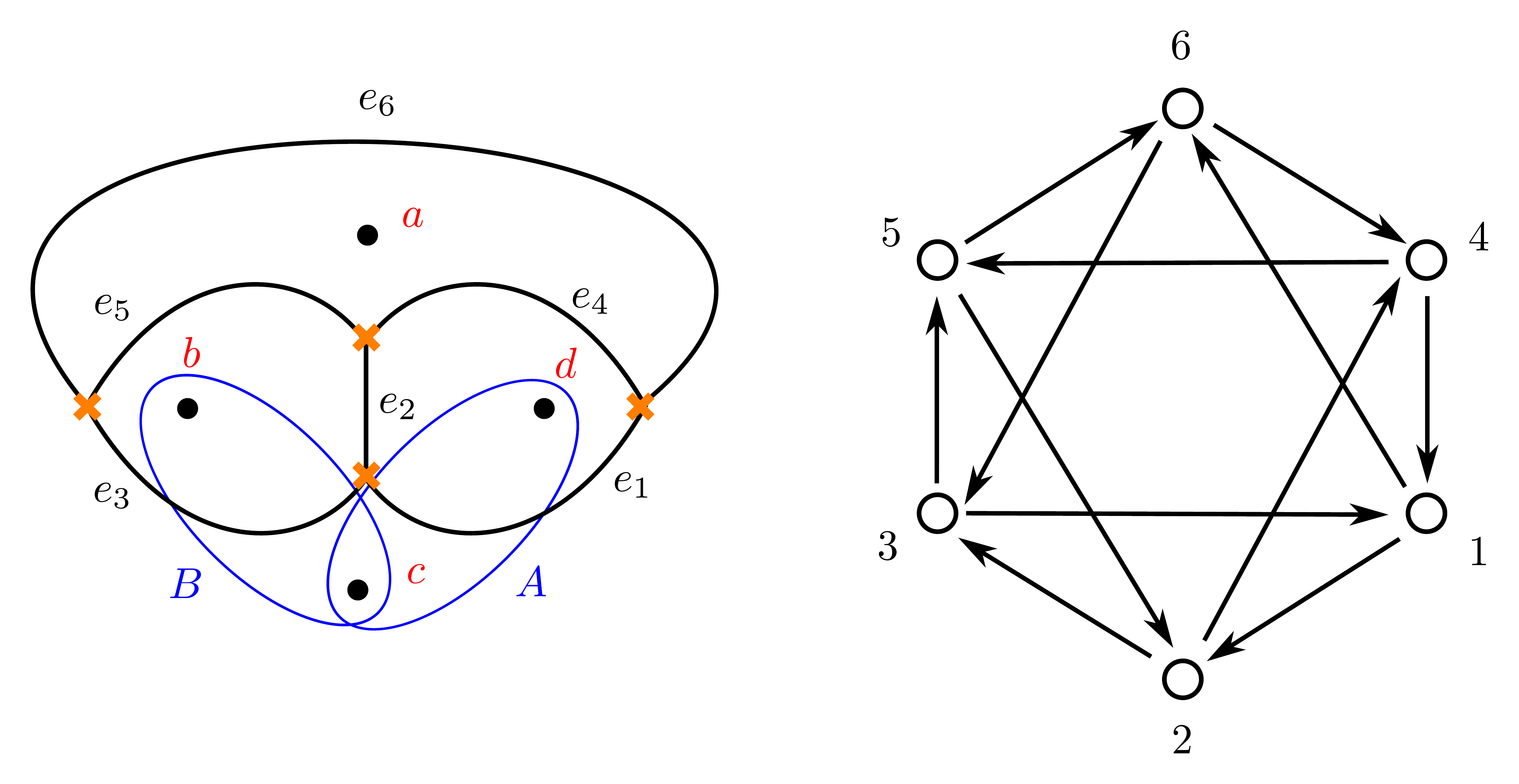}
\end{center}
\caption{The BPS graph of $SU(2)$ $N_f=4$ theory, and its quiver.}
\label{fig:su2nf4}
\end{figure}

The $SU(2)$ gauge theory with four fundamental hypermultplets is realized as an $A_1$ theory of class $\CS$ on a four-punctured sphere, with regular punctures.
Its BPS graph is shown in Figure \ref{fig:su2nf4}.
$\CG$ has six edges, each is mapped to a generator $\gamma_i=h(e_i)$ of $H_1(\Sigma,\IZ)$ with intersection pairing
\be\label{eq:su2nf4-quiver}
	\langle\gamma_i,\gamma_j\rangle = 
\left(
\begin{array}{cccccc}
 0 & -1 & 1 & 1 & 0 & -1 \\
 1 & 0 & -1 & -1 & 1 & 0 \\
 -1 & 1 & 0 & 0 & -1 & 1 \\
 -1 & 1 & 0 & 0 & -1 & 1 \\
 0 & -1 & 1 & 1 & 0 & -1 \\
 1 & 0 & -1 & -1 & 1 & 0 \\
\end{array}
\right)\,.
\ee

To construct the mapping class group, it will be convenient to adopt a slightly different approach than we did with the torus. Instead of working with homology cycles on $C$, it is  simpler to introduce a system of arcs connecting the punctures, and to keep track of how they evolve under the action of $\MCG(C)$.

\begin{figure}[htbp]
\begin{center}
\includegraphics[width=.9\textwidth]{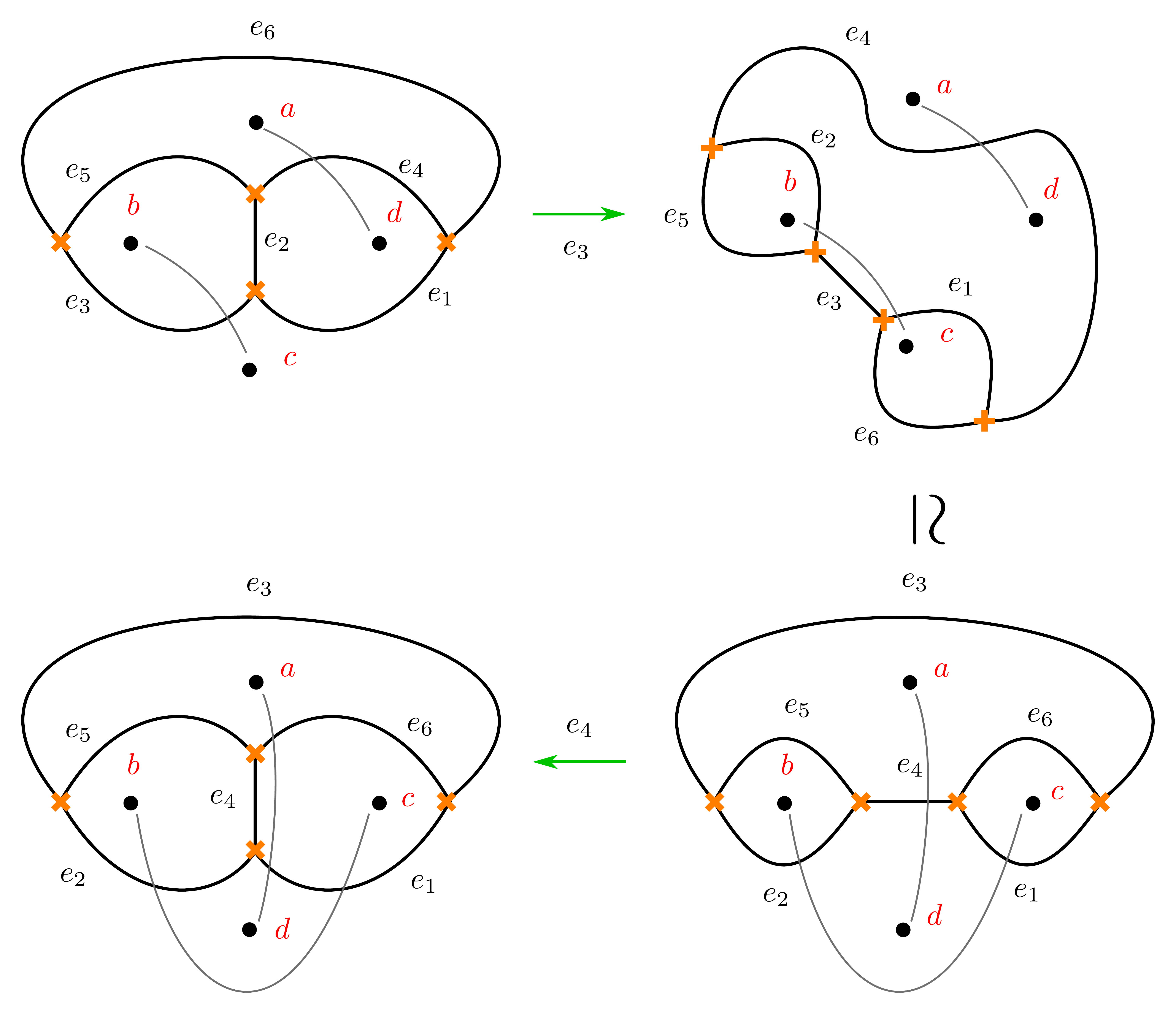}
\end{center}
\caption{A half Dehn twist along the $A$ cycle on $C$ exchanges punctures $c\leftrightarrow d$. In our setup however this should be thought as keeping punctures fixed, and acting the the inverse half-twist on $\CG$.}
\label{fig:su2nf4-DehnA}
\end{figure}

Performing a flip on edge $e_3$ followed by a flip on edge $e_4$, then relabeling according to 
\be\label{eq:su2nf4-dehnA-perm}
	\kappa_r : 2\to 3\to 6\to 4\to2
\ee
produces the BPS graph of Figure \ref{fig:su2nf4-DehnA} (to avoid clutter, we draw only two arcs). 
The picture is slightly deceiving, at first it appears that the BPS graph went back to itself while the Riemann surface $C$ was acted upon with a Dehn half-twist along the $A$ cycle. 
However just the opposite happened. We should regard $C$ as fixed, like in subsection \ref{subsec:A1-torus}, while the BPS graph was acted upon by an \emph{inverse} Dehn half-twist.
To obtain the full (inverse) Dehn twist we can simply apply the move twice.
The action on cluster variables corresponds to a mutation on node $3$, followed by a mutation on node $4$ and by a permutation of nodes as in (\ref{eq:su2nf4-dehnA-perm}). The result is
\be\label{eq:cluster-dehn-A}
\begin{split}
	y_1' &=  y_1  \(1+ y_3\) \left(1+ y_4\right) \,,\quad
	y_2' = y_4^{-1} \,,\quad
	y_3' = {y_2 y_3 y_4}\(1+ y_3\)^{-1}\( 1+ y_4\)^{-1}\,,\\
	y_4' &= {y_3 y_4 y_6}\(1+ y_3\)^{-1}\( 1+ y_4\)^{-1}\,,\quad
	y_5' = y_5  \(1+ y_3\) \left(1+ y_4\right) \,,\quad
	y_6' = y_3^{-1} \,.
\end{split}
\ee

\begin{figure}[htbp]
\begin{center}
\includegraphics[width=.9\textwidth]{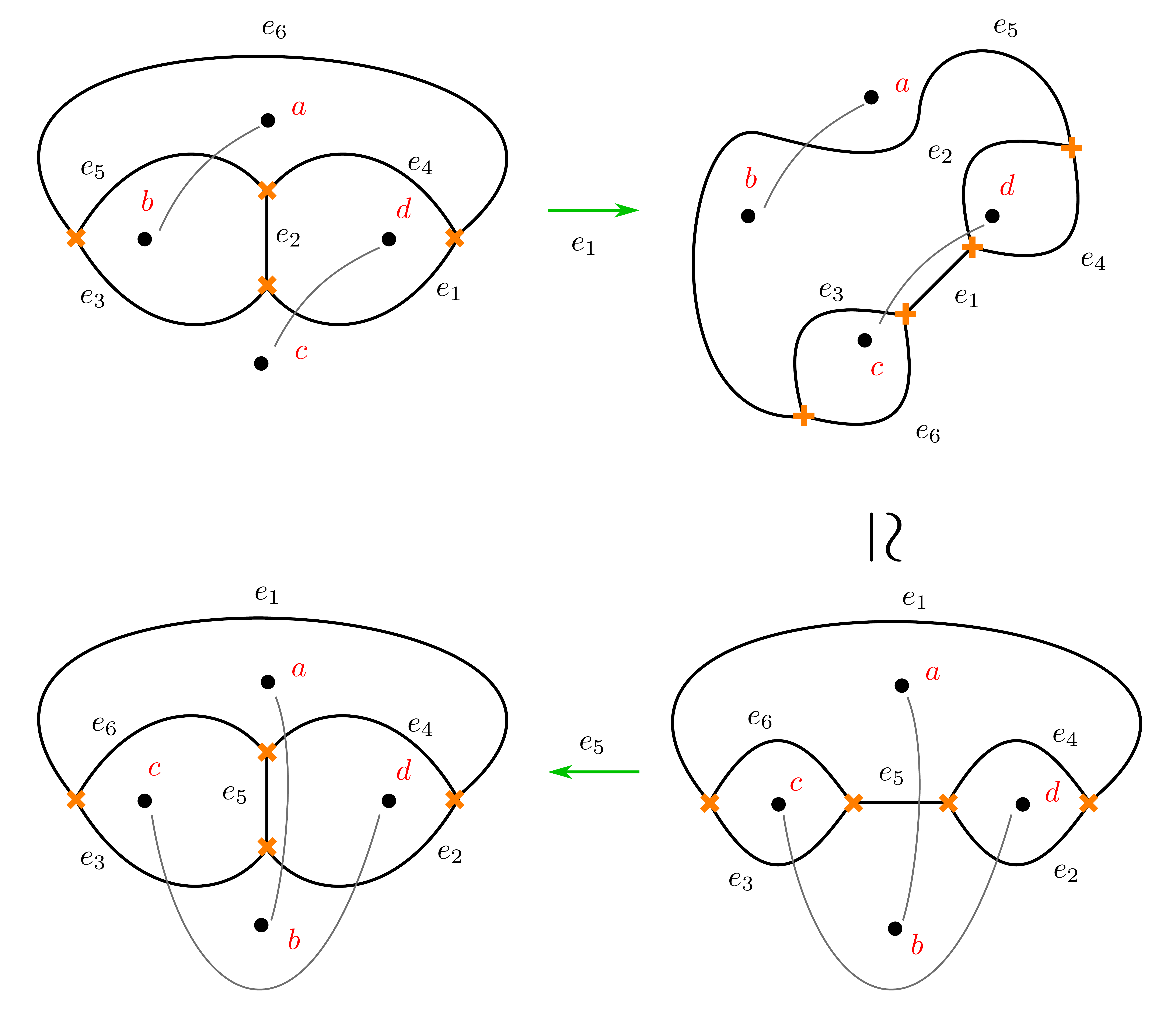}
\end{center}
\caption{An inverse half Dehn twist along the $B$ cycle exchanges punctures $b\leftrightarrow c$. 
In our setup however this should be thought as keeping punctures fixed, and acting the the positive half-twist on $\CG$.
}
\label{fig:su2nf4-DehnB}
\end{figure}

It is straightforward to generate other Dehn twists, by using the symmetries of $\CG$ viewed as a tetrahedron.
For example, a twist around the $B$-cycle can be generated by flipping edge $e_1$ and subsequently edge $e_5$, then relabeling according to 
\be
	\kappa_r : 1\to 6\to 5\to 2\to 1\,.
\ee
This produces the inverse of a half-Dehn twist along the $B$-cycle, see Figure \ref{fig:su2nf4-DehnB}.
The action on cluster variables in this case is
\be\label{eq:cluster-dehn-B}
\begin{split}
	y_1' &=  y_2  \(1+ y_1\) \left(1+ y_5\right) \,,\quad
	y_2' = y_5^{-1} \,,\quad
	y_3' = {y_1 y_3 y_5}\(1+ y_1\)^{-1}\( 1+ y_5\)^{-1} \,,\\
	y_4' &= {y_1 y_4 y_5}\(1+ y_1\)^{-1}\( 1+ y_5\)^{-1}\,,\quad
	y_5' = y_6  \(1+ y_1\) \left(1+ y_5\right) \,,\quad
	y_6' = y_1^{-1} \,.
\end{split}
\ee

\subsubsection{Line operators}

\begin{figure}[htbp]
\begin{center}
\includegraphics[width=.99\textwidth]{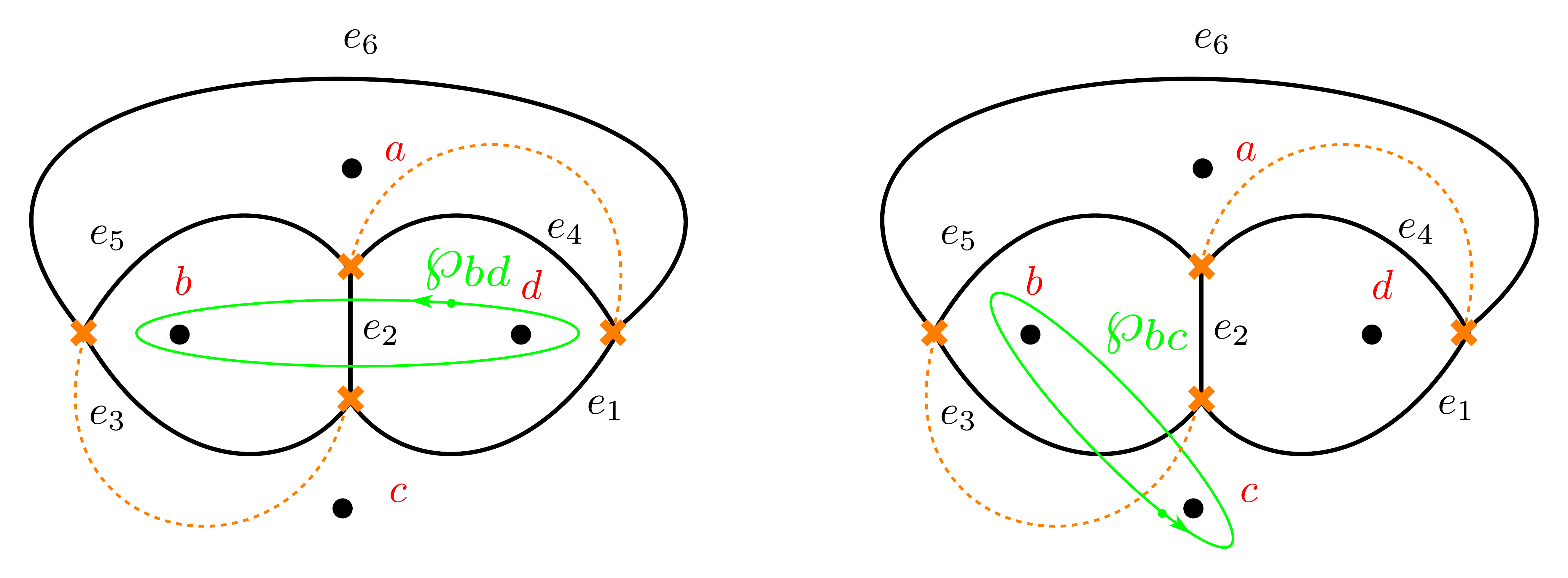}
\end{center}
\caption{Left: The path $\wp_{bd}$ labeling a line operator. Right: The path $\wp_{bc}$ obtained from $\wp_{bd}$ after a half Dehn twist around the A-cycle.}
\label{fig:su2nf4-line}
\end{figure}

Consider a line operator  labeled by a path $\wp_{bd}$ circling counterclockwise punctures $b$ and $d$, as shown in Figure \ref{fig:su2nf4-line}. 
The cycle $\wp_{bd}$ crosses only edge $e_2$, therefore we only need the generating functions $\Upsilon_2, \Delta_2$ of solitons running up/down on this edge. 
The resolved spectral network underlying $\CG$ will be denoted by $\bCG$ as usual.
Since we are working in the American resolution, $\Upsilon_2$ is the soliton data of the oriented edge that runs upward \emph{on the right}.
Let $a_2, b_2$ be the solitons sourced at the upper/lower branch points of edge $e_2$, and define $\hat\Upsilon, \hat\Delta$ by 
\be
	\Upsilon = X_{b_2} \hat\Upsilon_2\,,\qquad \Delta = X_{b_2} \hat\Delta_2\,.
\ee
These reduced generating functions are readily obtained\footnote{
The computation of soliton data is nearly identical to the one detailed in  \cite[Section 4.7]{Longhi:2016wtv}, with one important difference: here we are working with the \emph{American} resolution of the BPS graph.
}
using the soliton rules for the spectral network $\bCG$ obtained by resolving $\CG$ \cite{Longhi:2016wtv}
\be
\begin{split}
	\hat\Upsilon_2 & = \frac{1 + X_{\gamma_1} \( 1+ X_{\gamma_4} \)}{1-X_{\gamma_1+\gamma_2 + \gamma_4}} \,,\\
	\hat\Delta_2 & = \frac{1 + X_{\gamma_5} \(  1+ X_{\gamma_3} \)}{1-X_{\gamma_2+\gamma_3 + \gamma_5}}  \,.
\end{split}
\ee

\begin{figure}[htbp]
\begin{center}
\includegraphics[width=.99\textwidth]{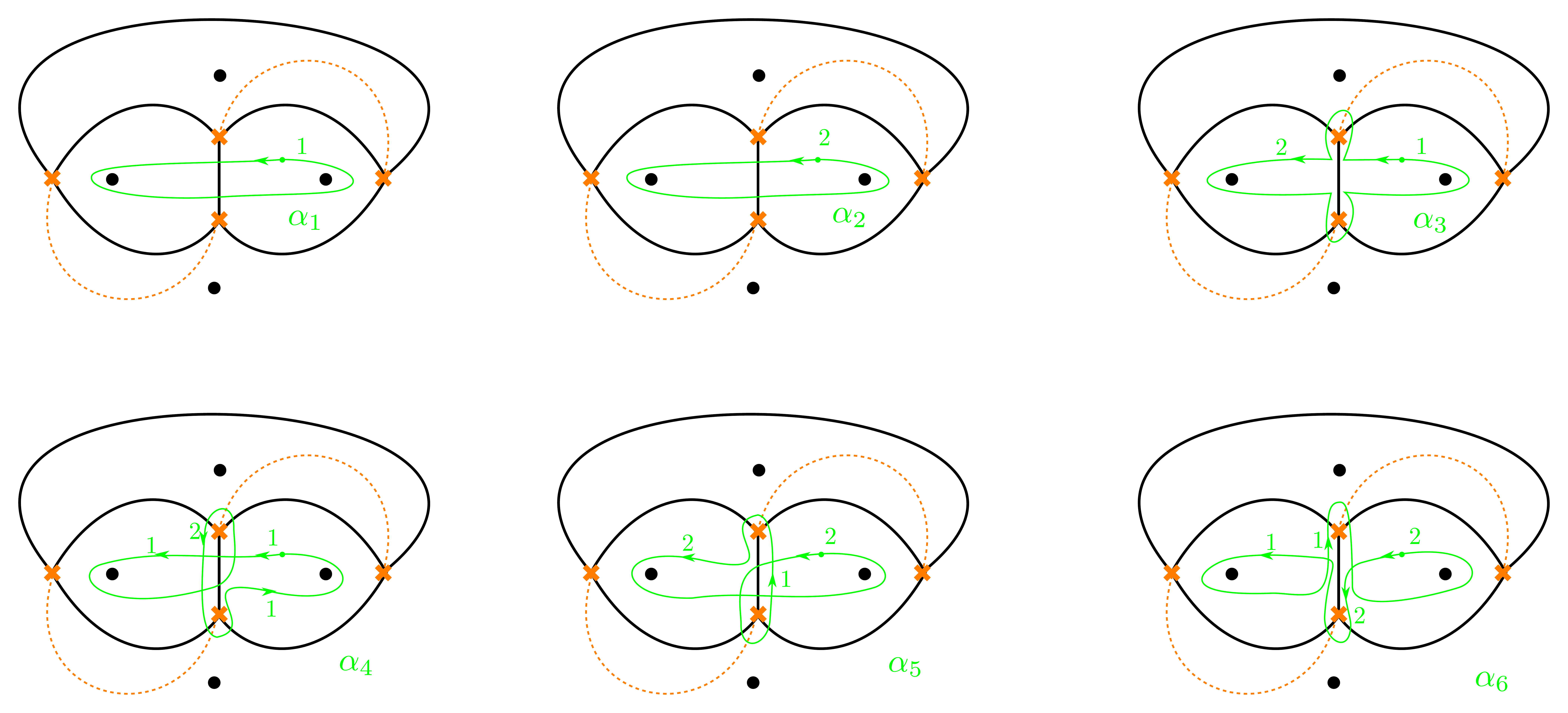}
\end{center}
\caption{The cycles $\alpha_1\dots\alpha_6$. The numbers $1,2$ near each path specify on which sheet of $\Sigma$ the path is running.}
\label{fig:su2nf4-cycles}
\end{figure}

With the choice of trivialization of Figure \ref{fig:su2nf4-line}, the solitons encoded by $\Upsilon_2$ are of type $21$, while those in $\Delta_2$ are of type $12$. Let us split $\wp_{bd}=\wp'\wp''\wp'''$ into three segments, running respectively from the basepoint to the first intersection with $e_2$, between the two intersections with $e_2$, and from the second intersection with $e_2$  back to the basepoint.
The formal parallel transport around $\wp_{bd}$ is
\be\label{eq:su2nf4-F}
\begin{split}
	\Tr F(\wp_{bd},\bCG)  & = %
	\Tr \left[ %
	\(\begin{array}{cc}	 X_{\wp'_1}	&	\\  	&	X_{\wp'_2}           \end{array}\) %
	\(\begin{array}{cc}	1 	&		\\ 	\Upsilon_2	&	1	\end{array}\) %
	\(\begin{array}{cc}	1 	&	\Delta_2	\\ 	&	1	\end{array}\) %
	\(\begin{array}{cc}	 X_{\wp''_1}	&	\\  	&	X_{\wp''_2}           \end{array}\) 
	\right. \\
	& 
	\qquad\quad \left.
	\(\begin{array}{cc}	1 	&	\Delta_2	\\ 	&	1	\end{array}\) %
	\(\begin{array}{cc}	1 	&		\\ 	\Upsilon_2	&	1	\end{array}\) %
	\(\begin{array}{cc}	 X_{\wp'''_1}	&	\\  	&	X_{\wp'''_2}           \end{array}\) 
	\right] \\%
	& = X_{\alpha_1} + X_{\alpha_2} + \hat\Upsilon_2 \hat\Delta_2 \(  X_{\alpha_3} - X_{\alpha_4} - X_{\alpha_5} + X_{\alpha_6}  \)\,,
\end{split}
\ee
where the negative signs arise from from spectral networks rules, see \cite[eq. (4.1)]{Gaiotto:2012rg}.\footnote{These signs count the winding of the tangent vector to each path $\alpha_i$ mod $2$. 
In our example, they show up in $\alpha_4, \alpha_5$ as can be seen from Figure \ref{fig:su2nf4-cycles}.}
The cycles $\alpha_1\dots\alpha_6$ are shown in Figure \ref{fig:su2nf4-cycles}, a tedious but simple inspection of each of them shows that
\be
\begin{array}{ll}
	\alpha_1 = \frac{1}{2} \(\gamma_3 + \gamma_5 - \gamma_1 -\gamma_4\) \,,&
	\alpha_2 = \frac{1}{2} \( \gamma_1 +\gamma_4 -\gamma_3 - \gamma_5 \) \,,\\
	\alpha_3 = -\frac{1}{2} \(\gamma_1+\gamma_4 +\gamma_3+\gamma_5\) \,,&
	\alpha_4 = \frac{1}{2} \(2\gamma_2 + \gamma_3 + \gamma_5 - \gamma_1 -\gamma_4\)\,,\\
	\alpha_5 = \frac{1}{2} \( 2\gamma_2 + \gamma_1 +\gamma_4 -\gamma_3 - \gamma_5 \)\,,\qquad  &
	\alpha_6 = \frac{1}{2}\( 4\gamma_2 + \gamma_1 + \gamma_4 +\gamma_3+\gamma_5 \) \,.
\end{array}
\ee
Plugging these into (\ref{eq:su2nf4-F}) we find 
\be\label{eq:su2nf4-hol}
\begin{split}
	\Tr F(\wp_{bd},\bCG)  
	& = X_{-\frac{1}{2}(\gamma_1+\gamma_3+\gamma_4+\gamma_5)} \\
	& \times \(
	1 + X_{\gamma_1} + X_{\gamma_5} + X_{\gamma_1+\gamma_5} + X_{\gamma_1+\gamma_4+\gamma_5} + X_{\gamma_1+\gamma_3+\gamma_5} + X_{\gamma_1+\gamma_3+\gamma_4+\gamma_5} 
	\)\,.
\end{split}
\ee
This expression matches with previous computations, see \cite[eq. (10.44)]{Gaiotto:2010be}.

Next we would like to check the action of the mapping class group on $\Tr F(\wp_{bd},\bCG)$. 
Acting with a positive Dehn half-twist around the A-cycle turns $\wp_{bd}$ into $\wp_{bc}$ shown in Figure \ref{fig:su2nf4-line}.
It is now important to recall the subtlety involved in Figure \ref{fig:su2nf4-DehnA}: the sequence of moves depicted there corresponds to the \emph{inverse} half-twist.
In order to get the positive half twist we act on the holonomy with the inverse of (\ref{eq:cluster-dehn-A}), that is
\be\label{eq:positive-half-twist}
	\kappa_{D_A} \equiv    \mu_3 \circ \mu_4 \circ  \kappa_r^{-1}\,.
\ee
This map acts on cluster variables as follows
\be\label{eq:cluster-dehn-A-positive}
\begin{split}
 y_1'& = {y_1 y_2 y_6}{\left(1+y_2\right)^{-1} \left(1+y_6\right)^{-1}} \,,\quad
 y_2'= y_3\left(1+y_2\right)  \left(1+y_6\right) \,,\quad
 y_3'= {y_6^{-1}} \,,\\
 y_4'& = {y_2^{-1}} \,,\quad
 y_5'= y_2 y_5 y_6 \left(1+y_2\right)^{-1} \left(1+y_6\right)^{-1} \,,\quad
 y_6'= y_4 \left(1+y_2\right)  \left(1+y_6\right)\,. 
\end{split}
\ee
Using the map (\ref{eq:rho-map}) allows to rewrite the holonomy in terms of cluster variables, then acting with the coordinate transformation (\ref{eq:cluster-dehn-A-positive}) gives
\be
\begin{split}
	\kappa_{D_A} \circ \rho \(\Tr F(\wp_{bd} ,\bCG) \)& =  %
	\kappa_{D_A} \( 
	\frac{1}{\sqrt{y_1 y_3 y_4 y_5}} 
	+ \sqrt{\frac{y_1}{ y_3 y_4 y_5}}
	+ \sqrt{\frac{y_5}{y_1  y_3 y_4}}
	+ \sqrt{\frac{y_1 y_5}{ y_3 y_4}}\right.
	\\
	&
	\left.\qquad \quad + \sqrt{\frac{y_1 y_4 y_5}{ y_3}}
	+ \sqrt{\frac{y_1 y_3 y_5}{ y_4}}
	+ \sqrt{{y_1 y_3 y_4 y_5}}
	\) \\
	& = \frac{1}{\sqrt{y_1 y_2 y_5 y_6}} 
	+ \sqrt{\frac{y_6}{y_1 y_2 y_5}}
	+ \sqrt{\frac{y_2}{y_1  y_5 y_6}}
	+ \sqrt{\frac{y_2 y_6}{ y_1 y_5}}
	\\
	&
	\qquad+ \sqrt{\frac{y_1 y_2 y_6}{ y_5}}
	+ \sqrt{\frac{y_2 y_5 y_6}{ y_1}}
	+ \sqrt{{y_1 y_2 y_5 y_6}}
	\\
	& =\rho \(\Tr F(\wp_{bc} ,\bCG) \)\,.
\end{split}
\ee
The last expression clearly coincides with the holonomy around $\wp_{bc}$: this is evident by simply acting on (\ref{eq:su2nf4-hol}) with the discrete symmetries of the BPS graph, viewed as a tetrahedron.

As a further check, we may use the symmetries of the BPS graph to write down the holonomy for a cycle $\wp_{cd}$ around punctures $c$ and $d$
\be
\begin{split}
	\Tr F(\wp_{cd},\bCG)  %
	& = X_{-\frac{1}{2}(\gamma_2+\gamma_3+\gamma_4+\gamma_6)} \\
	& \times \(
	1 + X_{\gamma_3} + X_{\gamma_4} + X_{\gamma_3+\gamma_4} + X_{\gamma_2+\gamma_3+\gamma_4} + X_{\gamma_3+\gamma_4+\gamma_6} + X_{\gamma_2+\gamma_3+\gamma_4+\gamma_6} 
	\)\,.
\end{split}
\ee
It is straightforward to check that (\ref{eq:cluster-dehn-A-positive}) leaves this holonomy invariant
\be
\begin{split}
	\kappa_{D_A} \circ \rho \(\Tr F(\wp_{cd} ,\bCG) \)
	& =\rho \(\Tr F(\wp_{cd} ,\bCG) \)\,,
\end{split}
\ee
as expected from the fact that the twist around the $A$-cycle leaves $\wp_{cd}$ invariant.

\section*{Acknowledgements}
We are grateful to Mauricio Romo for very helpful discussions and for collaboration during part of this project.
The work of D.G. was supported by Samsung Science and Technology Foundation under Project Number SSTBA1402-08.
The work of P.L. is supported by the grants ``Geometry and Physics'' and
``Exact Results in Gauge and String Theories'' from the Knut and Alice Wallenberg Foundation.
P.L. thanks the Kavli IPMU and the organizers of the Simons Summer Workshop for hospitality during completion of this work.
The work of M.Y. is partially supported by WPI program (MEXT, Japan), by JSPS Program No.\ R2603, by JSPS KAKENHI Grant No.\ 15K17634, by JSPS-NRF Joint Research Project, and by NSF under Grant No.\ PHY-1125915. 
M.Y. thanks the KITP, UCSB for hospitality during the final stages of this work.

\appendix

\section{\texorpdfstring{Cluster mutations and $\CK$-walls}{Cluster mutations and K-walls}}
\label{app:contravariance}

In this section we give a proof of the identity (\ref{eq:contravariance}). This statement is a version of the known statement in the literature, namely
the decomposition of the cluster transformation for the cluster variables into the so-called monomial part and the 
automorphism part \cite{FockGoncharovHigher}. We nevertheless present here the self-contained proof for completeness.

To begin with, suppose $\kappa$ consists of a single flip on edge $e_k$ of a BPS graph $\CG$. 
The corresponding resolved spectral network undergoes a single $\CK$-wall jump, which results in the following transformation on the $X_\gamma$ variables
\be
	\CK_{\gamma'_k}^{-1}(X_\gamma) =  X_\gamma (1+ X_{\gamma'_k})^{-\langle \gamma'_k,\gamma\rangle} =  X_\gamma (1+ X_{-\gamma_k})^{\langle \gamma_k,\gamma\rangle} \;.
\ee
Note the appearance of $\gamma'_k$ as opposed to $\gamma_k$, this is because the $\CK$-wall jump involves the edge of the new BPS graph \emph{after} the flip, see Figures \ref{fig:flip-resolved} and \ref{fig:K-wall}.
Now we would like to compare this coordinate transformation for $X_\gamma$ with the cluster transformation on the variables of the dual quiver, which is given in (\ref{eq:yi-mutation}).
Using the map $\rho$ introduced in (\ref{eq:rho-map}), the action of a single flip is, respectively, on the cluster variables and on $X_\gamma$
\be\begin{split}
	\mu_k\circ \rho (X_{\gamma_i}) & = \mu_k (y_i) 
	= y_i y_k^{[-b_{ik}]_+ } \(1+ y^{-1}_k\)^{-b_{ik}} \,,\\
	\rho\circ \CK_{\gamma'_k}(X_{\kappa(\gamma_i)}) & = \rho\(X_{\gamma'_i} (1+ X_{\gamma'_k})^{\langle\gamma'_k,\gamma_i\rangle}\) \\
	& = \rho\(X_{\gamma_i + \gamma_k \left[\langle \gamma_i,\gamma_k\rangle\right]_+  } (1+ X_{-\gamma_k})^{-\langle\gamma_k,\gamma_i\rangle}\) \\
	& = y_i y_k^{[-b_{ik}]_+ } \(1+ y^{-1}_k\)^{-b_{ik}}\,,
\end{split}
\ee
where we used the following reformulation of the jump of basis charges (\ref{eq:h-jump})
\be
	\gamma_i '  =\gamma_i + \gamma_k \left[\langle \gamma_i,\gamma_k\rangle\right]_+ \,,\qquad\qquad
	\gamma_k' = -\gamma_k\,,
\ee
compatibly with (\ref{eq:bij-jump}) through the relation $b_{ij} = -\langle\gamma_i,\gamma_j\rangle$.
This proves (\ref{eq:contravariance}) in the case when $\kappa$ is a single flip.

Next we consider two flips. The first one, denoted by $\lambda$, is performed on edge $e_\ell$, then a second one $\kappa$ is performed on edge $e_k$.
Let $\gamma_i, y_i$ be the charges and cluster variables associated to the quiver nodes before acting with $\lambda,\kappa$. Likewise, let $\gamma_i', y_i'$ be the charges and cluster variables after the flip $\lambda$, and $\gamma_i'', y_i''$ the charges and cluster variables after $\kappa\circ\lambda$.
Applying definitions, 
the cluster variables $y_i$ with $i\neq k,\ell$ transform as
\be\label{eq:composed-cluster-tmns}
\begin{split}
	\kappa\circ\lambda(y_i)  %
	& = y_i' (y_k')^{[-b'_{ik}]_+} (1+(y'_k)^{-1})^{-b_{ik}'} \\
	& = y_i y_\ell^{[-b_{i\ell}]_+} (1+y^{-1}_\ell)^{-b_{i\ell}}  \\
	& \qquad\times y_k^{[-b_{ik}']_+} y_\ell^{[-b_{ik}']_+[-b_{k\ell}]_+} (1+y_\ell^{-1})^{-b_{k\ell}[-b'_{ik}]_+}\\
	&\qquad \times \(1 + y_k^{-1} y_\ell^{-[-b_{k\ell}]_+} (1+ y_\ell^{-1})^{b_{k\ell}}\)^{-b_{ik}'} \,.
\end{split}
\ee
The $X_\gamma$ transform as
\be\label{eq:temp-yi}
	\CK_{\gamma_\ell'}\CK_{\gamma_k''}(X_{\gamma''_i}) = X_{\gamma_i''} (1+X_{\gamma_\ell'})^{\langle\gamma_\ell',\gamma_i''\rangle} (1+ X_{\gamma_k''}(1+ X_{\gamma_\ell'})^{\langle\gamma_\ell',\gamma_k''\rangle})^{\langle\gamma_\ell'',\gamma_i''\rangle}\,.
\ee
The charges are related as follows
\be
\begin{split}
	\gamma_i' & = \gamma_i +[\langle\gamma_i,\gamma_\ell\rangle]_+ \gamma_\ell \,,\qquad
	\gamma_k' = \gamma_k + [\langle\gamma_k,\gamma_\ell\rangle]_+\gamma_\ell \,,\\
	\gamma_i'' & = 
	\gamma_i 
	+ [\langle\gamma_i,\gamma_\ell\rangle]_+ \gamma_\ell
	+ [\langle\gamma'_i,\gamma'_k\rangle]_+ \gamma_k 
	+  [\langle\gamma_i',\gamma_k'\rangle]_+  [\langle\gamma_k,\gamma_\ell\rangle]_+ \gamma_\ell\,.
\end{split}
\ee
Therefore we can match (\ref{eq:composed-cluster-tmns}) and (\ref{eq:temp-yi}), piece by piece as follows
\be
	\rho(X_{\gamma_i''})= y_i'' = y_i y_\ell^{[-b_{i\ell}]_+} y_\ell^{ [-b_{k\ell}]_+[-b'_{ik}]_+}   y_k^{[-b'_{ik}]_+}\,,
\ee
\be
	\rho\(1+ X_{\gamma_\ell'}\)^{\langle\gamma_\ell',\gamma_i''\rangle} 
	= (1+ y_\ell^{-1})^{-b_{i\ell} - b_{k\ell} [-b'_{ik}]_+} \,,
\ee
and the last big parenthesis in (\ref{eq:temp-yi}) by noting
\be
	\langle\gamma_k'',\gamma_i''\rangle = -b_{ik}'' = -b_{ik}' \,, 
	\quad
	\rho\(X_{\gamma_{k}''}\) = y_{k}^{-1} y_\ell^{-[-b_{k\ell}]_+} \,,
	\quad 
	\rho\((1+X_{\gamma_\ell'})^{\langle\gamma_\ell',\gamma_k''\rangle} \)= (1+ y_\ell^{-1})^{b_{k\ell}}\,.
\ee
This proves (\ref{eq:contravariance}) for the case when $\kappa$ consists of two flips. By induction, this proof can be extended to the case of $n$ consecutive flips.

Last, we must deal with relabelings of edges of $\CG$, by our conventions these are performed after all flips and cooties.
We consider $\kappa=\kappa_r\circ\kappa_s$, with a permutation $\kappa_r$. For illustration, we take $\kappa_s = \mu_k$ to be a single flip on edge $e_k$.
$\kappa_r$ acts on the basis charges associated with elementary webs  simply as a permutation
\be
	\kappa_r : \ \gamma_i' \mapsto \gamma_i'' = \gamma'_{\kappa_r(i)}\,,
\ee
where $\kappa_r(i)$ denotes the image of node $i$ of the dual BPS quiver (cf. (\ref{eq:21-quiver-mcg-action})).
On cluster variables the overall action is
\be
	\kappa_r\circ\kappa_s (y_i) 
	= \kappa_r(y_i') 
	= y_{\kappa_r(i)}' 
	= y_{\kappa_r(i)} y_k^{[-b_{{\kappa_r(i)},k}]_+} (1+y^{-1}_k)^{-b_{{\kappa_r(i)},k}}\,.
\ee
On the $X_\gamma$, the ordering is reversed, so $\kappa_r$ must be applied before all $\CK$-wall transformation
\be
\begin{split}
	\CK_{\gamma'_k} \( X_{\kappa_r\circ\kappa_s(\gamma_i)} \)& = X_{\gamma'_{\kappa_r(i)}} (1+ X_{\gamma_k'})^{\langle\gamma'_{k},\gamma'_{\kappa_r(i)}\rangle} \\
	& \mathop{\to}^{\rho}_{} y_{\kappa_r(i)} y_k^{[-b_{\kappa_r(i),k}]_+} (1+y_k^{-1})^{-b_{\kappa_r(i),k}}\,.
\end{split}
\ee
The two clearly coincide, proving that (\ref{eq:contravariance}) behaves well also under composition of mutations with a permutation.

\section{BPS graphs on once-punctured torus with a partial puncture}\label{app.degeneration}

In this appendix
we present a heuristic procedure to derive the BPS graph of the once punctured torus $C_{1,1}$, with a partial puncture of the type $[k,1\ldots,1]$. 
Related construction has appeared previously in \cite{Xie:2012dw} in more combinatorial context.

Let us start by recalling the puncture reduction proposed in \cite{Gabella:2017hpz}. For our purposes it will suffice to review the 
case of the $A_{N-1}$ theory on a sphere with 3 punctures $C_{0,3}$, with two full punctures and one partial puncture of the type $[k,1\ldots,1]$. 

\begin{figure}[h]
\centering
\includegraphics[width=0.7\textwidth]{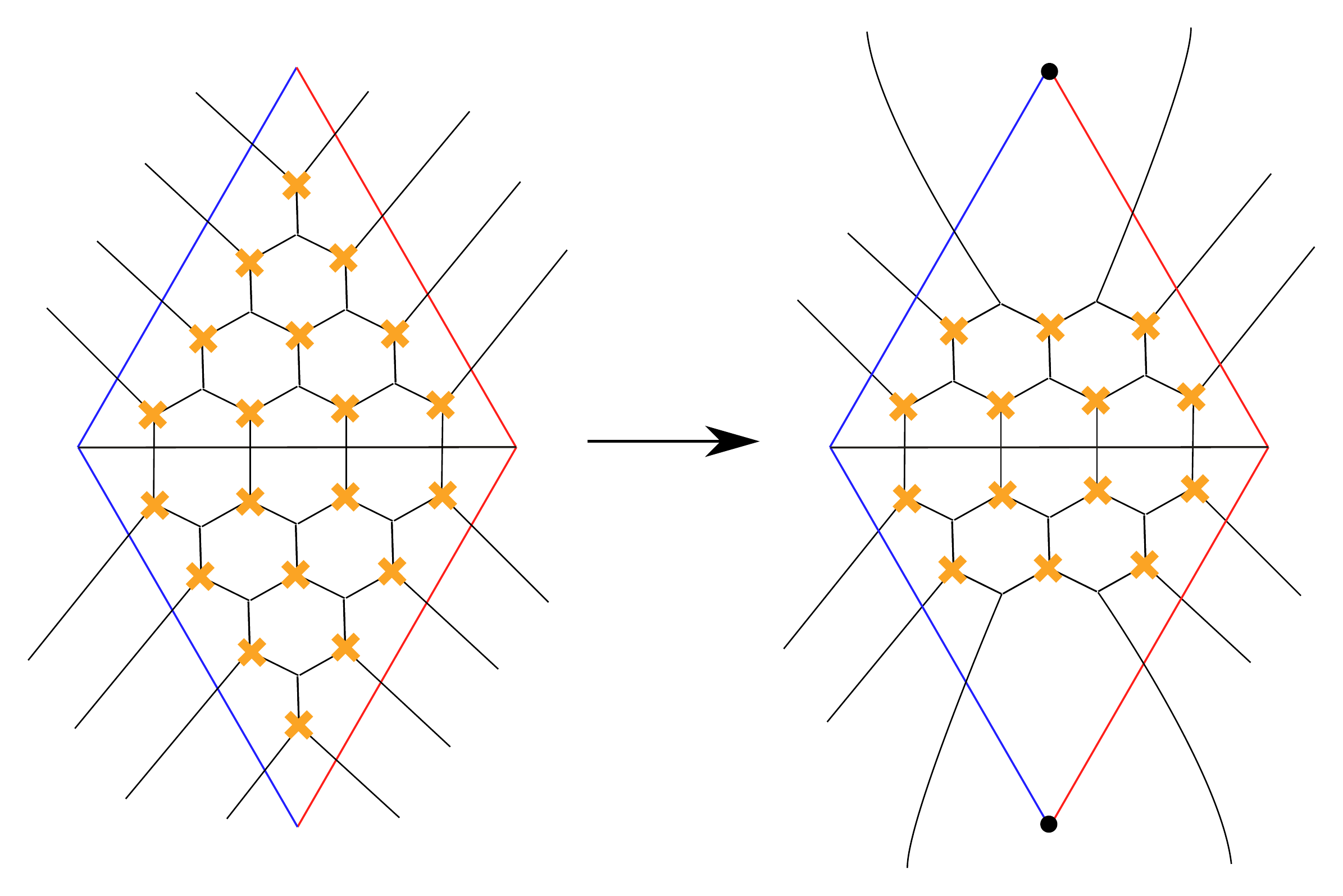}
\caption{Reduction from the maximal puncture $[1,1,1,1,1]$ to a partial puncture $[3,1,1]$, for $A_4$ theory
on a three punctured sphere $C_{0,3}$. The  highlighted puncture degenerated to a partial puncture, while
the other two punctures stay maximal. The two triangles are glued according to the colors of the edges.
}
\label{fig:punct_reduction}
\end{figure}

We first start with the case where all the punctures are maximal. The sphere $C_{0,3}$ can be triangulated by two ideal triangles,
and each triangle can be triangulated with $N(N-1)/2$ branch points, as in the left of Figure \ref{fig:punct_reduction} (shown there for the case $N=5$, where only one of the 
two triangles are shown).

The reduction from the maximal puncture to a $[k,1\ldots,1]$ is done in two steps: one starts from the BPS graph corresponding to an ideal triangulation of $C_{0,3}$ with three full punctures, and then remove $k(k-1)/2$ branch points around the puncture we want to reduce. This is illustrated in Figure \ref{fig:punct_reduction}, for the case $N=5$ and $k=3$.

\begin{figure}[h]
\centering
\includegraphics[width=2in,trim=0 0 0 10]{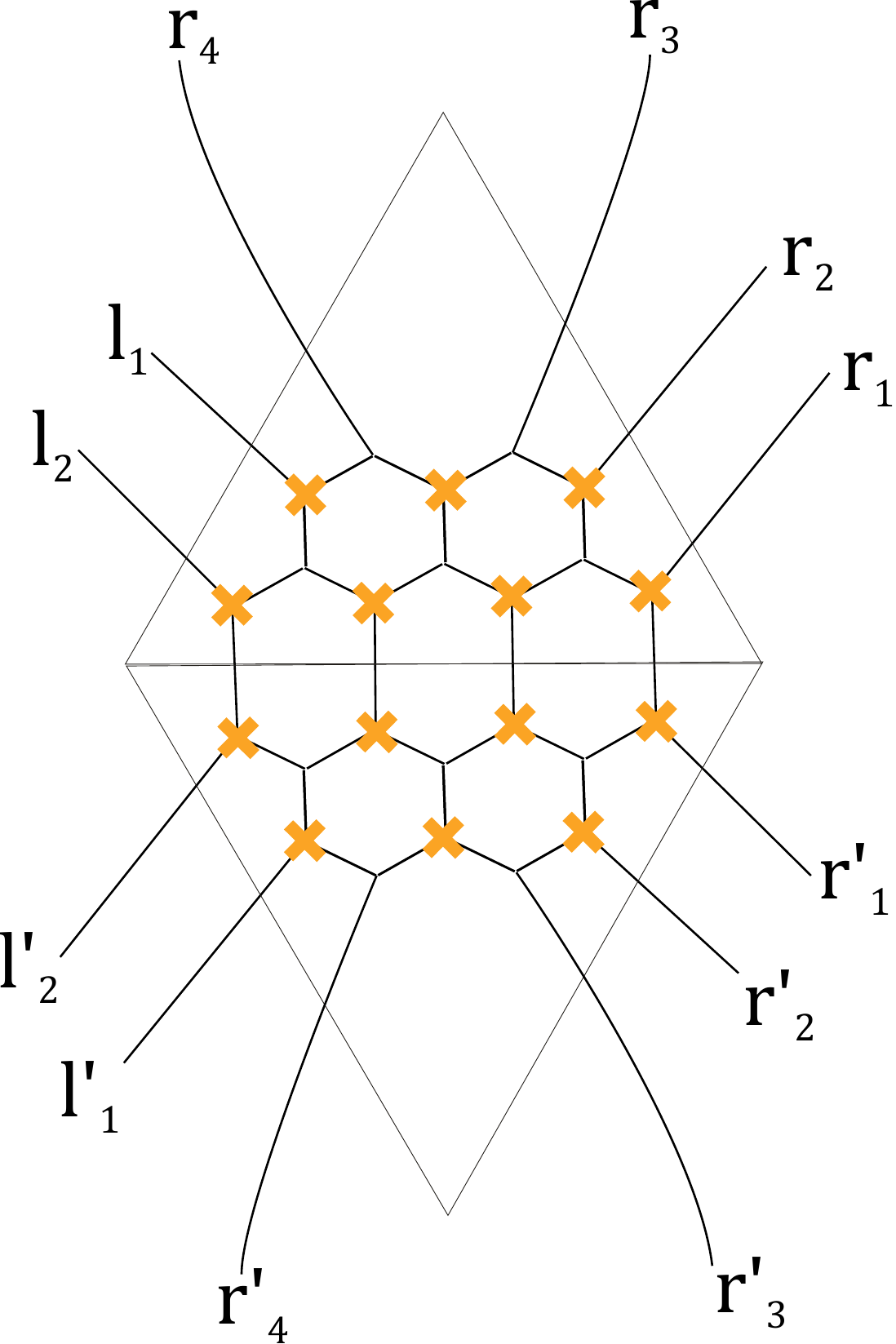}
\caption{Ideal triangulation of $C_{0,3}$ with a $[3,1,1]$ puncture and two full punctures with labels on the external edges.}
\label{fig:reds3}
\end{figure}

Now, in order to apply this to obtain the torus with one partial puncture, we start from an ideal triangulation of $C_{0,3}$ with two full punctures and one reduced puncture of type $[k,1,\ldots,1]$ as above. 
We write the triangulated surface in a plane, and starting from one of the full punctures, we label the external edges as $r_{1},\ldots, r_{k+1}$, $l_{1},\ldots,l_{k-1},l'_{k-1},\ldots,l'_{1}$ and $r'_{k+1},\ldots,r'_{1}$, see Figure \ref{fig:reds3} for an example. The identifications for the $C_{0,3}$ triangulations are
\begin{eqnarray}\label{idents3}
r_{i}\approx r'_{i} \quad (i=1,\ldots,k+1) \,, \qquad l'_{j}\approx l'_{j}\quad (j=1,\ldots,k-1) \,.
\end{eqnarray}

\begin{figure}[h]
\centering
\includegraphics[width=6in, trim=0 0 0 60]{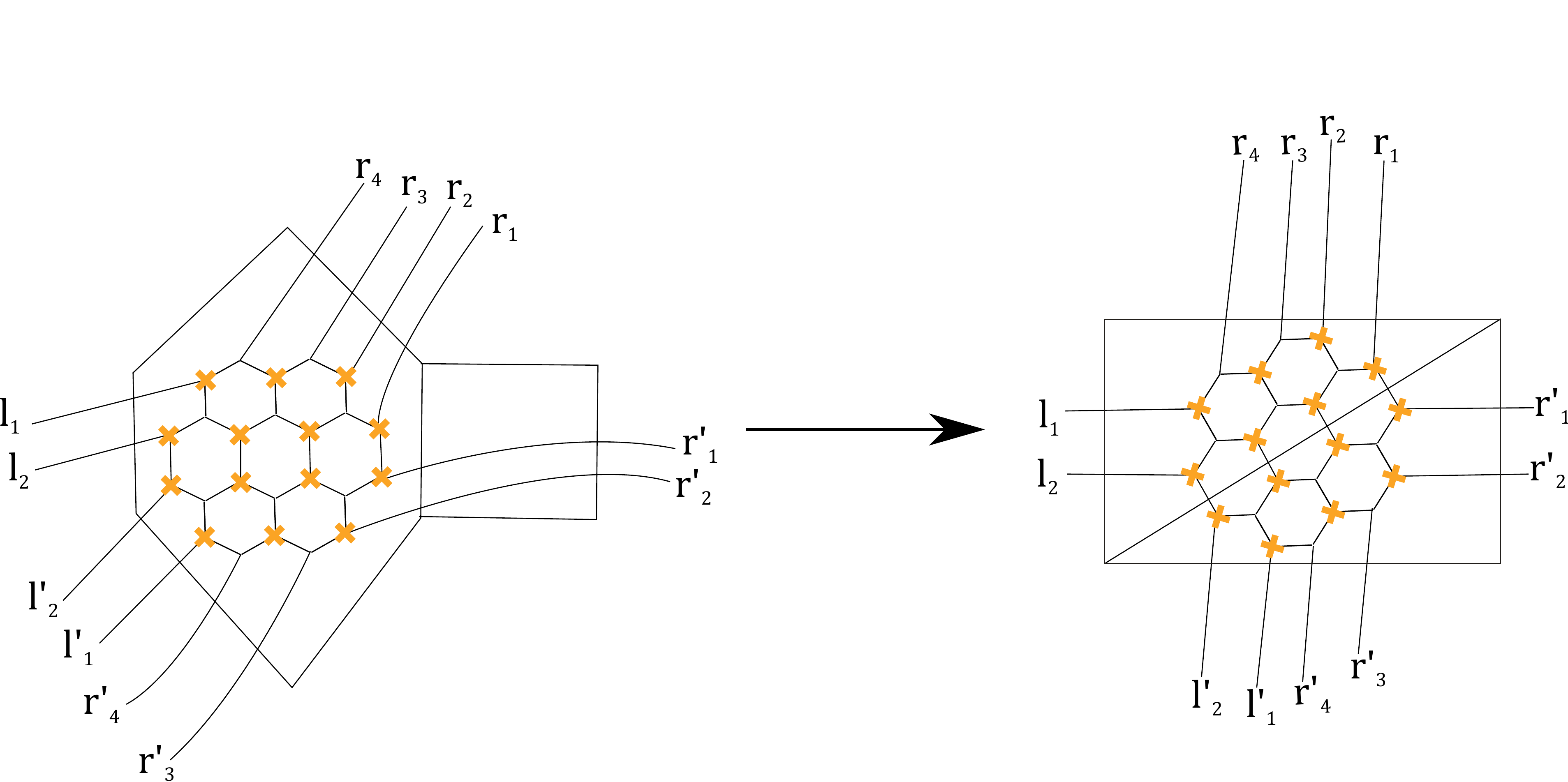}
\caption{Adding a handle to $C_{0,3}$ by gluing two full punctures.}
\label{fig:reds3torus}
\end{figure}

As pointed out in \cite{Gabella:2017hpz} $\lambda^{(i)}-\lambda^{(j)}$ remains finite at a partial puncture, for certain pairs of  sheets $i,j$.\footnote{This holds for the pairs of sheets $i,j$ whose corresponding eigenvalues $m_i, m_j$ coincide, in the residue matrix of the pole of the Higgs field of the Hitchin sysytem.} This implies that an $ij$ trajectory of the spectral network behaves as it would at a generic point over $C$, i.e. it doesn't feel the presence of a puncture at all, and in particular this implies that we can move an $ij$ edge across the puncture. 
We undo the identifications (\ref{idents3}) and we move the edges labeled
$r_{k+1}$ and $r'_{}$ across the $[k,1,\ldots,1]$ puncture.
Then we make the following identifications:
\begin{align}\label{identstorusred}
r_{j+2}\approx l'_{j} \,, \quad r'_{j} \approx l_{j}  \quad (j=1 ,\ldots,k-1)\,,
 \qquad r_{j}\approx r'_{k+1-j}\quad ( j=1,2) \,, 
\end{align}
see Figure \ref{fig:reds3torus}. Our proposal is that this is the BPS graph of $C_{1,1}$ with a puncture of type $[k,1,\ldots,1]$.

\cleardoublepage

\bibliography{mcg-bib}{}

\providecommand{\href}[2]{#2}\begingroup\raggedright\begin{thebibliography}{10}

\bibitem{Gaiotto:2009we}
D.~Gaiotto, {\it {$\mathcal{N}$=2 dualities}},  {\em JHEP} {\bf 1208} (2012)
  034, [\href{http://arxiv.org/abs/0904.2715}{{\tt arXiv:0904.2715}}].

\bibitem{Alday:2009aq}
L.~F. Alday, D.~Gaiotto, and Y.~Tachikawa, {\it {Liouville Correlation
  Functions from Four-dimensional Gauge Theories}},  {\em Lett.Math.Phys.} {\bf
  91} (2010) 167--197, [\href{http://arxiv.org/abs/0906.3219}{{\tt
  arXiv:0906.3219}}].

\bibitem{Drukker:2009tz}
N.~Drukker, D.~R. Morrison, and T.~Okuda, {\it {Loop operators and S-duality
  from curves on Riemann surfaces}},  {\em JHEP} {\bf 09} (2009) 031,
  [\href{http://arxiv.org/abs/0907.2593}{{\tt arXiv:0907.2593}}].

\bibitem{Terashima:2011qi}
Y.~Terashima and M.~Yamazaki, {\it {SL(2,R) Chern-Simons, Liouville, and Gauge
  Theory on Duality Walls}},  {\em JHEP} {\bf 1108} (2011) 135,
  [\href{http://arxiv.org/abs/1103.5748}{{\tt arXiv:1103.5748}}].

\bibitem{Dimofte:2011ju}
T.~Dimofte, D.~Gaiotto, and S.~Gukov, {\it {Gauge Theories Labelled by
  Three-Manifolds}},  {\em Commun. Math. Phys.} {\bf 325} (2014) 367--419,
  [\href{http://arxiv.org/abs/1108.4389}{{\tt arXiv:1108.4389}}].

\bibitem{Gaiotto:2010be}
D.~Gaiotto, G.~W. Moore, and A.~Neitzke, {\it {Framed BPS States}},  {\em
  Adv.Theor.Math.Phys.} {\bf 17} (2013) 241--397,
  [\href{http://arxiv.org/abs/1006.0146}{{\tt arXiv:1006.0146}}].

\bibitem{Cordova:2013bza}
C.~C\'ordova and A.~Neitzke, {\it {Line Defects, Tropicalization, and
  Multi-Centered Quiver Quantum Mechanics}},  {\em JHEP} {\bf 09} (2014) 099,
  [\href{http://arxiv.org/abs/1308.6829}{{\tt arXiv:1308.6829}}].

\bibitem{Gaiotto:2012rg}
D.~Gaiotto, G.~W. Moore, and A.~Neitzke, {\it {Spectral networks}},  {\em
  Annales Henri Poincare} {\bf 14} (2013) 1643--1731,
  [\href{http://arxiv.org/abs/1204.4824}{{\tt arXiv:1204.4824}}].

\bibitem{Gabella:2017hpz}
M.~Gabella, P.~Longhi, C.~Y. Park, and M.~Yamazaki, {\it {BPS Graphs: From
  Spectral Networks to BPS Quivers}},  {\em JHEP} {\bf 07} (2017) 032,
  [\href{http://arxiv.org/abs/1704.04204}{{\tt arXiv:1704.04204}}].

\bibitem{Longhi:2016wtv}
P.~Longhi, {\it {Wall-Crossing Invariants from Spectral Networks}},
  \href{http://arxiv.org/abs/1611.00150}{{\tt arXiv:1611.00150}}.

\bibitem{FockGoncharovHigher}
V.~Fock and A.~Goncharov, {\it Moduli spaces of local systems and higher
  {T}eichm\"uller theory},  {\em Publ. Math. Inst. Hautes \'Etudes Sci.}
  (2006), no.~103 1--211.

\bibitem{FST1}
S.~Fomin, M.~Shapiro, and D.~Thurston, {\it Cluster algebras and triangulated
  surfaces. {I}. {C}luster complexes},  {\em Acta Math.} {\bf 201} (2008),
  no.~1 83--146.

\bibitem{Kapustin:2005py}
A.~Kapustin, {\it {Wilson-'t Hooft operators in four-dimensional gauge theories
  and S-duality}},  {\em Phys. Rev.} {\bf D74} (2006) 025005,
  [\href{http://arxiv.org/abs/hep-th/0501015}{{\tt hep-th/0501015}}].

\bibitem{Alday:2009fs}
L.~F. Alday, D.~Gaiotto, S.~Gukov, Y.~Tachikawa, and H.~Verlinde, {\it {Loop
  and surface operators in N=2 gauge theory and Liouville modular geometry}},
  {\em JHEP} {\bf 01} (2010) 113, [\href{http://arxiv.org/abs/0909.0945}{{\tt
  arXiv:0909.0945}}].

\bibitem{Drukker:2009id}
N.~Drukker, J.~Gomis, T.~Okuda, and J.~Teschner, {\it {Gauge Theory Loop
  Operators and Liouville Theory}},  {\em JHEP} {\bf 02} (2010) 057,
  [\href{http://arxiv.org/abs/0909.1105}{{\tt arXiv:0909.1105}}].

\bibitem{Terashima:2013fg}
Y.~Terashima and M.~Yamazaki, {\it {3d N=2 Theories from Cluster Algebras}},
  {\em PTEP} {\bf 2014} (2014) 023B01,
  [\href{http://arxiv.org/abs/1301.5902}{{\tt arXiv:1301.5902}}].

\bibitem{Gang:2015wya}
D.~Gang, N.~Kim, M.~Romo, and M.~Yamazaki, {\it {Aspects of Defects in 3d-3d
  Correspondence}},  {\em JHEP} {\bf 10} (2016) 062,
  [\href{http://arxiv.org/abs/1510.05011}{{\tt arXiv:1510.05011}}].

\bibitem{Dimofte:2011py}
T.~Dimofte, D.~Gaiotto, and S.~Gukov, {\it {3-Manifolds and 3d Indices}},  {\em
  Adv. Theor. Math. Phys.} {\bf 17} (2013), no.~5 975--1076,
  [\href{http://arxiv.org/abs/1112.5179}{{\tt arXiv:1112.5179}}].

\bibitem{Lee:2013ida}
S.~Lee and M.~Yamazaki, {\it {3d Chern-Simons Theory from M5-branes}},  {\em
  JHEP} {\bf 12} (2013) 035, [\href{http://arxiv.org/abs/1305.2429}{{\tt
  arXiv:1305.2429}}].

\bibitem{Yagi:2013fda}
J.~Yagi, {\it {3d TQFT from 6d SCFT}},  {\em JHEP} {\bf 08} (2013) 017,
  [\href{http://arxiv.org/abs/1305.0291}{{\tt arXiv:1305.0291}}].

\bibitem{Cordova:2013cea}
C.~Cordova and D.~L. Jafferis, {\it {Complex Chern-Simons from M5-branes on the
  Squashed Three-Sphere}},  \href{http://arxiv.org/abs/1305.2891}{{\tt
  arXiv:1305.2891}}.

\bibitem{Romo:2017ruv}
M.~Romo, {\it {Cluster Partition Function and Invariants of 3-manifolds}},
  2017.
\newblock \href{http://arxiv.org/abs/1704.00933}{{\tt arXiv:1704.00933}}.

\bibitem{future}
D.~Gang, P.~Longhi, M.~Romo, and M.~Yamazaki, {\it {To Appear}}, .

\bibitem{Hollands:2013qza}
L.~Hollands and A.~Neitzke, {\it {Spectral Networks and Fenchel-Nielsen
  Coordinates}},  {\em Lett. Math. Phys.} {\bf 106} (2016), no.~6 811--877,
  [\href{http://arxiv.org/abs/1312.2979}{{\tt arXiv:1312.2979}}].

\bibitem{Gabella:2017hxq}
M.~Gabella, {\it {BPS spectra from BPS graphs}},
  \href{http://arxiv.org/abs/1710.08449}{{\tt arXiv:1710.08449}}.

\bibitem{Alim:2011ae}
M.~Alim, S.~Cecotti, C.~Cordova, S.~Espahbodi, A.~Rastogi, and C.~Vafa, {\it
  {BPS Quivers and Spectra of Complete N=2 Quantum Field Theories}},  {\em
  Commun. Math. Phys.} {\bf 323} (2013) 1185--1227,
  [\href{http://arxiv.org/abs/1109.4941}{{\tt arXiv:1109.4941}}].

\bibitem{Alim:2011kw}
M.~Alim, S.~Cecotti, C.~Cordova, S.~Espahbodi, A.~Rastogi, and C.~Vafa, {\it
  {$\mathcal{N} = 2$ quantum field theories and their BPS quivers}},  {\em Adv.
  Theor. Math. Phys.} {\bf 18} (2014), no.~1 27--127,
  [\href{http://arxiv.org/abs/1112.3984}{{\tt arXiv:1112.3984}}].

\bibitem{FominZelevinsky1}
S.~Fomin and A.~Zelevinsky, {\it Cluster algebras {I}: {Foundations}},  {\em J.
  Amer. Math. Soc.} {\bf 15} (2002), no.~2 497--529.

\bibitem{FominZelevinsky4}
S.~Fomin and A.~Zelevinsky, {\it Cluster algebras. {IV}. {C}oefficients},  {\em
  Compos. Math.} {\bf 143} (2007), no.~1 112--164.

\bibitem{Nagao:2011aa}
K.~Nagao, Y.~Terashima, and M.~Yamazaki, {\it {Hyperbolic 3-manifolds and
  Cluster Algebras}},  \href{http://arxiv.org/abs/1112.3106}{{\tt
  arXiv:1112.3106}}. To appear in Nagoya J. Math.

\bibitem{Terashima:2011xe}
Y.~Terashima and M.~Yamazaki, {\it {Semiclassical Analysis of the 3d/3d
  Relation}},  {\em Phys. Rev.} {\bf D88} (2013), no.~2 026011,
  [\href{http://arxiv.org/abs/1106.3066}{{\tt arXiv:1106.3066}}].

\bibitem{KitayamaTerashima}
T.~Kitayama and Y.~Terashima, {\it Torsion functions on moduli spaces in view
  of the cluster algebra},  {\em Geom. Dedicata} {\bf 175} (2015) 125--143.

\bibitem{2016arXiv160705228G}
A.~B. {Goncharov}, {\it {Ideal webs, moduli spaces of local systems, and 3d
  Calabi-Yau categories}},  {\em ArXiv e-prints} (July, 2016)
  [\href{http://arxiv.org/abs/1607.05228}{{\tt arXiv:1607.05228}}].

\bibitem{Xie:2012dw}
D.~Xie, {\it {Network, Cluster coordinates and N=2 theory I}},
  \href{http://arxiv.org/abs/1203.4573}{{\tt arXiv:1203.4573}}.

\bibitem{Coman:2015lna}
I.~Coman, M.~Gabella, and J.~Teschner, {\it {Line operators in theories of
  class $\mathcal{S}$, quantized moduli space of flat connections, and Toda
  field theory}},  {\em JHEP} {\bf 10} (2015) 143,
  [\href{http://arxiv.org/abs/1505.05898}{{\tt arXiv:1505.05898}}].

\bibitem{Moore:2014jfa}
G.~W. Moore, A.~B. Royston, and D.~Van~den Bleeken, {\it {Parameter counting
  for singular monopoles on $\mathbb R^3$}},  {\em JHEP} {\bf 10} (2014) 142,
  [\href{http://arxiv.org/abs/1404.5616}{{\tt arXiv:1404.5616}}].

\bibitem{Moore:2014gua}
G.~W. Moore, A.~B. Royston, and D.~Van~den Bleeken, {\it {Brane bending and
  monopole moduli}},  {\em JHEP} {\bf 10} (2014) 157,
  [\href{http://arxiv.org/abs/1404.7158}{{\tt arXiv:1404.7158}}].

\bibitem{Moore:2015szp}
G.~W. Moore, A.~B. Royston, and D.~Van~den Bleeken, {\it {Semiclassical framed
  BPS states}},  {\em JHEP} {\bf 07} (2016) 071,
  [\href{http://arxiv.org/abs/1512.08924}{{\tt arXiv:1512.08924}}].

\bibitem{Moore:2015qyu}
G.~W. Moore, A.~B. Royston, and D.~Van~den Bleeken, {\it {$L^2$-Kernels Of
  Dirac-Type Operators On Monopole Moduli Spaces}},
  \href{http://arxiv.org/abs/1512.08923}{{\tt arXiv:1512.08923}}.

\bibitem{Brennan:2016znk}
T.~D. Brennan and G.~W. Moore, {\it {A note on the semiclassical formulation of
  BPS states in four-dimensional $N=$ 2 theories}},  {\em PTEP} {\bf 2016}
  (2016), no.~12 12C110, [\href{http://arxiv.org/abs/1610.00697}{{\tt
  arXiv:1610.00697}}].

\bibitem{Dimofte:2011jd}
T.~Dimofte and S.~Gukov, {\it {Chern-Simons Theory and S-duality}},  {\em JHEP}
  {\bf 05} (2013) 109, [\href{http://arxiv.org/abs/1106.4550}{{\tt
  arXiv:1106.4550}}].

\bibitem{Gang:2013sqa}
D.~Gang, E.~Koh, S.~Lee, and J.~Park, {\it {Superconformal Index and 3d-3d
  Correspondence for Mapping Cylinder/Torus}},  {\em JHEP} {\bf 01} (2014) 063,
  [\href{http://arxiv.org/abs/1305.0937}{{\tt arXiv:1305.0937}}].

\bibitem{Gaiotto:2009hg}
D.~Gaiotto, G.~W. Moore, and A.~Neitzke, {\it Wall-crossing, {H}itchin systems,
  and the {WKB} approximation},  {\em Adv. Math.} {\bf 234} (2013) 239--403,
  [\href{http://arxiv.org/abs/0907.3987}{{\tt arXiv:0907.3987}}].

\end{thebibliography}\endgroup
\bibliographystyle{JHEP}

\end{document}